\documentclass[11pt]{article}
\usepackage{cite}
\usepackage{amssymb}
\usepackage{amsfonts}
\usepackage{mathrsfs}
\usepackage{graphicx}
\usepackage{amsmath}
\usepackage{color}
\usepackage{amsthm}
\usepackage{epstopdf}
\usepackage{booktabs}
\usepackage{pifont,bm}
\usepackage{tikz}
\usepackage{enumerate}
\usepackage{threeparttable}
\usepackage{diagbox}

\theoremstyle{definition}

\makeatletter
\def\@biblabel#1{[#1]}
\makeatother

\makeatletter \@addtoreset{equation}{section}

\begin{document}
%\begin{CJK*}{GBK}{song}

\begin{titlepage}
\title{\bf{$N$-double poles solutions for nonlocal Hirota equation with nonzero boundary conditions using Riemann-Hilbert method and PINN algorithm%via the Fokas  method
\footnote{%Project supported by the Fundamental Research Fund for the Central Universities under the Grant no. 2017XKQY101.\protect\\
Corresponding authors.\protect\\
\hspace*{3ex} E-mail addresses: ychen@sei.ecnu.edu.cn (Y. Chen)}
}}
\author{Wei-Qi Peng$^{a}$, Yong Chen$^{a,b,*}$\\
%%%%%%%%%%%%%%%%%%%%%%%%%%%%%%%%%%%%%%%%%%%%%%%%%%%%%%%%%%%%%%%%%%%%%%%%%%%%%%%%%%%%%%%%%
%%%%%              以下两行为作者单位
%%%%%%%%%%%%%%%%%%%%%%%%%%%%%%%%%%%%%%%%%%%%%%%%%%%%%%%%%%%%%%%%%%%%%%%%%%%%%%%%%%%%%%%%%
\small \emph{$^{a}$School of Mathematical Sciences, Shanghai Key Laboratory of PMMP} \\
\small \emph{East China Normal University, Shanghai, 200241, China} \\
\small \emph{$^{b}$College of Mathematics and Systems Science, Shandong University }\\
\small \emph{of Science and Technology, Qingdao, 266590, China} \\
\date{}}
\thispagestyle{empty}
\end{titlepage}
\maketitle

\vspace{-0.5cm}
\begin{center}
\rule{15cm}{1pt}\vspace{0.3cm}

\parbox{15cm}{\small
{\bf Abstract}\\
\hspace{0.5cm}  In this paper, we systematically investigate the nonlocal Hirota equation with nonzero boundary conditions via Riemann-Hilbert method and multi-layer physics-informed neural networks algorithm. Starting from the Lax pair of nonzero nonlocal Hirota equation, we first give out the Jost function, scattering matrix, their symmetry and asymptotic behavior. Then, the Riemann-Hilbert problem with nonzero boundary conditions are constructed and the precise formulaes of $N$-double poles solutions  and $N$-simple poles solutions are written by determinants. Different from the local Hirota equation, the symmetry of scattering data for nonlocal Hirota equation is completely different, which results in disparate discrete spectral distribution. In particular, it could be more complicated and difficult to obtain the symmetry of scattering data under the circumstance of double poles. Besides, we also analyse the asymptotic state of one-double poles solution as $t\rightarrow \infty$. Whereafter, the multi-layer physics-informed neural networks algorithm is applied to research the data-driven soliton solutions of the nonzero nonlocal Hirota equation by using the training data obtained from the Riemann-Hilbert method. Most strikingly, the integrable nonlocal equation is firstly solved via multi-layer physics-informed neural networks algorithm. As we all know, the nonlocal equations contain the $\mathcal{PT}$ symmetry $\mathcal{P}:x\rightarrow -x,$ or $\mathcal{T}:t\rightarrow -t,$ which are different with local ones. Adding the nonlocal term
into the NN, we can successfully solve the integrable nonlocal Hirota equation by multi-layer physics-informed neural networks algorithm.  The numerical results indicate the algorithm  can well recover the data-driven soliton solutions of the integrable nonlocal equation. Noteworthily, the inverse problems of the
integrable nonlocal equation are discussed for the first time through applying the physics-informed neural networks algorithm to
discover the parameters of the equation in terms of its soliton solution.
}

\vspace{0.5cm}
\parbox{15cm}{\small{

\vspace{0.3cm} \emph{Key words:} Nonlocal Hirota equation; Riemann-Hilbert method; Nonzero boundary
conditions; Simple/double poles solutions; Physics-informed neural networks algorithm.\\

 } }
\end{center}
\vspace{0.3cm} \rule{15cm}{1pt} \vspace{0.2cm}

\section{Introduction}
It is well-known that nonlinear partial differential equation (PDE) plays a prominent  role in the subjects of mathematical physics, such as fluid mechanics, nonlinear optics, ocean communication, etc. In the history of soliton theory,  finding exact solutions of  integrable PDE is still a crucial issue. During the past several decades, in order to solve nonlinear evolution equations, more and more methods and techniques have been presented including inverse scattering transformation (IST) \cite{Peng-3}, Hirota bilinear method \cite{Peng-1}, Darboux transformation (DT) \cite{Peng-2},  etc.  Among them, IST method is one of the most powerful techniques to analyse the Cauchy problems of integrable nonlinear evolution equations. In 1967, Gardner et al. came up with this method first to handle the KdV equation with initial value problem\cite{Peng-5}.
The critical point of classic IST method is to solve the Gel'fand-Levitan-Marchenko(GLM) integral
equations. Later on, Zakharov et al. simplifies the IST method by developing a Riemann-Hilbert(RH) formulation to replace the GLM equation\cite{Peng-6}. After that, the RH formulation has been constantly applied to numerous integrable equations with zero boundary conditions(ZBCs), such as coupled  nonlinear Schr\"{o}dinger (NLS) equation, derivative Schr\"{o}dinger equation,
Sasa-Satsuma equation, and so on\cite{Peng-7,Peng-jia1,Peng-8,Peng-9,Peng-10,Peng-12,Peng-13,Peng-14}. Recently, the RH method was also applied to construct the soliton solutions  \cite{Peng-15,Peng-16,Peng-17,Peng-18,Peng-19} and the rogue waves\cite{Peng-20,Peng-21,Peng-11} for the integrable equations with nonzero boundary conditions(NZBCs).

With the rapid development of machine learning methods, deep learning has become a powerful tool for solving PDE. Recently,  a new neural network (NN), named  physics-informed neural networks (PINN) \cite{Peng-4},  was recently proposed, which can be used to accomplish the high-dimensional network tasks with fewer data sets\cite{Peng-4}. Besides, it is verified that PINN is very effective for solving and inverting equations controlled by mathematical physical systems. Later on, using PINN method to generate data-driven solutions and reveal the dynamic behavior of nonlinear partial differential equations with physical constraints  has attracted extensive attention and raised a hot wave of research.
Very recently, in terms of PINN  method, Chen group constructed data-driven soliton solution, high-order
breather wave, rogue wave,  rogue periodic wave for several types of nonlinear evolution equations including KdV equation, NLS equation, KN equation, KP equation, Manakov system,  etc. \cite{Peng-CNSN47,Peng-CNSN50,Peng-CNSN51,Peng-CNSN52,Peng-CNSN53,Peng-CNSN54}. Particularly, based on conserved quantities, a two-stage PINN method is used to derive some data-driven localized wave solutions\cite{Peng-CNSN55}.  Besides, other scholars have also got some important results on data-driven solutions for the defocusing NLS equation with a potential, high-order NLS equation and coupled NLS equation\cite{Peng-CNSN56,Peng-CNSN57,dai-chos,lin-pla}.  Thus it is natural to consider how to apply the PINN in the nonlocal integrable system. In this paper, we devote to construct a new PINN for solving the nonlocal integrable  system via adding the nonlocal term into the NN, and the data-driven solution will be simulated through using the new PINN.

The parity-time ($\mathcal{PT}$) symmetry, first proposed in quantum mechanics by Bender and his coworkers, plays a important  role in characterizing the wave propagation for the NLS equation in mechanical systems, optical fibers and magnetism\cite{Rao7,Rao8,Rao9}. Up to now, the research of $\mathcal{PT}$ system has made great progress in both theory and application\cite{Rao10,Rao11,Rao12}.  For instance,  Ablowitz and Musslimani first introduced the $\mathcal{PT}$ symmetry to the well-known
AKNS system by raising a nonlocal (also named reverse-space) NLS equation\cite{Rao13}
\begin{align}\label{1}
u_{t}\left(x, t\right)-iu_{xx}\left(x, t\right)\pm 2i Q\left(x, t\right)u \left(x, t\right)=0,\quad Q\left(x, t\right)=u\left(x, t\right)u^{\ast}\left(-x, t\right),
\end{align}
which includes the $\mathcal{PT}$ symmetric potential $Q(x, t)$ and satisfies the $\mathcal{PT}$ symmetry restriction $Q(x, t)=Q^{\ast}(-x, t)$. The $\ast$ represents complex conjugate, and $Q(x, t)$ denotes electric field envelope and complex refractive index distribution of beam. Since then,  Fokas extended the nonlocal NLS equation to a higher dimensional form\cite{Rao14}.
Moreover, other nonlocal integrable equations were also investigated, such as nonlocal
Davey-Stewartson equations, nonlocal modified KdV equation, nonlocal sine-Gordon equation, nonlocal derivative NLS equation , etc. \cite{zhang-33,zhang-34,zhang-35,zhang-36,zhang-37,zhang-38,zhang-39}. It's worth mentioning that
several nonlocal integrable equations with NZBCs have been researched through developing the IST\cite{zhang-40,zhang-42}. After that, soliton solutions and high-order pole
solutions for both focusing and defocusing nonlocal  mKdV equations with NZBCs at infinity have been presented by a systematical inverse scattering transform \cite{Zhang-PD, Tian-yang}.

Recently, starting with
\begin{align}\label{1.1}
iq_{t}+\kappa\left[q_{xx}-2q^{2}r\right]+i\beta\left[q_{xxx}-6qrq_{x}\right]=0,\notag\\
ir_{t}-\kappa\left[r_{xx}-2qr^{2}\right]+i\beta\left[r_{xxx}-6qrr_{x}\right]=0,
\end{align}
an integrable nonlocal (also named reverse-space-time) Hirota equation
\begin{align}\label{2}
q_{t}+\delta\left[q_{xx}-2q^{2}q^{\ast}(-x, -t)\right]+\beta\left[q_{xxx}-6qq^{\ast}(-x,-t)q_{x}\right]=0,
\end{align}
was reduced in Ref.\cite{Liu} for $r=q^{\ast}(-x, -t), \kappa=i\delta$.  Of which $\delta, \beta\in \mathbb{R}$ are arbitrary parameters. Eq.\eqref{2} changes into the reverse-space-time nonlocal complex mKdV equation at the case of $\delta=0, \beta=1$.  Besides, for the $\mathcal{PT}$ symmetric case $q^{\ast}(-x, -t)=q(x, t)$, Eq.\eqref{2} becomes the usual Hirota equation. The explicit multi-soliton solutions were generated for Eq.\eqref{2}  by employing Hirota's direct method as well as Darboux-Crum transformations\cite{Liu}.  In what follows, we would like to write out its Lax pair firstly based on the early results\cite{Liu}, given by
\begin{gather}
\Psi_{x}=U\Psi,\qquad U\equiv\left(\begin{array}{cc}
    -i\lambda  &  q(x,t) \\
    q^{\ast}(-x, -t)  &  i\lambda \\
\end{array}\right),\notag\\
\Psi_{t}=V\Psi,\qquad V\equiv \left(\begin{array}{cc}
    \tilde{A}(x, t)  &  \tilde{B}(x, t) \\
    \tilde{C}(x, t)  &  -\tilde{A}(x, t) \\
\end{array}\right),\notag\\
\tilde{A}=\delta qq^{\ast}(-x, -t)+2\delta\lambda^{2}+\beta\left[q^{\ast}(-x, -t)q_{x}-qq^{\ast}_{x}(-x, -t)-4i\lambda^{3}-2i\lambda qq^{\ast}(-x, -t)\right],\notag\\
\tilde{B}=-\delta q_{x}+2i\delta\lambda q+\beta\left[2q^{2}q^{\ast}(-x, -t)-q_{xx}+2i\lambda q_{x}+4\lambda^{2}q\right],\notag\\
\tilde{C}=\delta q^{\ast}_{x}(-x, -t)+2i\delta\lambda q^{\ast}(-x, -t)\notag\\
+\beta\left[2qq^{\ast}(-x, -t)^{2}-q^{\ast}_{xx}(-x, -t)-2i\lambda q^{\ast}_{x}(-x, -t)+4\lambda^{2}q^{\ast}(-x, -t)\right].\label{6}
\end{gather}

To the best of our knowledge, although some people have studied the nonlocal Hirota equation with ZBCs\cite{Tian-LY1,Tian-LY2,Tian-LY3},
the studies of the nonlocal Hirota Eq.\eqref{2} with NZBCs have been rarely reported via using RH method, as well as the RH method with NZBCs is more complicated than one with ZBCs. In what follows, we would systematically  consider the matrix RH problem for the nonlocal Hirota Eq.\eqref{2} with following NZBCs at infinity
\begin{align}\label{7}
\lim_{x\rightarrow \pm \infty}q(x, t)=q_{\pm}e^{2\delta q_{0}^{2}t},
\end{align}
where $|q_{\pm}|=q_{0}>0$, and $q_{+}=q_{-}$ are constant. Different from previous work about the usual Hirota equation, the nonlocal Hirota equation involves different symmetry reductions and disparate discrete spectra distribution.  Especially, for the case of double poles, symmetry could be more complicated. It is worth mentioning that the asymptotic state of one-double poles solution is also given out as $t\rightarrow \infty$. Furthermore, we also find that PINN deep learning for solving nonlocal integrable equation has not been researched so far.  Therefore, in this paper, we also commit to propose a scheme to solve integrable nonlocal equation in terms of PINN method. As an example, we choose the integrable nonzero nonlocal Hirota equation to highlight the ability of our strategy to handle the integrable nonlocal equation, and successfully predict the
data-driven soliton solution of the nonzero nonlocal Hirota equation. On the other hand, we also discuss the data-driven parameter discovery
for the nonzero nonlocal Hirota equation.

The outline of this paper is organized as follows: In section 2, we carry out the spectral analysis for nonlocal Hirota equation under NZBCs, and derive the corresponding Lax pair, Jost solutions, scattering matrix and their symmetry reductions. In section 3 and 4, through analysing the inverse scattering problem for nonlocal Hirota equation under NZBCs, we establish the RH problem for the nonlocal Hirota equation under NZBCs and obtain the explicit $N$-simple poles solution and $N$-double poles solution for the  reflectionless coefficients under NZBCs. In section 5, the data-driven soliton solutions and data-driven
parameter discovery for the nonlocal Hirota equation are researched via PINN method. Finally, some summaries are given in the last section.

\section{Spectral analysis for nonlocal Hirota equation under NZBCs}

\subsection{Lax pair and Jost solutions}
In order to facilitate the later calculation, we deal with Lax pair \eqref{6} with the boundary condition
\eqref{7} at the beginning. Let's make an appropriate transformation
\begin{align}\label{8}
q\rightarrow q e^{2\delta q_{0}^{2}t},\qquad \Psi\rightarrow e^{\delta q_{0}^{2}t\sigma_{3}}\Psi,
\end{align}
where $\sigma_{3}$ is  one of the following Pauli matrices
\begin{align}\label{9}
\sigma_{1}=\left(\begin{array}{cc}
    0  &  1\\
    1 &  0\\
\end{array}\right), \quad \sigma_{2}=\left(\begin{array}{cc}
    0  &  -i\\
    i &  0\\
\end{array}\right), \quad \sigma_{3}=\left(\begin{array}{cc}
    1  &  0\\
    0 &  -1\\
\end{array}\right).
\end{align}
Then, the nonlocal Hirota equation \eqref{2} changes into
\begin{align}\label{10}
q_{t}+\delta\left[q_{xx}-2q^{2}q^{\ast}(-x, -t)+2q_{0}^{2}q\right]+\beta\left[q_{xxx}-6qq^{\ast}(-x,-t)q_{x}\right]=0,
\end{align}
with following boundary
\begin{align}\label{11}
\lim_{x\rightarrow \pm \infty}q(x, t)=q_{\pm},
\end{align}
which admits the Lax pair
\begin{align}\label{12}
&\Psi_{x}=U\Psi,\qquad U\equiv\left(\begin{array}{cc}
    -i\lambda  &  q(x,t) \\
    q^{\ast}(-x, -t)  &  i\lambda \\
\end{array}\right),\notag\\
&\Psi_{t}=V\Psi,\qquad V\equiv \left(\begin{array}{cc}
    \tilde{A}(x, t)-\delta q_{0}^{2}  &  \tilde{B}(x, t) \\
    \tilde{C}(x, t)  &  -\tilde{A}(x, t)+\delta q_{0}^{2} \\
\end{array}\right).
\end{align}
As $x \rightarrow \pm \infty$, the Lax pair \eqref{12} under the boundary \eqref{11} becomes
\begin{align}\label{13}
\Psi_{x}=U_{\pm}\Psi=(-i\lambda\sigma_{3}+Q_{\pm})\Psi, \qquad \Psi_{t}=V_{\pm}\Psi=\left[2i\delta\lambda+\beta(4\lambda^{2}+2q_{0}^{2})\right]U_{\pm}\Psi,
\end{align}
where
\begin{align}\label{14}
Q_{\pm}=\left(\begin{array}{cc}
    0  &  q_{\pm}\\
    q_{\mp}^{\ast} &  0\\
\end{array}\right).
\end{align}
The system \eqref{13} can be solved by
\begin{align}\label{15}
\Psi_{\pm}^{bg}(x, t, \lambda)=\left\{
\begin{array}{lr}
Y_{\pm}(\lambda)e^{-i\theta(x, t, \lambda)\sigma_{3}}, \quad \lambda\neq\pm q_{0}\\
I+\left(x+\left(2i\delta\lambda+\beta(4\lambda^{2}+2q_{0}^{2})\right)t\right)U_{\pm}(\lambda), \quad \lambda=\pm q_{0},
\end{array}
\right.
\end{align}
where
\begin{align}\label{16}
Y_{\pm}=\left(\begin{array}{cc}
    1  &  -\frac{iq_{\pm}}{\lambda+k}\\
    \frac{iq_{\mp}^{\ast}}{\lambda+k} &  1\\
\end{array}\right),\quad \theta(x, t, k)=k[x+\left(2i\delta\lambda+\beta(4\lambda^{2}+2q_{0}^{2})\right)t], \quad k^{2}=\lambda^{2}-q_{0}^{2}.
\end{align}
In order to analyse the scattering problem on a standard $z$-plane instead of the two-sheeted
Riemann surface, we introduce a uniformization variable $z=k+\lambda$, given by
\begin{align}\label{17}
k=\frac{1}{2}(z-\frac{q_{0}^{2}}{z}), \qquad \lambda=\frac{1}{2}(z+\frac{q_{0}^{2}}{z}).
\end{align}

Setting $D_{+}$, $D_{-}$ and $\Sigma$ on $z$-plane as $D_{\pm}= \left\{z\in \mathbb{C} | \mbox{Im}z\gtrless 0\right\}, \Sigma=\mathbb{R}, $ the Jost solutions $\Psi_{\pm}(x, t, z)$ are defined by
\begin{align}\label{18}
\Psi_{\pm}(x, t, z)=Y_{\pm}e^{-i\theta(x, t, z)\sigma_{3}}+o(1), \quad z\in \Sigma, \quad \mbox{as} \quad x\rightarrow \pm \infty.
\end{align}
Using the variable transformation
\begin{align}\label{19}
\mu_{\pm}(x, t, z)=\Psi_{\pm}(x, t, z)e^{i\theta(x, t, z)\sigma_{3}},
\end{align}
we get the modified Jost solutions $\mu_{\pm}(x, t, z)\rightarrow Y_{\pm}(z)$ as $x\rightarrow \pm\infty$, which satisfies the following Volterra integral equations
\begin{align}\label{20}
\mu_{\pm}(x,t,z)=\left\{
\begin{array}{lr}
Y_{\pm}+\int_{\pm\infty}^{x}Y_{\pm}e^{-ik(x-y)\hat{\sigma}_{3}}
\left[Y_{\pm}^{-1}\Delta U_{\pm}(y, t)\mu_{\pm}(y,t,z)\right]dy, \quad z\neq\pm q_{0},\\
Y_{\pm}+\int_{\pm\infty}^{x}
\left[I+(x-y)U_{\pm}(z)\right]\Delta U_{\pm}(y, t)\mu_{\pm}(y,t,z)dy, \quad z=\pm q_{0},
\end{array}
\right.
\end{align}
where $\Delta U_{\pm}=U-U_{\pm}$.

\noindent \textbf{Proposition 2.1.} \emph{
Assume $q-q_{\pm}\in L^{1} (\mathbb{R}^{\pm})$, then $\mu_{\pm}(x,t,z)$ defined in Eq.\eqref{19} uniquely solve the Volterra integral equation \eqref{20} in $\Sigma_{0}=\Sigma\setminus\{\pm q_{0}\}$, and $\mu_{\pm}(x,t,z)$ yield:}

\emph{$\bullet$ $\mu_{-1}(x, t, z)$ and $\mu_{+2}(x, t, z)$ is analytical in $D_{+}$ and
continuous in $D_{+}\cup \Sigma_{0}$,}

\emph{$\bullet$ $\mu_{+1}(x, t, z)$ and $\mu_{-2}(x, t, z)$  is analytical in $D_{-}$ and
continuous in $D_{-}\cup \Sigma_{0}$,}

\emph{$\bullet$ $\mu_{\pm}(x,t,z)\rightarrow I$ \mbox{as}  $z\rightarrow \infty$,}

\emph{$\bullet$ $\mu_{\pm}(x,t,z) \rightarrow -\frac{i}{z}\sigma_{3}Q_{\pm}$ \mbox{as}  $z\rightarrow 0$,}

\emph{$\bullet$ $\det \mu_{\pm}(x, t, z)=\det Y_{\pm}=\gamma=1-\frac{q_{0}^{2}}{z^{2}}, \quad x, t\in \mathbb{R}, \quad z\in \Sigma_{0}$.}

\centerline{\begin{tikzpicture}[scale=1.8]
\path [fill=gray] (2.5,0) -- (0.5,0) to
(0.5,2) -- (2.5,2);
\path [fill=gray] (4.5,0) -- (2.5,0) to
(2.5,2) -- (4.5,2);
\draw[-][thick](0.5,0)--(0.75,0);
\draw[<-][thick](0.75,0)--(1,0);
\draw[-][thick](1,0)--(2,0);
\draw[<-][thick](2,0)--(2.5,0);
\draw[fill] (2.5,0) circle [radius=0.03];
\draw[->][thick](2.5,0)--(3,0);
\draw[->][thick](3,0)--(4,0);
\draw[-][thick](4,0)--(4.5,0)node[above]{$\mbox{Re}z$};
\draw[-][thick](2.5,2)node[right]{$\mbox{Im}z$}--(2.5,0);
\draw[-][thick](2.5,0)--(2.5,-2);
\draw[->][thick](2.5,-1.5)--(2.5,-0.5);
\draw[->][thick](2.5,-2)--(2.5,-1.5);
\draw[->][thick](2.5,1.5)--(2.5,0.5);
\draw[->][thick](2.5,2)--(2.5,1.5);
\draw[fill] (2.5,-0.1) node[right]{$0$};
\draw[fill] (1.5,0) circle [radius=0.03];
\draw[fill] (1.3,0) node[below]{$-q_{0}$};
\draw[fill] (3.5,0) circle [radius=0.03];
\draw[fill] (3.6,0) node[below]{$q_{0}$};
\draw[fill](3.8,1.5) circle [radius=0.03] node[right]{$z_{n}$};
\draw[fill] (1.2,1.5) circle [radius=0.03] node[left]{$-z^{*}_{n}$};
\draw[fill] (2,-0.5) circle [radius=0.03] node[right]{$-\frac{q^{2}_{0}}{z^{*}_{n}}$};
\draw[fill] (3,-0.5) circle [radius=0.03] node[left]{$\frac{q^{2}_{0}}{z_{n}}$};
\draw[-][thick](3.5,0) arc(0:360:1);
\draw[-][thick](3.5,0) arc(0:30:1);
\draw[-][thick](3.5,0) arc(0:150:1);
\draw[-][thick](3.5,0) arc(0:210:1);
\draw[-][thick](3.5,0) arc(0:330:1);
\end{tikzpicture}}
\noindent { \small \textbf{Figure 1.} (Color online) Distribution of the discrete spectrum and jumping curves for the RH problem on complex $z$-plane, Region $D_{+}=\left\{z\in \mathbb{C} | \mbox{Im}z> 0\right\}$ (gray region), region $D_{-}=\left\{z\in \mathbb{C} | \mbox{Im}z< 0\right\}$ (white region).}\\

Since the Jost solutions $\Psi_{\pm}(x, t, z)$ are the simultaneous solutions of Lax pair \eqref{12}, we can establish the linear relation by a scattering matrix $S(z)=(s_{i j} (z))_{2\times 2}$, given by
\begin{align}\label{21}
\Psi_{+}(x, t, z)=\Psi_{-}(x, t, z)S(z), \quad z\in \Sigma_{0}.
\end{align}
We can write the scattering coefficients into the form of Wronskians determinant
\begin{align}\label{22}
s_{11}(z)=\frac{Wr(\Psi_{+,1},\Psi_{-,2})}{\gamma(z)}, \quad s_{12}(z)=\frac{Wr(\Psi_{+,2},\Psi_{-,2})}{\gamma(z)},\notag\\
s_{21}(z)=\frac{Wr(\Psi_{-,1},\Psi_{+,1})}{\gamma(z)}, \quad s_{22}(z)=\frac{Wr(\Psi_{-,1},\Psi_{+,2})}{\gamma(z)}.
\end{align}

\noindent \textbf{Proposition 2.2.} \emph{ Suppose $q-q_{\pm}\in L^{1} (\mathbb{R}^{\pm})$, then
the scattering  matrix $S(z)$ has the following characteristics:
}

\emph{$\bullet$ $\det S(z)=1$ for $z\in \Sigma_{0}$,}

\emph{$\bullet$ $s_{22}(z)$ is analytical in $D_{+}$ and
continuous in $D_{+}\cup \Sigma_{0}$,}

\emph{$\bullet$ $s_{11}(z)$ is analytical in $D_{-}$ and
continuous in $D_{-}\cup \Sigma_{0}$,}

\emph{$\bullet$ $S(x,t,z)\rightarrow I$ \mbox{as}  $z\rightarrow \infty$,}

\emph{$\bullet$ $S(x,t,z) \rightarrow I$ \mbox{as}  $z\rightarrow 0$.}

\subsection{Symmetry reductions}
In this subsection, according to  the reduction conditions of the Lax pair on the complex $z$-plane, the symmetries of the Jost solutions $\Psi(x, t, z)$ and scattering matrix $S(z)$ are studied for the nonlocal Hirota equation with NZBCs. The details are as follows:

\noindent \textbf{Proposition 2.3.} \emph{The $X(x, t, z)$ and $T(x, t, z)$ in the Lax pair \eqref{12} meet the following
reduction conditions on $z$-plane:
}

\emph{$\bullet$ The first symmetry reduction}
\begin{align}\label{22.a}
U(x, t, z)=-\sigma_{2}U(-x, -t, -z^{\ast})^{\ast}\sigma_{2},\qquad V(x, t, z)=-\sigma_{2}V(-x, -t, -z^{\ast})^{\ast}\sigma_{2}.
\end{align}

\emph{$\bullet$ The second symmetry reduction}
\begin{align}\label{22.b}
U(x, t, z)=U(x, t, \frac{q_{0}^{2}}{z}), \qquad V(x, t, z)=V(x, t, \frac{q_{0}^{2}}{z}).
\end{align}

\noindent \textbf{Proposition 2.4.} \emph{The Jost solutions $\Psi(x, t, z)$ and scattering matrix $S(z)$ possess the following
reduction conditions on $z$-plane:
}

\emph{$\bullet$ The first symmetry reduction}
\begin{align}\label{22.1}
\Psi_{\pm}(x, t, z)=\sigma_{2}\Psi_{\mp}(-x, -t, -z^{\ast})^{\ast}\sigma_{2},\qquad S(z)=\sigma_{2}S^{\ast}(-z^{\ast})^{-1}\sigma_{2}.
\end{align}

\emph{$\bullet$ The second symmetry reduction}
\begin{align}\label{22.2}
\Psi_{\pm}(x, t, z)=-\frac{i}{z}\Psi_{\pm}(x, t, \frac{q_{0}^{2}}{z})\sigma_{3}Q_{\pm}, \qquad S(z)=Q_{-}^{-1}\sigma_{3}S(\frac{q_{0}^{2}}{z})\sigma_{3}Q_{+}.
\end{align}

\subsection{Discrete spectrum with simple poles}
In this subsection,
assuming that $s_{22}(z)$ has $N$ simple zeros $z_{n}$ ($n=1, 2, \cdots, N$) in $D_{+}\cap\left\{z\in\mathbb{ C}: \mbox{Re} z>0\right\}$,  then we have $s_{22}(z_{0})=0$ and $s'_{22}(z_{0})\neq 0$ in the situations of $z_{0}$ being the simple zero of $s_{22}(z)$. In terms of the symmetries of the scattering matrix, the corresponding discrete spectrum is summarized into(see Fig. 1)
\begin{align}\label{22.3}
\Upsilon=\left\{z_{n}, -z_{n}^{\ast}, \frac{q_{0}^{2}}{z_{n}}, -\frac{q_{0}^{2}}{z_{n}^{\ast}}\right\}_{n=1}^{N}.
\end{align}

Considering $s_{22}(z_{0})=0$ ($z_{0}\in \Upsilon\cap D_{+}$), it is not hard to find that $\Psi_{-1}(x, t, z_{0})$ and $\Psi_{+2}(x, t, z_{0})$
are linearly dependent.  Homoplastically, $\Psi_{+1}(x, t, z_{0})$ and $\Psi_{-2}(x, t, z_{0})$
are linearly dependent because of $s_{11}(z_{0})=0$ ($z_{0}\in \Upsilon\cap D_{-}$). Then, one has
\begin{align}\label{22.4}
\Psi_{+2}(x, t, z_{0})=b[z_{0}]\Psi_{-1}(x, t, z_{0}),\quad z_{0}\in \Upsilon\cap D_{+},\notag\\
\Psi_{+1}(x, t, z_{0})=b[z_{0}]\Psi_{-2}(x, t, z_{0}),\quad z_{0}\in \Upsilon\cap D_{-},
\end{align}
where $b[z_{0}]$ is a undetermined parameter. Therefore, we arrive at
\begin{align}\label{22.5}
&\mathop{\mbox{Res}}_{z=z_{0}}\left[\frac{\Psi_{+2}(x, t; z)}{s_{22}(z)}\right]=A[z_{0}]\Psi_{-1}(x, t; z_{0}), \quad z_{0}\in \Upsilon\cap D_{+},\notag\\
&\mathop{\mbox{Res}}_{z=z_{0}}\left[\frac{\Psi_{+1}(x, t; z)}{s_{11}(z)}\right]=A[z_{0}]\Psi_{-2}(x, t; z_{0}), \quad z_{0}\in \Upsilon\cap D_{-},
\end{align}
where
\begin{align}\label{22.6}
A[z_{0}]=\left\{
\begin{array}{lr}
\frac{b[z_{0}]}{s'_{22}(z_{0})}, \quad z_{0}\in \Upsilon\cap D_{+}\\
\\
\frac{b[z_{0}]}{s'_{11}(z_{0})}, \quad z_{0}\in \Upsilon\cap D_{-}.
\end{array}
\right.
\end{align}

\noindent \textbf{Proposition 2.5.} \emph{Two relations for $b[z_{0}]$, $s_{22}'(z_{0})$
 and $s_{11}'(z_{0})$ are got below:
}

\emph{$\bullet$ The first relation}
\begin{align}\label{22.7}
b[z_{0}]=-\frac{1}{b[-z_{0}^{\ast}]^{\ast}}, \quad s_{11}'(z_{0})=-s_{11}'(-z_{0}^{\ast})^{\ast}, \quad s_{22}'(z_{0})=-s_{22}'(-z_{0}^{\ast})^{\ast}.
\end{align}

\emph{$\bullet$ The second relation}
\begin{align}\label{22.8}
&b[z_{0}]=-\frac{q_{+}}{q_{-}^{\ast}}b[\frac{q_{0}^{2}}{z_{0}}],\quad s_{22}'(z_{0})=-\frac{q_{0}^{2}}{z_{0}^{2}}s_{11}'(\frac{q_{0}^{2}}{z_{0}}),\quad z_{0}\in D_{+},\notag\\
&b[z_{0}]=-\frac{q_{+}^{\ast}}{q_{-}}b[\frac{q_{0}^{2}}{z_{0}}],\quad s_{11}'(z_{0})=-\frac{q_{0}^{2}}{z_{0}^{2}}s_{22}'(\frac{q_{0}^{2}}{z_{0}}),\quad z_{0}\in D_{-}.
\end{align}

\section{Inverse scattering problem for nonlocal Hirota equation under NZBCs: simple poles}

\subsection{The Riemann-Hilbert problem under NZBCs}

In terms of the analyticity of Jost solutions $\mu_{\pm}(x, t, z)$ in Proposition 2.1, the following sectionally meromorphic matrices can be defined
\begin{align}\label{23}
M_{-}(x, t, z)=(\frac{\mu_{+1}}{s_{11}},\mu_{-2}),\qquad M_{+}(x, t, z)=(\mu_{-1},\frac{\mu_{+2}}{s_{22}}),
\end{align}
where $\pm$ represent analyticity in $D_{+}$ and $D_{-}$, respectively. Subsequently, a matrix RH problem is constructed:

\noindent \textbf{Riemann-Hilbert Problem}  \emph{
$M(x, t, z)$ solves the following RHP:
\begin{align}\label{24}
\left\{
\begin{array}{lr}
M(x, t, z)\ \mbox{is analytic in} \ \mathbb{C }\setminus \Sigma,\\
M_{-}(x, t, z)=M_{+}(x, t, z)(I-G(x, t, z)), \qquad z\in \Sigma,\\
M(x, t, z)\rightarrow I,\qquad z\rightarrow \infty,\\
M(x, t, z)\rightarrow -\frac{i}{z}\sigma_{3}Q_{-},\qquad z\rightarrow 0,
  \end{array}
\right.
\end{align}
of which the jump matrix $G(x, t, z)$ is
\begin{align}\label{25}
G=\left(\begin{array}{cc}
    \rho(z)\tilde{\rho}(z)  &  e^{-2i\theta(x, t, z)}\tilde{\rho}(z)\\
  -e^{2i\theta(x, t, z)}\rho(z) &  0\\
\end{array}\right),
\end{align}
where $\rho(z)=\frac{s_{21}(z)}{s_{11}(z)}, \tilde{\rho}(z)=\frac{s_{12}(z)}{s_{22}(z)}$.
}
Let
\begin{align}\label{26}
M(x, t, z)=I+\frac{1}{z}M^{(1)}(x, t; z)+O(\frac{1}{z^{2}}),\qquad z\rightarrow \infty,
\end{align}
then the potential $q(x, t)$ of the nonlocal Hirota equation \eqref{10} with NZBCs is given by
\begin{align}\label{27}
q(x, t)=iM_{12}^{(1)}(x, t, z)=i\lim_{z\rightarrow\infty}zM_{12}(x, t, z).
\end{align}
As a matter of convenience, we take $\hat{\zeta}_{n}=\frac{q_{0}^{2}}{\zeta_{n}}$
and define
\begin{align}\label{28}
\zeta_{n}=\left\{
\begin{array}{lr}
z_{n}, \qquad n=1, 2, \cdots, N\\
-z_{n-N}^{\ast}, \qquad n=N+1, N+2, \cdots, 2N\\
\end{array}
\right.
\end{align}
Then the residue of $M(x, t, z)$ is
\begin{align}\label{29}
&\mathop{\mbox{Res}}_{z=\zeta_{n}}M_{+}=\left(0, A[\zeta_{n}]e^{-2i\theta(x, t, \zeta_{n})}\mu_{-1}(x, t, \zeta_{n})\right),\notag\\
&\mathop{\mbox{Res}}_{z=\hat{\zeta}_{n}}M_{-}=\left(A[\hat{\zeta}_{n}]e^{2i\theta(x, t; \hat{\zeta}_{n})}\mu_{-2}(x, t, \hat{\zeta}_{n}), 0\right).
\end{align}

through subtracting out the residue and the asymptotic values as $z\rightarrow\infty, z\rightarrow 0$ from the original non-regular RHP, the following regular RHP can be obtained
\begin{gather}
M_{-}+\frac{i}{z}\sigma_{3}Q_{-}-I-\sum_{n=1}^{2N}\left[\frac{\mathop{\mbox{Res}}_{z=\zeta_{n}}M_{+}}{z-\zeta_{n}}
+\frac{\mathop{\mbox{Res}}_{z=\hat{\zeta}_{n}}M_{-}}{z-\hat{\zeta}_{n}}\right]=\notag\\
M_{+}+\frac{i}{z}\sigma_{3}Q_{-}-I-\sum_{n=1}^{2N}\left[\frac{\mathop{\mbox{Res}}_{z=\zeta_{n}}M_{+}}{z-\zeta_{n}}
+\frac{\mathop{\mbox{Res}}_{z=\hat{\zeta}_{n}}M_{-}}{z-\hat{\zeta}_{n}}\right]-M_{+}G.\label{30}
\end{gather}
which can be solved by the Plemelj's formulae, given by
\begin{gather}
M(x, t; z)=I-\frac{i}{z}\sigma_{3}Q_{-}+\frac{1}{2\pi i}\int_{\Sigma}\frac{M_{+}(x, t; \zeta)G(x, t; \zeta)}{\zeta-z}d\zeta\notag\\
+\sum_{n=1}^{2N}\left[\frac{\mathop{\mbox{Res}}_{z=\zeta_{n}}M_{+}}{z-\zeta_{n}}
+\frac{\mathop{\mbox{Res}}_{z=\hat{\zeta}_{n}}M_{-}}{z-\hat{\zeta}_{n}}\right].\label{31}
\end{gather}
where
\begin{gather}
\frac{\mathop{\mbox{Res}}_{z=\zeta_{n}}M_{+}}{z-\zeta_{n}}+\frac{\mathop{\mbox{Res}}_{z=\hat{\zeta}_{n}}M_{-}}{z-\hat{\zeta}_{n}}=
\left(\hat{C}_{n}(z)\mu_{-2}(\hat{\zeta}_{n}),
C_{n}(z)\mu_{-1}(\zeta_{n})\right),\label{32}
\end{gather}
and
\begin{align}\label{33}
C_{n}(z)=\frac{A[\zeta_{n}]e^{-2i\theta(\zeta_{n})}}{z-\zeta_{n}}, \quad
\hat{C}_{n}(z)=\frac{A[\hat{\zeta}_{n}]e^{2i\theta(\hat{\zeta}_{n})}}{z-\hat{\zeta}_{n}}.
\end{align}
Furthermore, according to \eqref{26}, one has
\begin{gather}\label{34}
M^{(1)}(x, t, z)=-\frac{1}{2\pi i}\int_{\Sigma}M_{+}(x, t; \zeta)G(x, t; \zeta)d\zeta -i\sigma_{3}Q_{-} \notag\\
+\sum_{n=1}^{2N}\left(A[\hat{\zeta}_{n}]e^{2i\theta(\hat{\zeta}_{n})}\mu_{-2}(\hat{\zeta}_{n}),
A[\zeta_{n}]e^{-2i\theta(\zeta_{n})}\mu_{-1}(\zeta_{n})\right).
\end{gather}
Therefore, the potential $q(x, t)$ with simple poles for the  nonlocal Hirota equation with NZBCs is given by
\begin{gather}
q(x, t)=iM_{12}^{(1)}=q_{-}+i\sum_{n=1}^{2N}A[\zeta_{n}]e^{-2i\theta(\zeta_{n})}\mu_{-11}(\zeta_{n})
-\frac{1}{2\pi}\int_{\Sigma}(M_{+}(x, t; \zeta)G(x, t; \zeta))_{12}d\zeta.\label{35}
\end{gather}

\subsection{Trace formulae and theta condition}
The scattering coefficients $s_{22}(z)$ and $s_{11}(z)$ respectively have simple zeros $\zeta_{n}$ and $\hat{\zeta}_{n}$, thus we can take
\begin{align}\label{Ajia1}
\beta^{+}(z)=s_{22}(z)\prod_{n=1}^{2N}\frac{z-\hat{\zeta}_{n}}{z-\zeta_{n}},\ \beta^{-}(z)=s_{11}(z)\prod_{n=1}^{2N}\frac{z-\zeta_{n}}{z-\hat{\zeta}_{n}},
\end{align}
that means $\beta^{+}(z)$ is analytic and has no zeros in $D_{+}$, and $\beta^{-}(z)$ is analytic and has no zeros in $D_{-}$.
Also, $\beta^{\pm}(z)\rightarrow o(1)$ as $z\rightarrow\infty$. According to the Plemelj's formulae, $\beta^{\pm}(z)$ can be written as
\begin{align}\label{Ajia2}
\log\beta^{\pm}(z)=\mp\frac{1}{2\pi i}\int_{\Sigma}\frac{\log(1-\rho\tilde{\rho})}{s-z}ds,\quad z\in D^{\pm}.
\end{align}
Using Eq.\eqref{Ajia1}, we derive the following trace formulae
\begin{align}\label{Ajia3}
&s_{22}(z)=\mbox{exp}\left[-\frac{1}{2\pi i}\int_{\Sigma}\frac{\log(1-\rho\tilde{\rho})}{s-z}ds\right] \prod_{n=1}^{2N}\frac{z-\zeta_{n}}{z-\hat{\zeta}_{n}},\notag\\
&s_{11}(z)=\mbox{exp}\left[\frac{1}{2\pi i}\int_{\Sigma}\frac{\log(1-\rho\tilde{\rho})}{s-z}ds\right] \prod_{n=1}^{2N}\frac{z-\hat{\zeta}_{n}}{z-\zeta_{n}}.
\end{align}
Let $z\rightarrow0$ in the first formula of Eq.\eqref{Ajia3}, one has
\begin{align}\label{Ajia4}
\mbox{exp}\left[\frac{i}{2\pi}\int_{\Sigma}\frac{\log(1-\rho\tilde{\rho})}{s}ds\right] \prod_{n=1}^{N}\frac{\mid z_{n}\mid^{4}}{q_{0}^{4}}=1.
\end{align}
In addition,  taking the derivative of the Eq.\eqref{Ajia3} with respect to $z$, we obtain $s'_{22}(\zeta_{j})$ and $s'_{11}(\hat{\zeta}_{j})$, given by
\begin{align}\label{Ajia5}
&s'_{22}(\zeta_{j})=\mbox{exp}\left[-\frac{1}{2\pi i}\int_{\Sigma}\frac{\log(1-\rho\tilde{\rho})}{s-\zeta_{j}}ds\right] \frac{\prod_{m\neq j}(\zeta_{j}-\zeta_{m})}{\prod_{m=1}^{2N}(\zeta_{j}-\hat{\zeta}_{m})},\notag\\
&s'_{11}(\hat{\zeta}_{j})=\mbox{exp}\left[\frac{1}{2\pi i}\int_{\Sigma}\frac{\log(1-\rho\tilde{\rho})}{s-\hat{\zeta}_{j}}ds\right] \frac{\prod_{m\neq j}(\hat{\zeta}_{j}-\hat{\zeta}_{m})}{\prod_{m=1}^{2N}(\hat{\zeta}_{j}-\zeta_{m})}.
\end{align}

\subsection{The simple poles soliton solution of nonlocal Hirota equation under NZBCs}
In the case of no reflection,  i. e. $\rho(z)=\tilde{\rho}(z)=0$,  we can get the soliton solution. We first take the second column of Eq.\eqref{31}
\begin{align}\label{36}
\mu_{-2}(z)=\left(\begin{array}{c}
    -\frac{i}{z}q_{-}\\
  1\\
\end{array}\right)+\sum_{n=1}^{2N}C_{n}(z)\mu_{-1}(\zeta_{n}).
\end{align}
In terms of the symmetric relation, we easily get
\begin{align}\label{37}
\mu_{-2}(z)=-\frac{iq_{-}}{z}\mu_{-1}(\frac{q_{0}^{2}}{z}).
\end{align}
Substituting Eq.\eqref{37}  into Eq. \eqref{36},  and setting $z=\hat{\zeta}_{j}, j=1, 2,\cdots,
2N$, we generate a $2N$ linear system:
\begin{align}\label{38}
\sum_{n=1}^{2N}\left(C_{n}(\hat{\zeta}_{j})+\frac{iq_{-}}{\hat{\zeta}_{j}}\delta_{j,n}\right)\mu_{-1}(\zeta_{n})
+\left(\begin{array}{c}
    -\frac{i}{\hat{\zeta}_{j}}u_{-}\\
  1\\
\end{array}\right)=0.
\end{align}

\noindent \textbf{Theorem 3.1}  \emph{
The precise formulae of $N$-simple poles solutions for the nonlocal Hirota equation \eqref{10} with NZBCs \eqref{11} is expressed as
\begin{align}\label{39}
q(x,t)=q_{-}-i\frac{\det \left(\begin{array}{cc}
    \mathcal{H}  &  \varphi\\
  \chi^{T} &  0\\
\end{array}\right)}{\det (\mathcal{H})}.
\end{align}
}
\begin{proof}
From Eq.\eqref{38}, it is not hard to derive a $2N$ linear system with respect to $\mu_{-11}(\zeta_{n})$
\begin{align}\label{40}
\sum_{n=1}^{2N}\left(C_{n}(\hat{\zeta}_{j})+\frac{iq_{-}}{\hat{\zeta}_{j}}\delta_{j,n}\right)\mu_{-11}(\zeta_{n})
=\frac{iq_{-}}{\hat{\zeta}_{j}}, \quad j=1, 2, \cdots, 2N.
\end{align}
The linear system can be rewritten into a matrix form:
\begin{align}\label{41}
\mathcal{H}\alpha=\varphi,
\end{align}
where $\mathcal{H}=(h_{jn})_{2N\times 2N}$, $\alpha=(\alpha_{n})_{2N\times 1}$, $\varphi=(\varphi_{j})_{2N\times 1}$
with
\begin{align}\label{42}
h_{jn}=C_{n}(\hat{\zeta}_{j})+\frac{iq_{-}}{\hat{\zeta}_{j}}\delta_{j,n},\quad \alpha_{n}=\mu_{-11}(\zeta_{n}),\quad \varphi_{j}=\frac{iq_{-}}{\hat{\zeta}_{j}}.
\end{align}
In the case of reflectionless potential, Eq.\eqref{35} is rewritten as
\begin{align}\label{43}
q=q_{-}+i\chi^{T} \alpha,
\end{align}
where $\chi=(\chi_{n})_{2N\times 1}$ with $\chi_{n}=A[\zeta_{n}]e^{-2i\theta(\zeta_{n})}$.
Combining Eq. \eqref{41}, the expression of the soliton solution is  presented finally.
\end{proof}

As a example, through choosing some appropriate parameters, we discuss the dynamical behaviors for simple poles solution in the case of $N=1$ and  $N=2$, respectively. It is worth mentioning that the discrete spectrum can not to be pure imaginary number, which is verified from Proposition 2.5.\\
\textbf{Case 1:}$N=1$

Let $z_{1}=q_{0}e^{i\vartheta_{1}}, \vartheta_{1}\in(0, \frac{\pi}{2}).$ Then, from \eqref{28}, we have $\zeta_{1}=q_{0}e^{i\vartheta_{1}}, \zeta_{2}=-q_{0}e^{-i\vartheta_{1}}, \hat{\zeta}_{1}=q_{0}e^{-i\vartheta_{1}}, \hat{\zeta}_{2}=-q_{0}e^{i\vartheta_{1}}.$  Let $q_{-}=q_{0}e^{i\theta_{-}}, \theta_{-}\in \{0, \pi \}$, $b[\zeta_{1}]=b_{1}$, where $b_{1}$ is arbitrary parameter. From Proposition 2.5, we have $b[\zeta_{2}]=-\frac{1}{b_{1}^{\ast}}$. From Eq.\eqref{Ajia5}, one has
\begin{align}\label{44}
&s_{22}'(\zeta_{1})=\frac{\zeta_{1}-\zeta_{2}}{(\zeta_{1}-\hat{\zeta}_{1})(\zeta_{1}-\hat{\zeta}_{2})}
=\frac{\cos(\vartheta_{1})}{2q_{0}ie^{i\vartheta_{1}}\sin(\vartheta_{1})},\notag\\ &s_{22}'(\zeta_{2})=\frac{\zeta_{2}-\zeta_{1}}{(\zeta_{2}-\hat{\zeta}_{1})(\zeta_{2}-\hat{\zeta}_{2})}
=\frac{\cos(\vartheta_{1})}{2q_{0}ie^{-i\vartheta_{1}}\sin(\vartheta_{1})},
\end{align}
then we have
\begin{align}\label{45}
A[\zeta_{1}]=\frac{b[\zeta_{1}]}{s_{22}'(\zeta_{1})}=\frac{2q_{0}b_{1}ie^{i\vartheta_{1}}\sin(\vartheta_{1})}{\cos(\vartheta_{1})},\ A[\zeta_{2}]=\frac{b[\zeta_{2}]}{s_{22}'(\zeta_{2})}=-\frac{2q_{0}ie^{-i\vartheta_{1}}\sin(\vartheta_{1})}{b_{1}^{\ast}\cos(\vartheta_{1})}.
\end{align}
Thus the one-simple pole solution of the nonlocal Hirota equation \eqref{10} is deduced as
\begin{align}\label{46}
q(x, t)=q_{0}\frac{\Omega_{1}}{\Omega_{2}},
\end{align}
where
\begin{align}\label{46.1}
\Omega_{1}=&(2\sin(\vartheta_{1})+i)\cos(\vartheta_{1})\Theta_{1}
-(2\sin(\vartheta_{1})-i)b_{1}b_{1}^{\ast}\cos(\vartheta_{1})\Theta_{2}+b_{1}^{\ast}\cos(\vartheta_{1})^{2}e^{3i\theta_{-}}\notag\\
&+2ib_{1}\sin(\vartheta_{1})\Theta_{3}(e^{-i\vartheta_{1}+i\theta_{-}}
-e^{i\vartheta_{1}+i\theta_{-}})
+(3b_{1}\cos(\vartheta_{1})^{2}-4b_{1})\Theta_{3}e^{i\theta_{-}}\notag\\
\Omega_{2}=&i\cos(\vartheta_{1})\Theta_{1}e^{i\vartheta_{1}-i\theta_{-}}+
ib_{1}b_{1}^{\ast}\cos(\vartheta_{1})\Theta_{2}e^{-i\vartheta_{1}-i\theta_{-}}
-b_{1}\Theta_{3}\notag\\
&+(1-\cos(\vartheta_{1})^{2})b_{1}\Theta_{4}+b_{1}^{\ast}\cos(\vartheta_{1})^{2}
e^{2i\theta_{-}},\notag\\
\Theta_{1}=&e^{-4q_{0}\sin(\vartheta_{1})(-\frac{1}{2}x-\beta q_{0}^{2}t+i\delta q_{0}\cos(\vartheta_{1})t-2\beta q_{0}^{2}\cos(\vartheta_{1})^{2}t)-i\vartheta_{1}+2i\theta_{-}},\notag\\
\Theta_{2}=&e^{4q_{0}\sin(\vartheta_{1})(\frac{1}{2}x+\beta q_{0}^{2}t+i\delta q_{0}\cos(\vartheta_{1})t+2\beta q_{0}^{2}\cos(\vartheta_{1})^{2}t)+i\vartheta_{1}+2i\theta_{-}},\notag\\
\Theta_{3}=&e^{4q_{0}\sin(\vartheta_{1})(x+2\beta q_{0}^{2}t+4\beta q_{0}^{2}\cos(\vartheta_{1})^{2}t)},\notag\\
\Theta_{4}=&e^{4q_{0}(\sin(\vartheta_{1})x+\beta q_{0}^{2}\sin(3\vartheta_{1})t
+3\beta q_{0}^{2}\sin(\vartheta_{1})t)}.
\end{align}

Through choosing different  parameters, the different dynamic behaviors are presented in Figs. 2-5. Of which, the Fig. 2(a) (b) display a dark soliton, the Fig. 2(c) describes the peak of wave  along the line $t+0.25x=0$ which is the center trajectory. In Fig. 2(c), we find  the crest is less than 1(the height of background wave)  within a certain region, and beyond the region, the crest is greater than 1. As we can see from Fig. 3, as $|t|$ increases, the dark soliton turns into the anti-dark soliton, this fact can also be verified in Fig. 2(c). This phenomenon is quite interesting and original for the nonlocal Hirota equation\eqref{10}. Fig. 4 shows a bright-dark soliton, and Fig. 5  exhibits a breather wave. Meanwhile, comparing Fig. 2 and Fig. 4, we find that as the parameters $b_{1}$ increase, the dark soliton turns into the bright-dark soliton.  On the other hand, from Fig. 2 and Fig. 5, we find that as parameter $\delta$ increases and parameter $\beta$ decreases, the dark soliton changes into the breather wave.\\
{\rotatebox{0}{\includegraphics[width=4.6cm,height=4.0cm,angle=0]{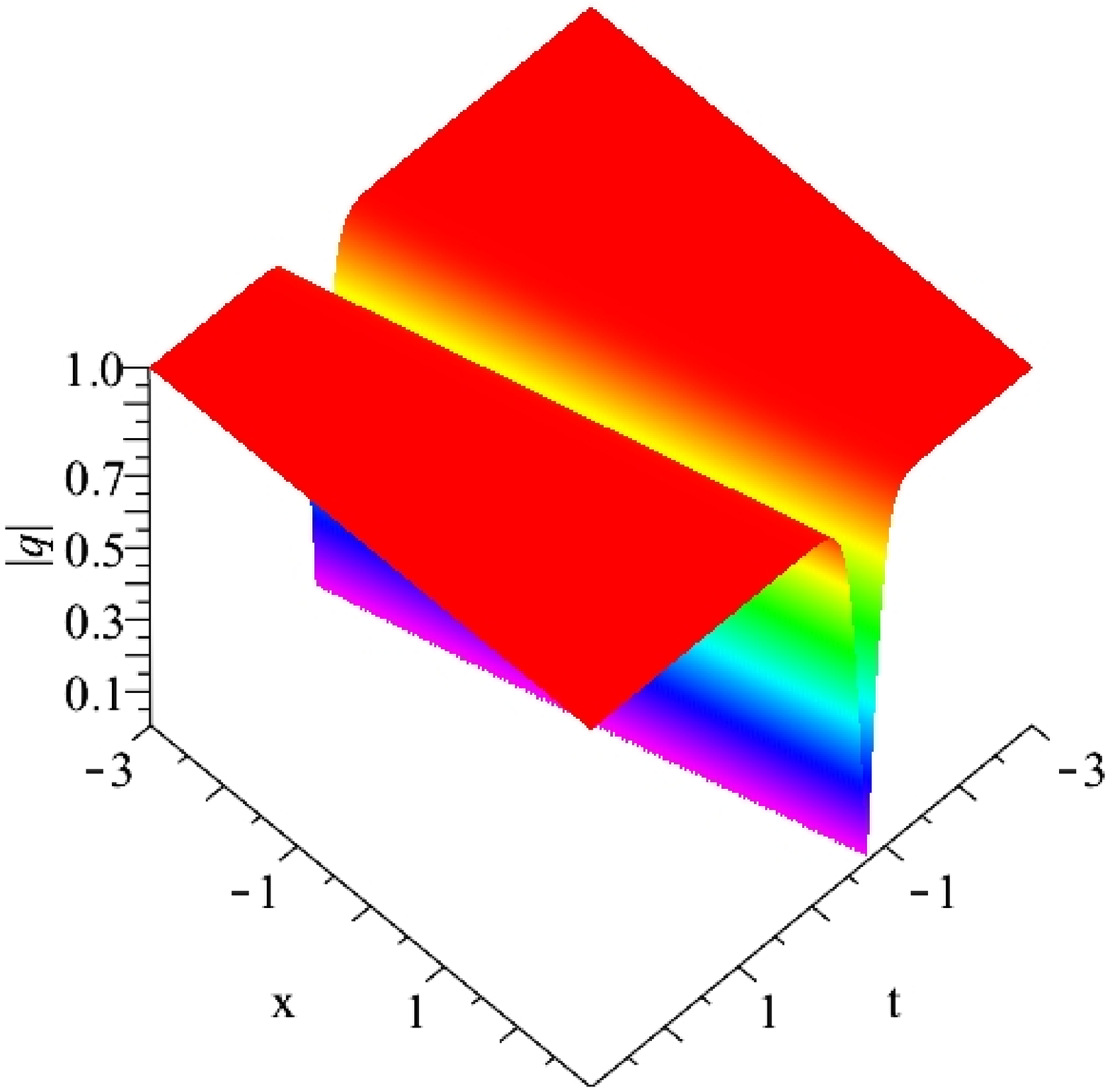}}}
~~~~
{\rotatebox{0}{\includegraphics[width=4.6cm,height=4.0cm,angle=0]{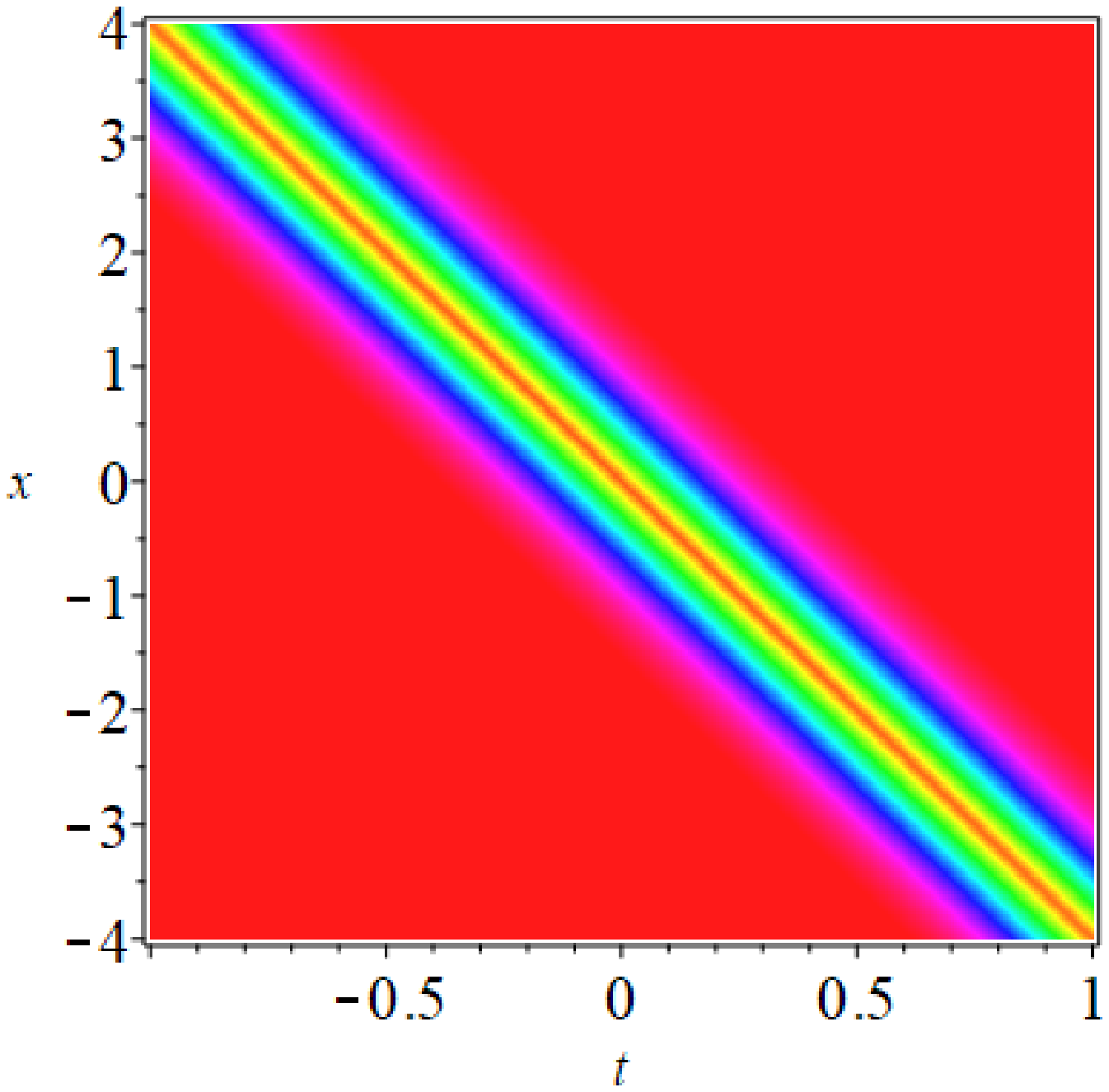}}}
~~~~
{\rotatebox{0}{\includegraphics[width=4.6cm,height=4.0cm,angle=0]{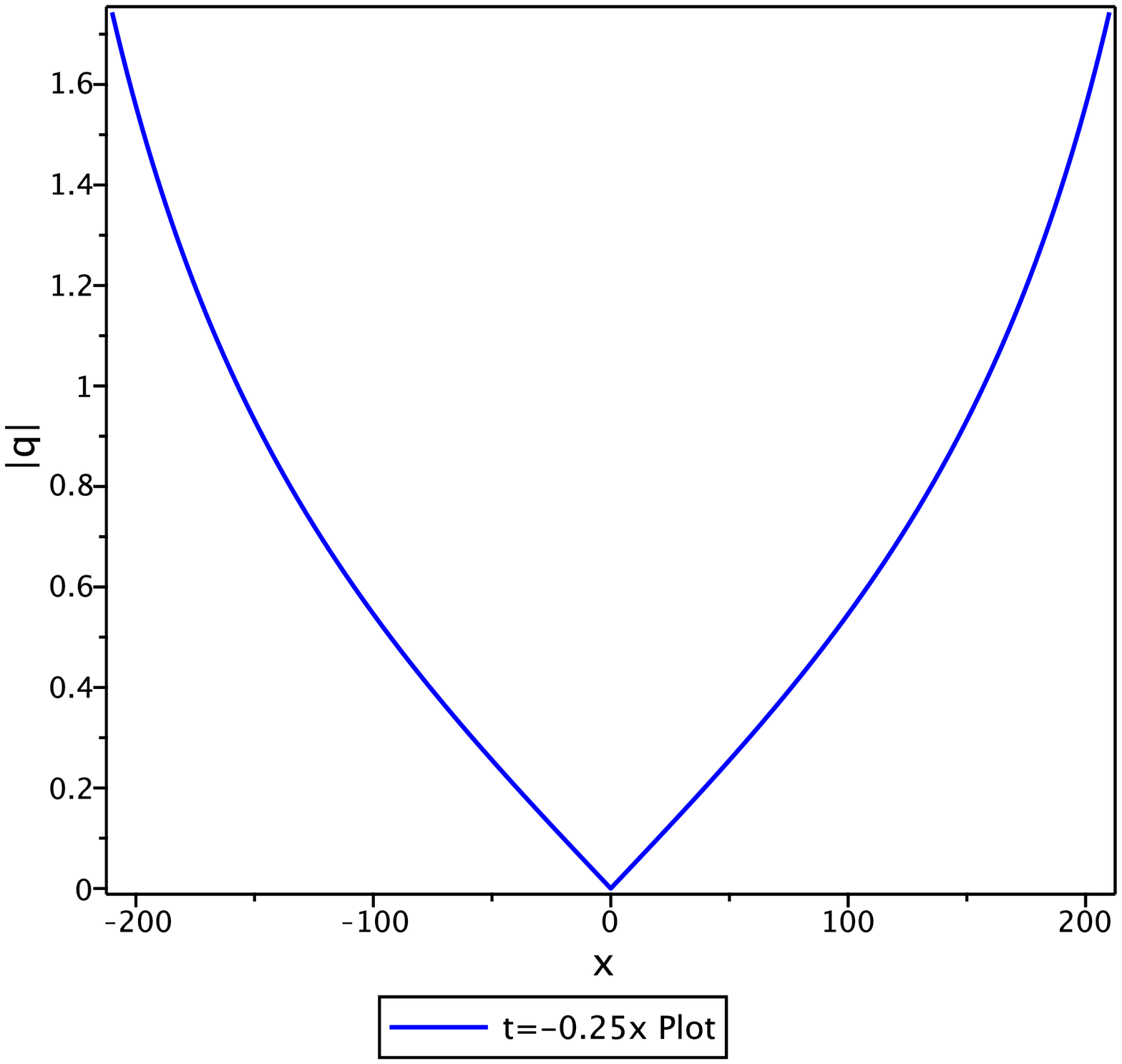}}}\\
$~~~~~~~~~~~~~~~(\textbf{a})~~~~~~~~~~~~~~~~
~~~~~~~~~~~~~~~~~~~~~~~~~(\textbf{b})~~~~~~~
~~~~~~~~~~~~~~~~~~~~~~~~~~~~(\textbf{c})$\\
\noindent { \small \textbf{Figure 2.} (Color online) one-simple pole solution for Eq.\eqref{10} with the parameters: $b_{1}=1, q_{0}=1, \vartheta_{1}=\frac{\pi}{4}, \theta_{-}=0, \delta=\frac{1}{100}, \beta=1$.
$\textbf{(a)}$ Three dimensional plot;
$\textbf{(b)}$ The density plot;
$\textbf{(c)}$ The wave propagation along the $x$-axis at $t=-5$(black), $t=0$(blue), $t=5$(red).}\\
\\
{\rotatebox{0}{\includegraphics[width=4.6cm,height=4.0cm,angle=0]{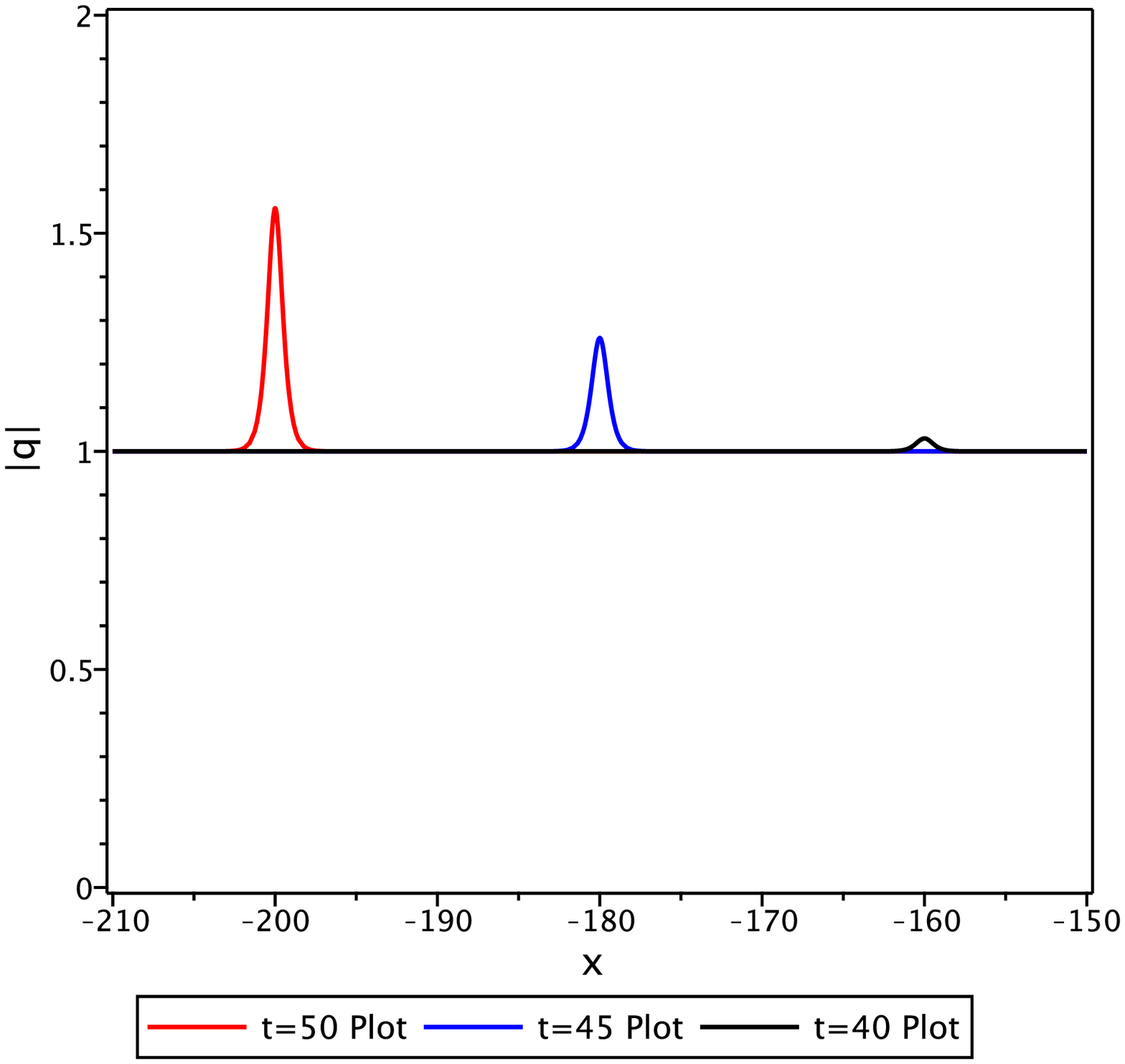}}}
~~~~
{\rotatebox{0}{\includegraphics[width=4.6cm,height=4.0cm,angle=0]{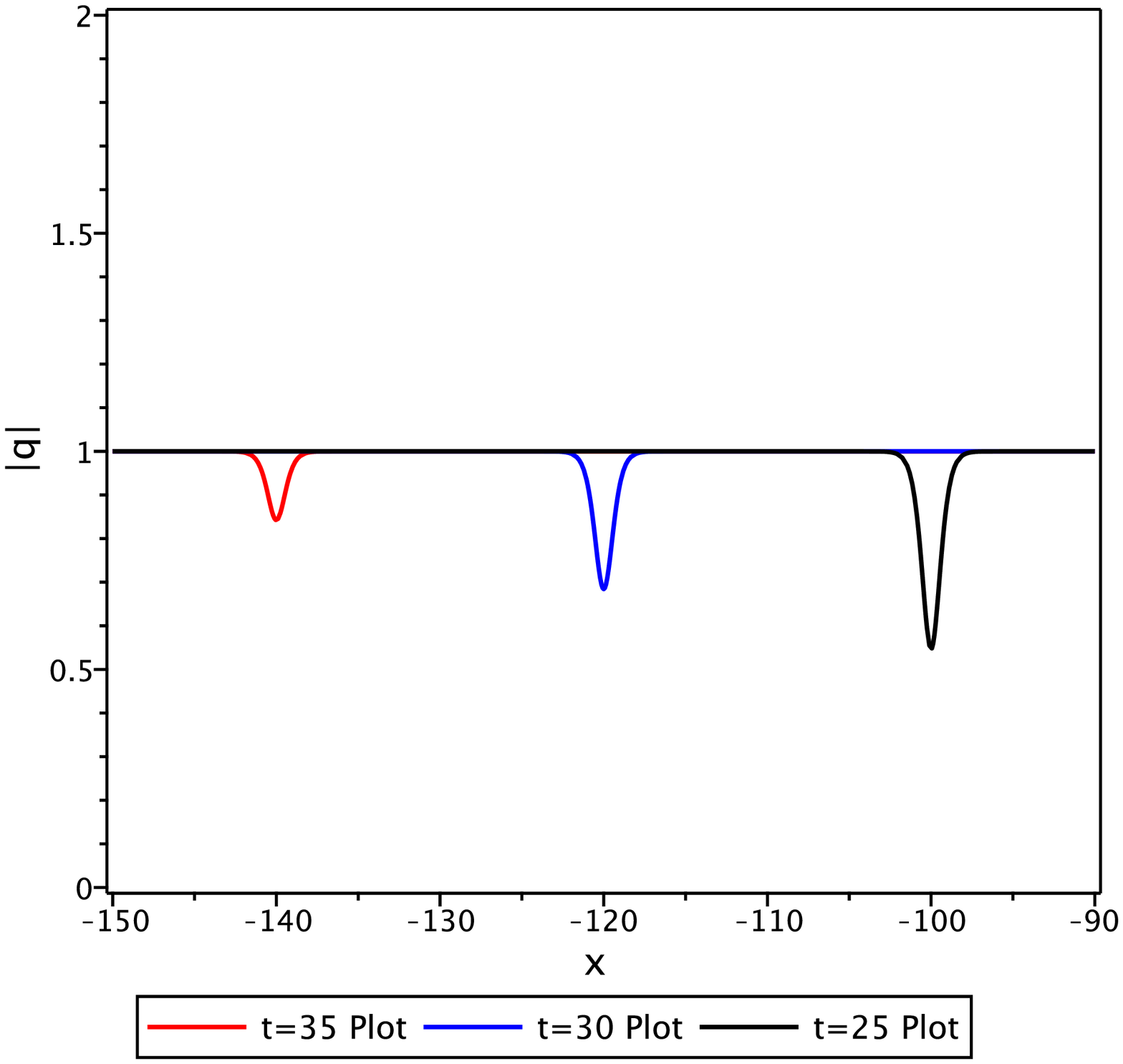}}}
~~~~
{\rotatebox{0}{\includegraphics[width=4.6cm,height=4.0cm,angle=0]{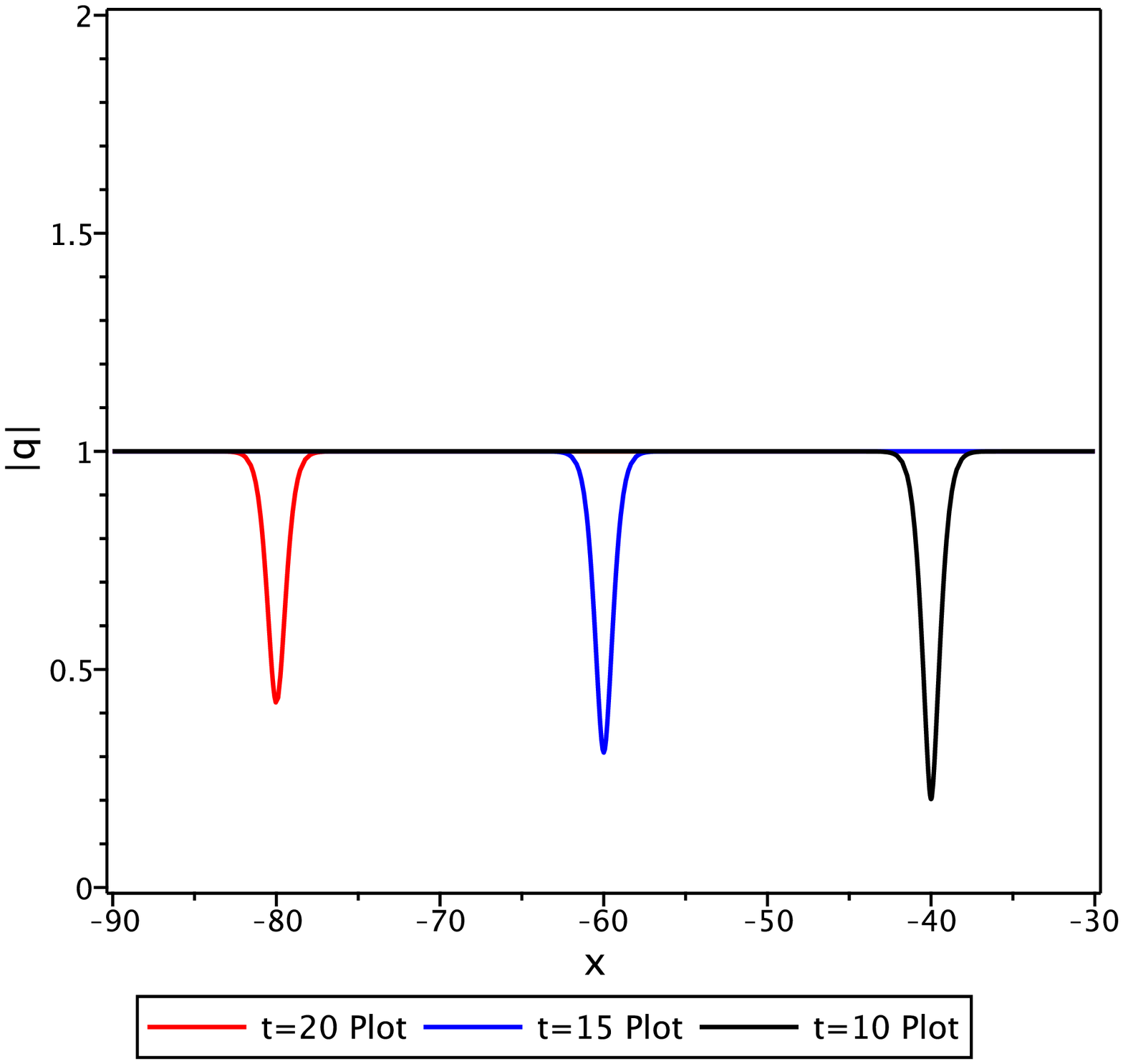}}}\\
$~~~~~~~~~~~~~~~(\textbf{a})~~~~~~~~~~~~~~~~
~~~~~~~~~~~~~~~~~~~~~~~~~(\textbf{b})~~~~~~~
~~~~~~~~~~~~~~~~~~~~~~~~~~~~(\textbf{c})$\\
{\rotatebox{0}{\includegraphics[width=4.6cm,height=4.0cm,angle=0]{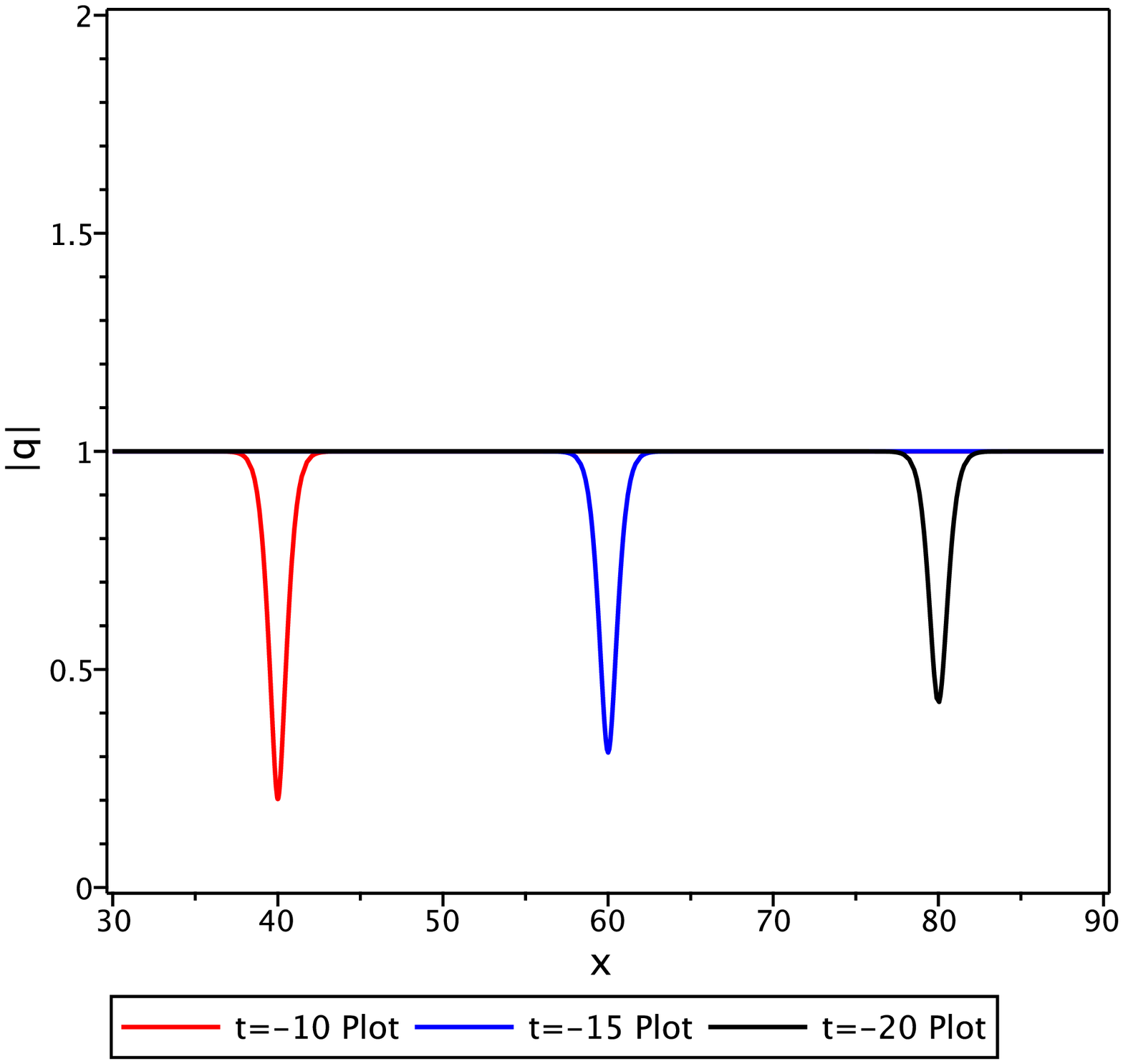}}}
~~~~
{\rotatebox{0}{\includegraphics[width=4.6cm,height=4.0cm,angle=0]{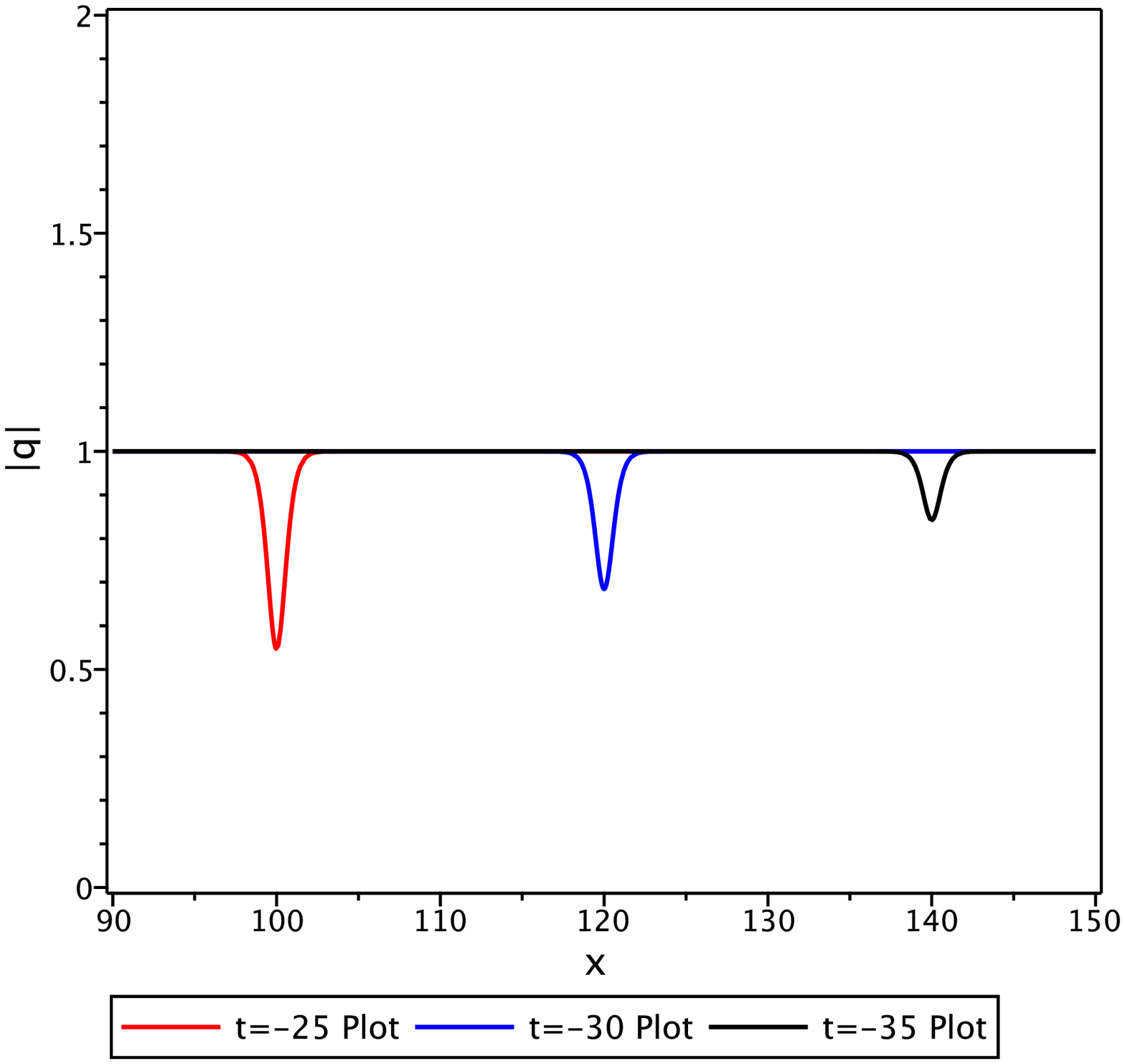}}}
~~~~
{\rotatebox{0}{\includegraphics[width=4.6cm,height=4.0cm,angle=0]{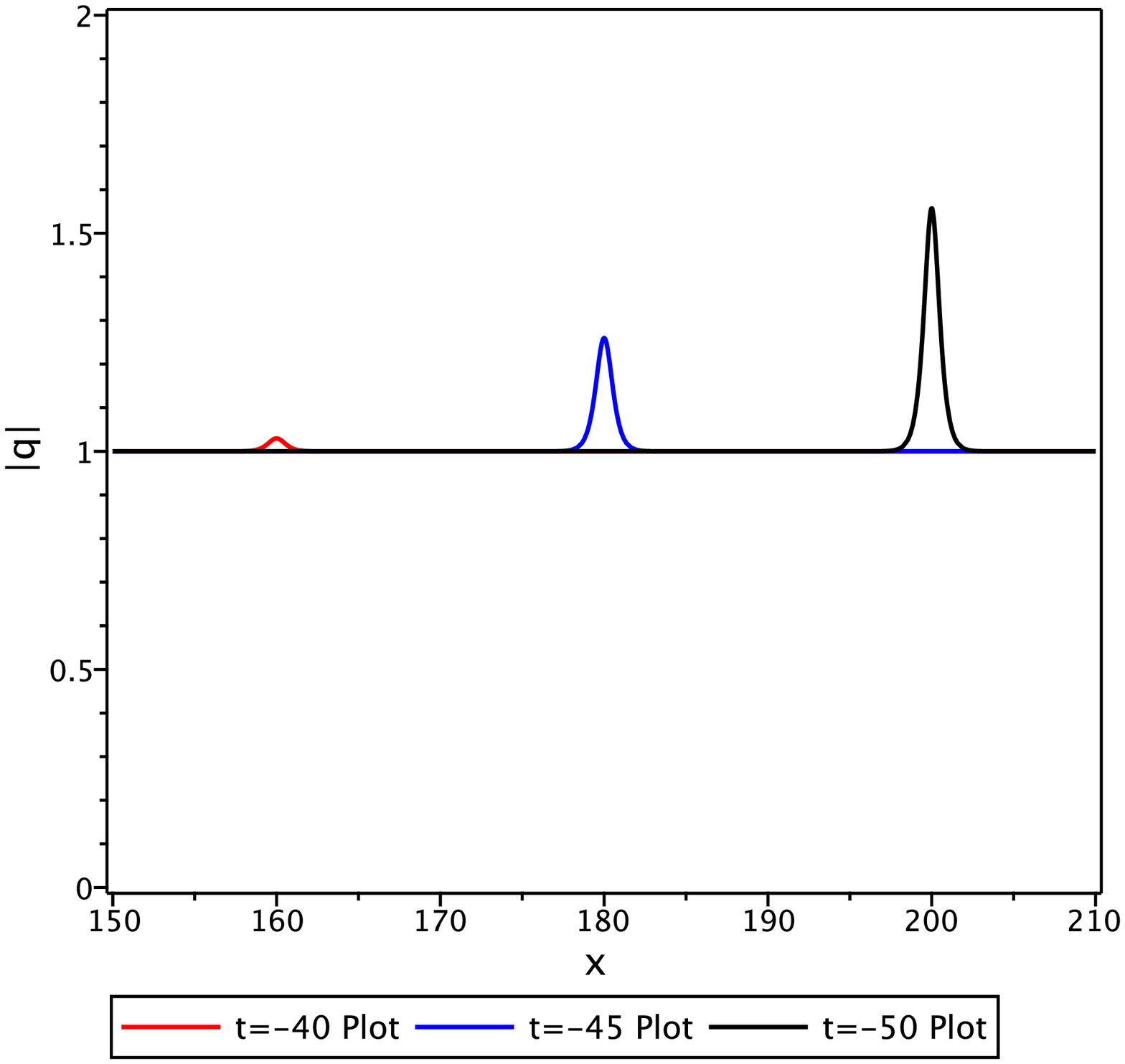}}}\\
$~~~~~~~~~~~~~~~(\textbf{d})~~~~~~~~~~~~~~~~
~~~~~~~~~~~~~~~~~~~~~~~~~(\textbf{e})~~~~~~~
~~~~~~~~~~~~~~~~~~~~~~~~~~~~(\textbf{f})$\\
\noindent { \small \textbf{Figure 3.} (Color online) The wave propagation of one-simple pole solution along the $x$-axis at different time. The parameters are $b_{1}=1, q_{0}=1, \vartheta_{1}=\frac{\pi}{4}, \theta_{-}=0, \delta=\frac{1}{100}, \beta=1$.}
\\
{\rotatebox{0}{\includegraphics[width=4.6cm,height=4.0cm,angle=0]{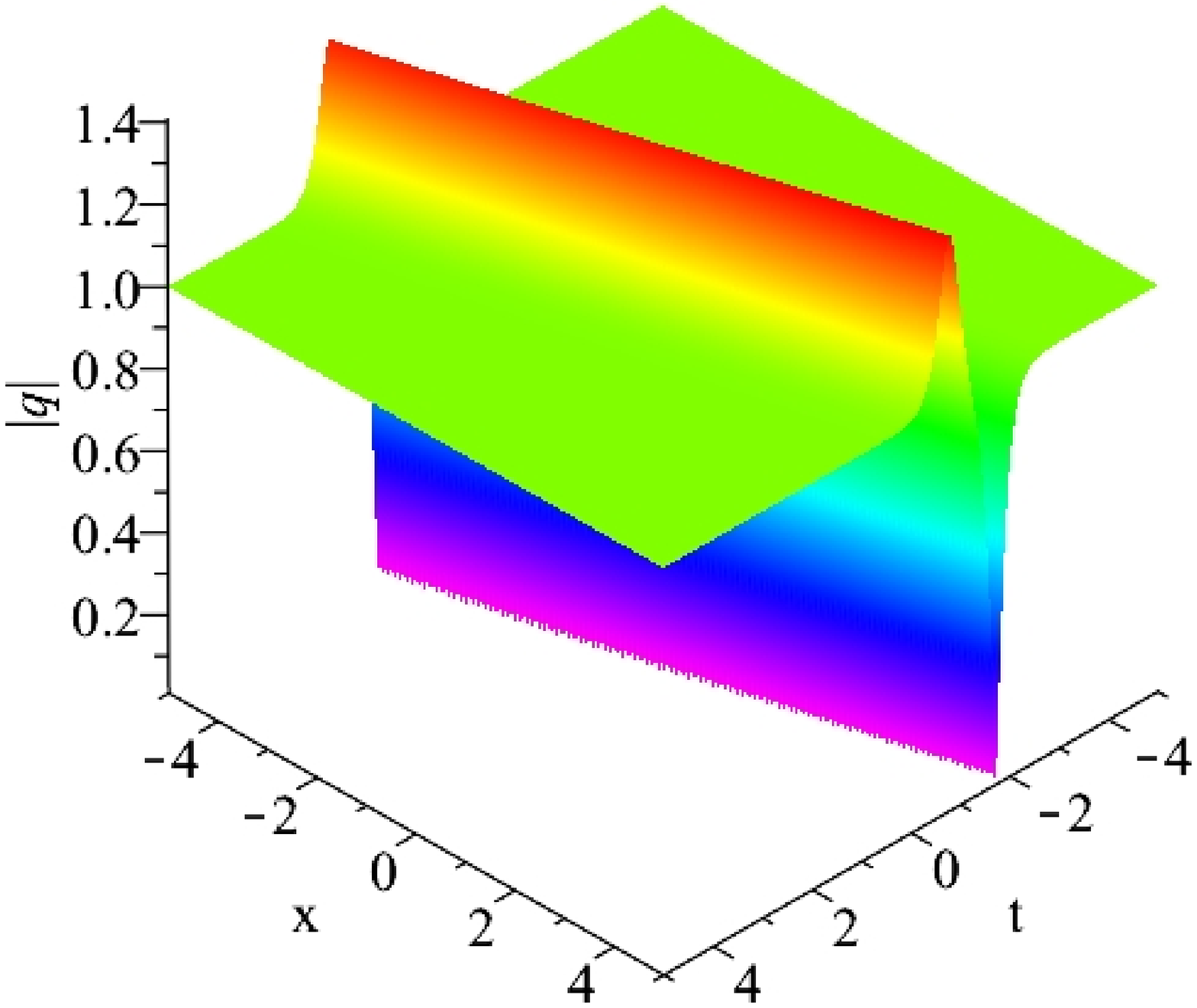}}}
~~~~
{\rotatebox{0}{\includegraphics[width=4.6cm,height=4.0cm,angle=0]{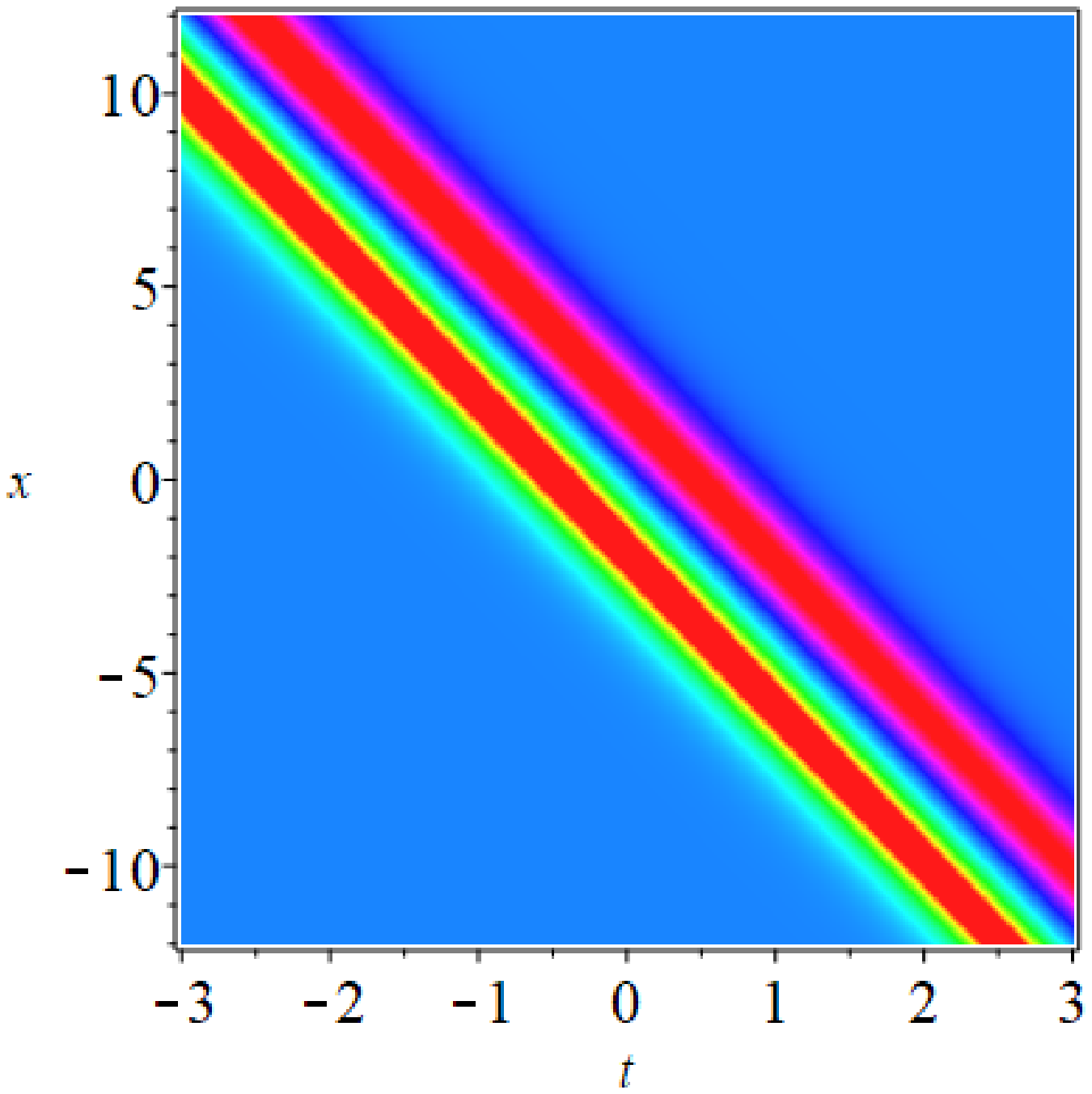}}}
~~~~
{\rotatebox{0}{\includegraphics[width=4.6cm,height=4.0cm,angle=0]{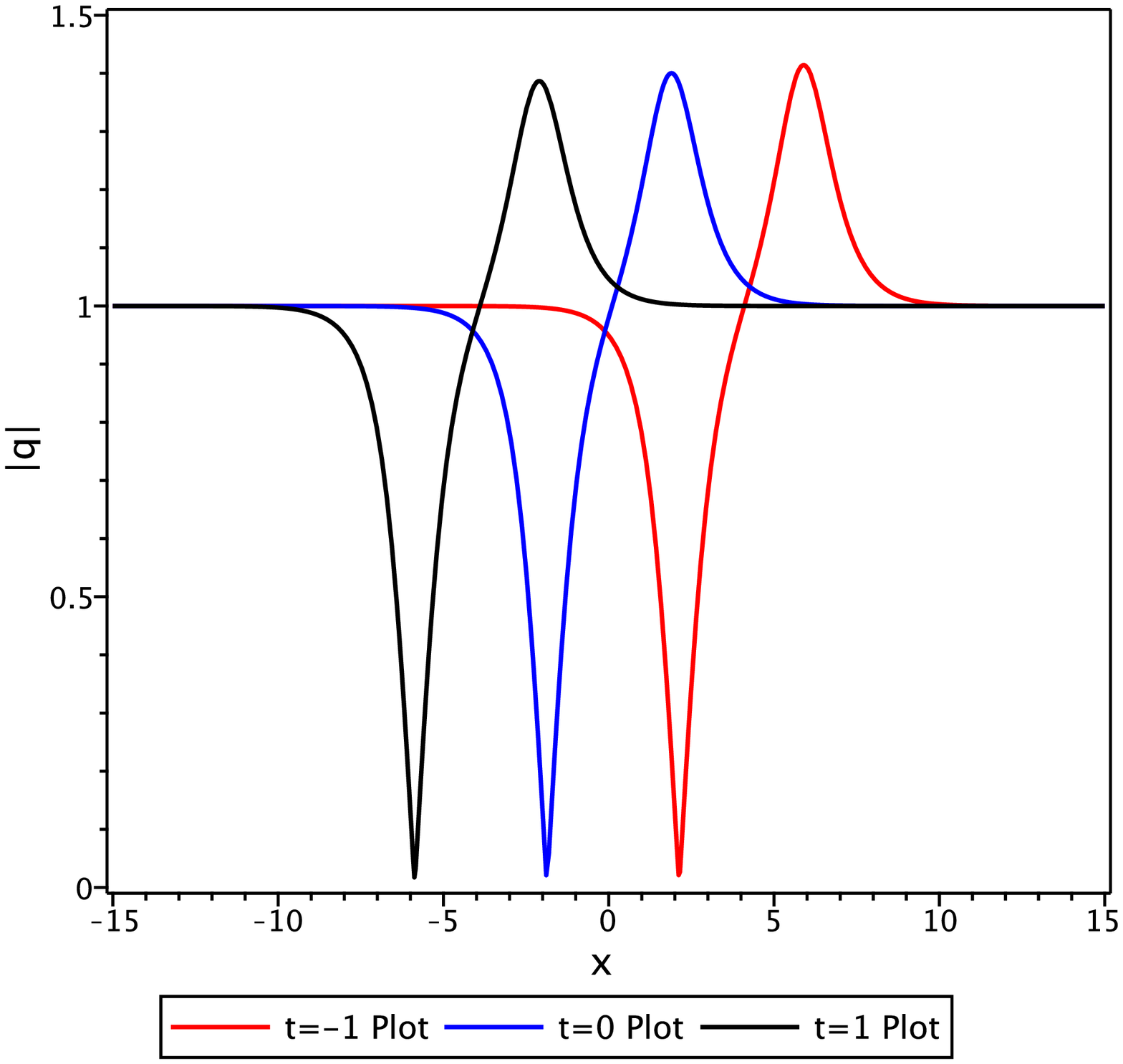}}}\\
$~~~~~~~~~~~~~~~(\textbf{a})~~~~~~~~~~~~~~~~
~~~~~~~~~~~~~~~~~~~~~~~~~(\textbf{b})~~~~~~~
~~~~~~~~~~~~~~~~~~~~~~~~~~~~(\textbf{c})$\\
\noindent { \small \textbf{Figure 4.} (Color online) one-simple pole solution for Eq.\eqref{10} with the parameters: $b_{1}=10, q_{0}=1, \vartheta_{1}=\frac{\pi}{4}, \theta_{-}=0, \delta=\frac{1}{100}, \beta=1$.
$\textbf{(a)}$ Three dimensional plot;
$\textbf{(b)}$ The density plot;
$\textbf{(c)}$ The wave propagation along the $x$-axis at $t=-1$(red), $t=0$(blue), $t=1$(black).}\\
{\rotatebox{0}{\includegraphics[width=4.6cm,height=4.0cm,angle=0]{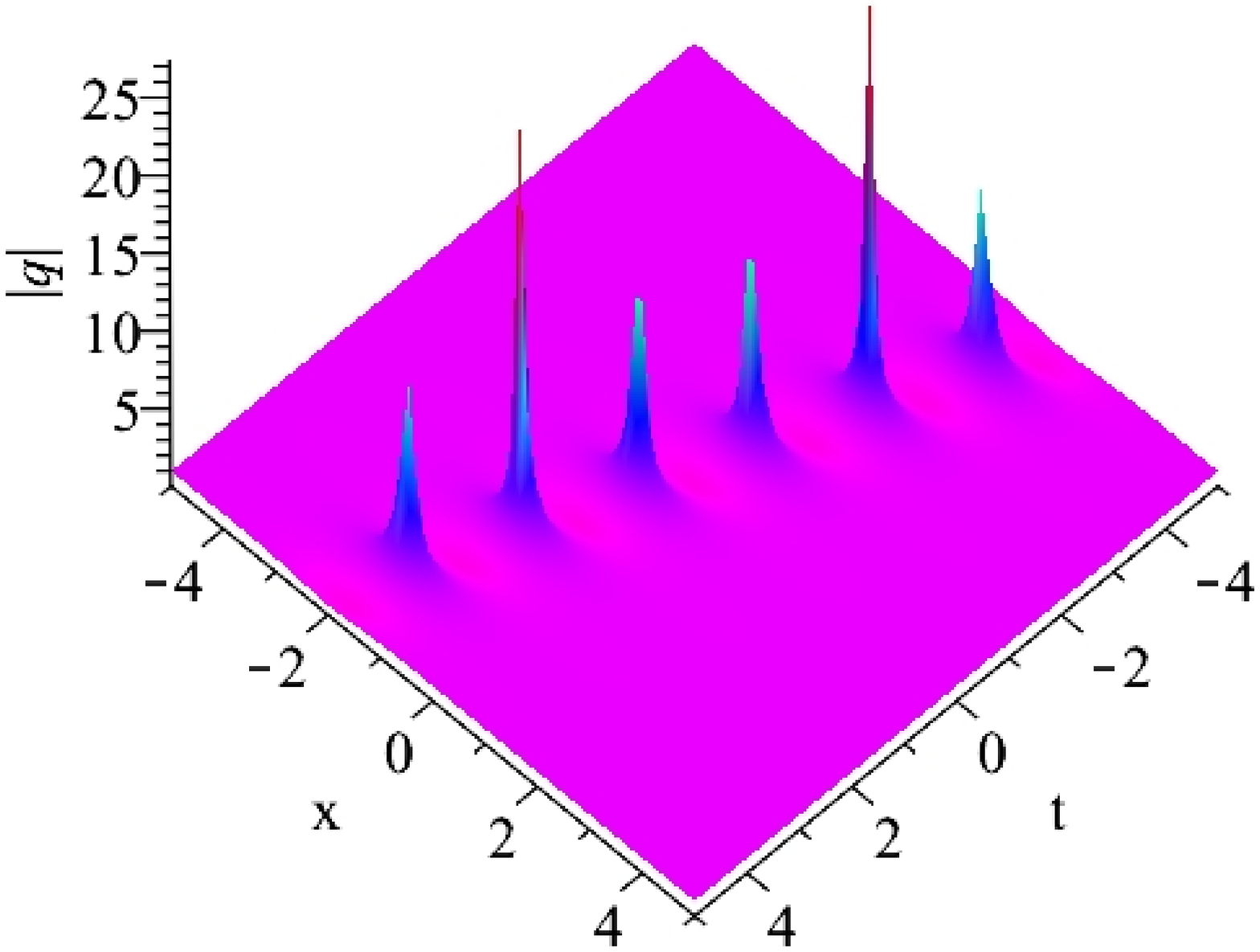}}}
~~~~
{\rotatebox{0}{\includegraphics[width=4.6cm,height=4.0cm,angle=0]{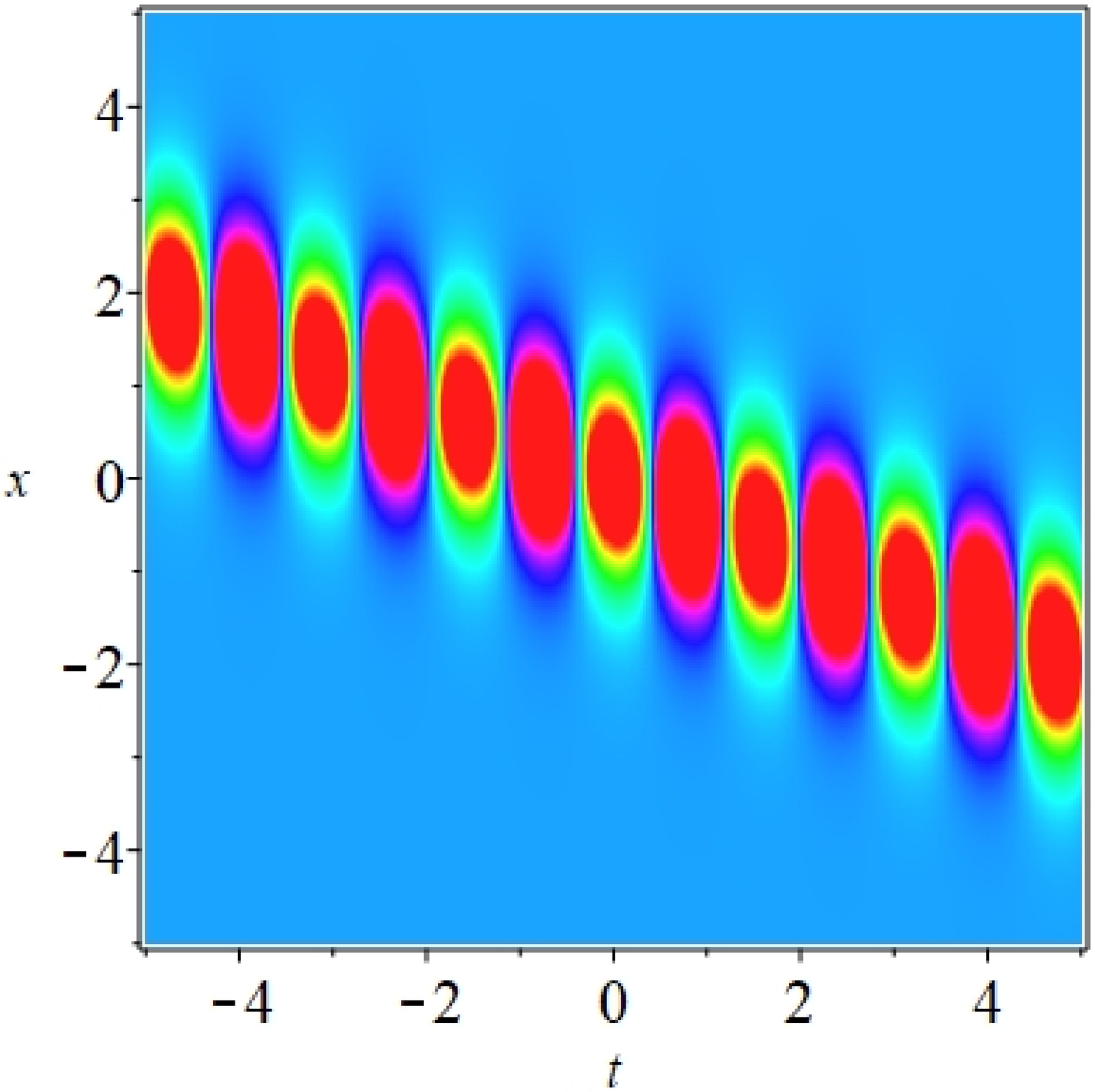}}}
~~~~
{\rotatebox{0}{\includegraphics[width=4.6cm,height=4.0cm,angle=0]{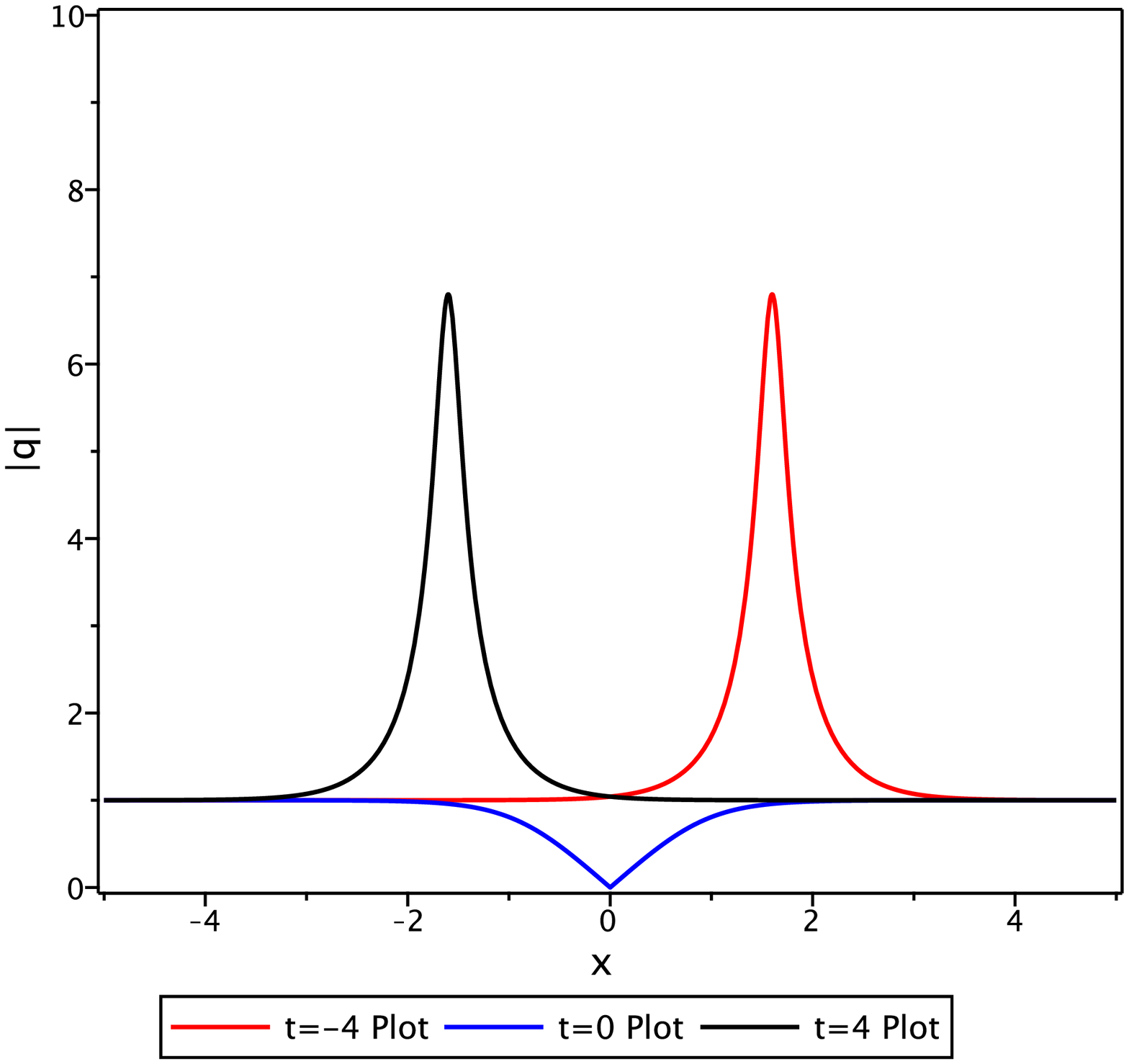}}}\\
$~~~~~~~~~~~~~~~(\textbf{a})~~~~~~~~~~~~~~~~
~~~~~~~~~~~~~~~~~~~~~~~~~(\textbf{b})~~~~~~~
~~~~~~~~~~~~~~~~~~~~~~~~~~~~(\textbf{c})$\\
\noindent { \small \textbf{Figure 5.} (Color online) one-simple pole solution for Eq.\eqref{10} with the parameters: $b_{1}=1, q_{0}=1, \vartheta_{1}=\frac{\pi}{4}, \theta_{-}=0, \delta=1, \beta=\frac{1}{10}$.
$\textbf{(a)}$ Three dimensional plot;
$\textbf{(b)}$ The density plot;
$\textbf{(c)}$ The wave propagation along the $x$-axis at $t=-4$(red), $t=0$(blue), $t=4$(black).}\\

\textbf{Case 2:}$N=2$

Let $z_{1}=q_{0}e^{i\vartheta_{1}}, z_{2}=q_{0}e^{i\vartheta_{2}}, \vartheta_{1}, \vartheta_{2} \in(0, \frac{\pi}{2}).$  Let $q_{-}=q_{0}e^{i\theta_{-}}, \theta_{-}\in \{0, \pi \}$, $b[\zeta_{1}]=b_{1}, b[\zeta_{2}]=b_{2},$ where $b_{1}, b_{2}$ are all arbitrary parameter. We obtain the two-simple pole solution, which displays the interaction of two bright-dark soliton solutions(see Fig. 6). As shown in Fig. 6, we can easily see  when $t=-5$ (before the interaction), the wave profile consists of two
bright-dark solitons, when $t=0$ (they have the strong interaction),
the wave profile creates two spikes. When
$t=5$ (after the interaction), the wave profile restores the original shape. Moreover, in Fig. 7, when $|t|$ increases, the bright-dark soliton degrades into the two anti-dark soliton.\\
{\rotatebox{0}{\includegraphics[width=4.6cm,height=4.0cm,angle=0]{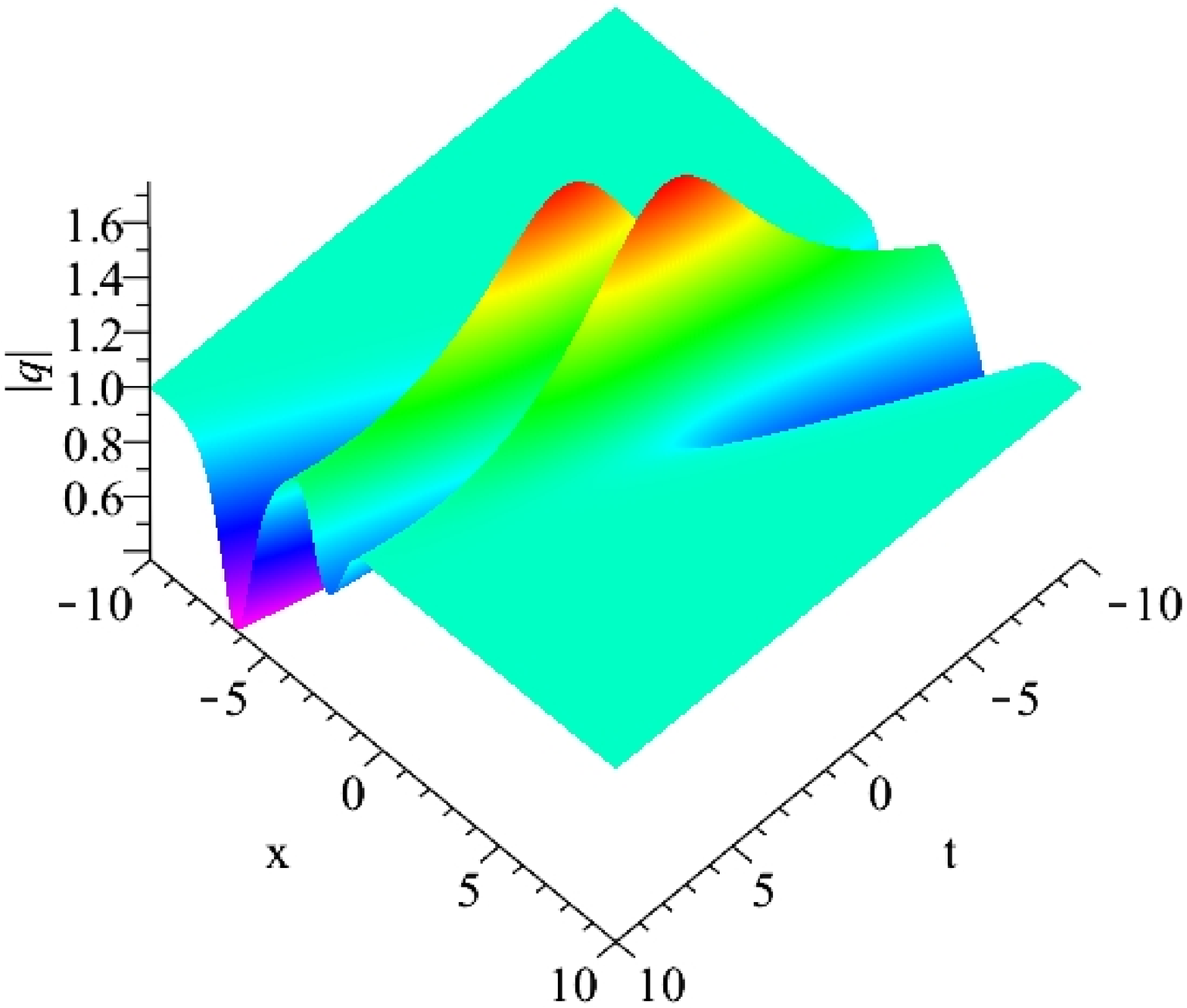}}}
~~~~
{\rotatebox{0}{\includegraphics[width=4.6cm,height=4.0cm,angle=0]{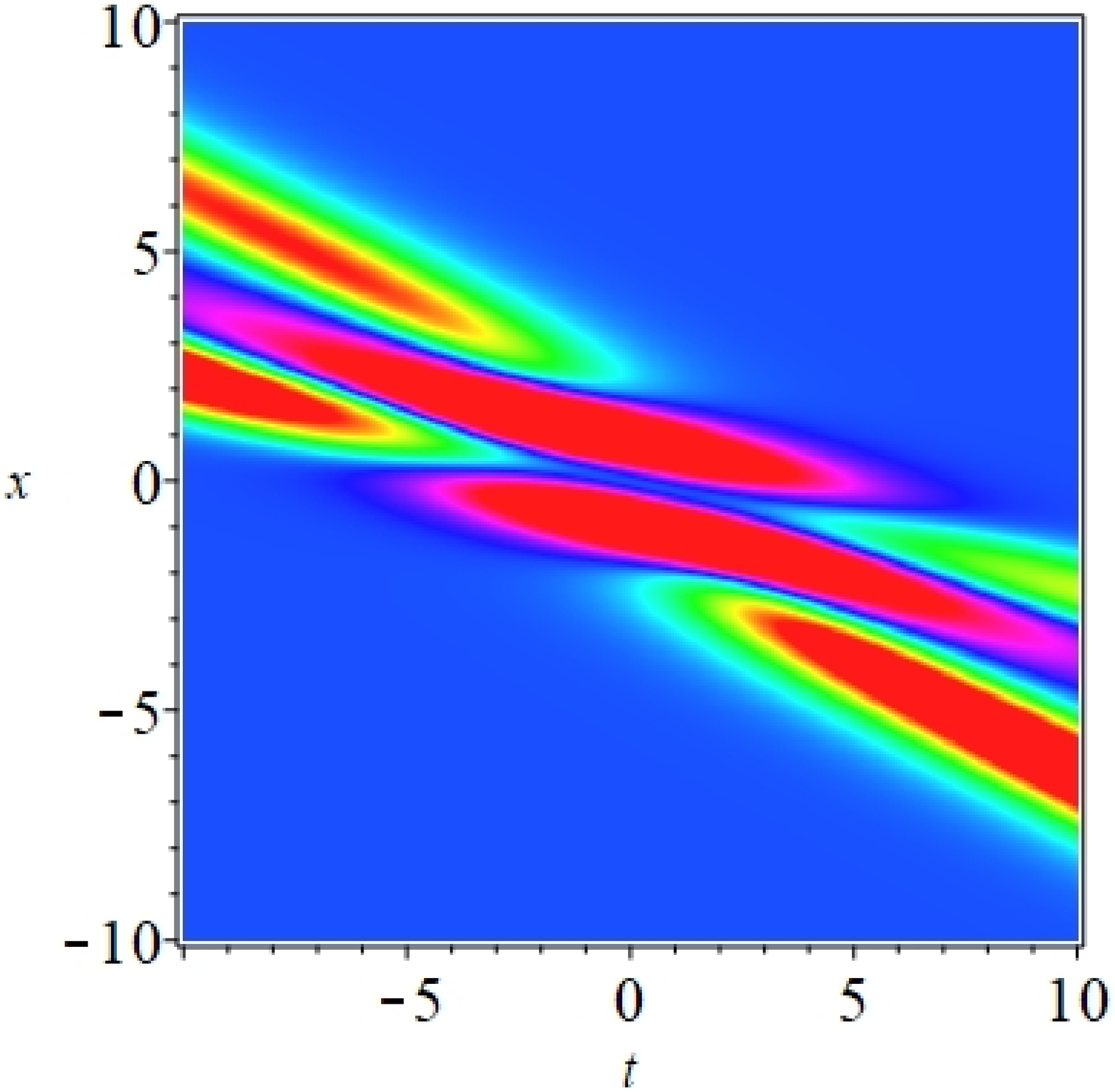}}}
~~~~
{\rotatebox{0}{\includegraphics[width=4.6cm,height=4.0cm,angle=0]{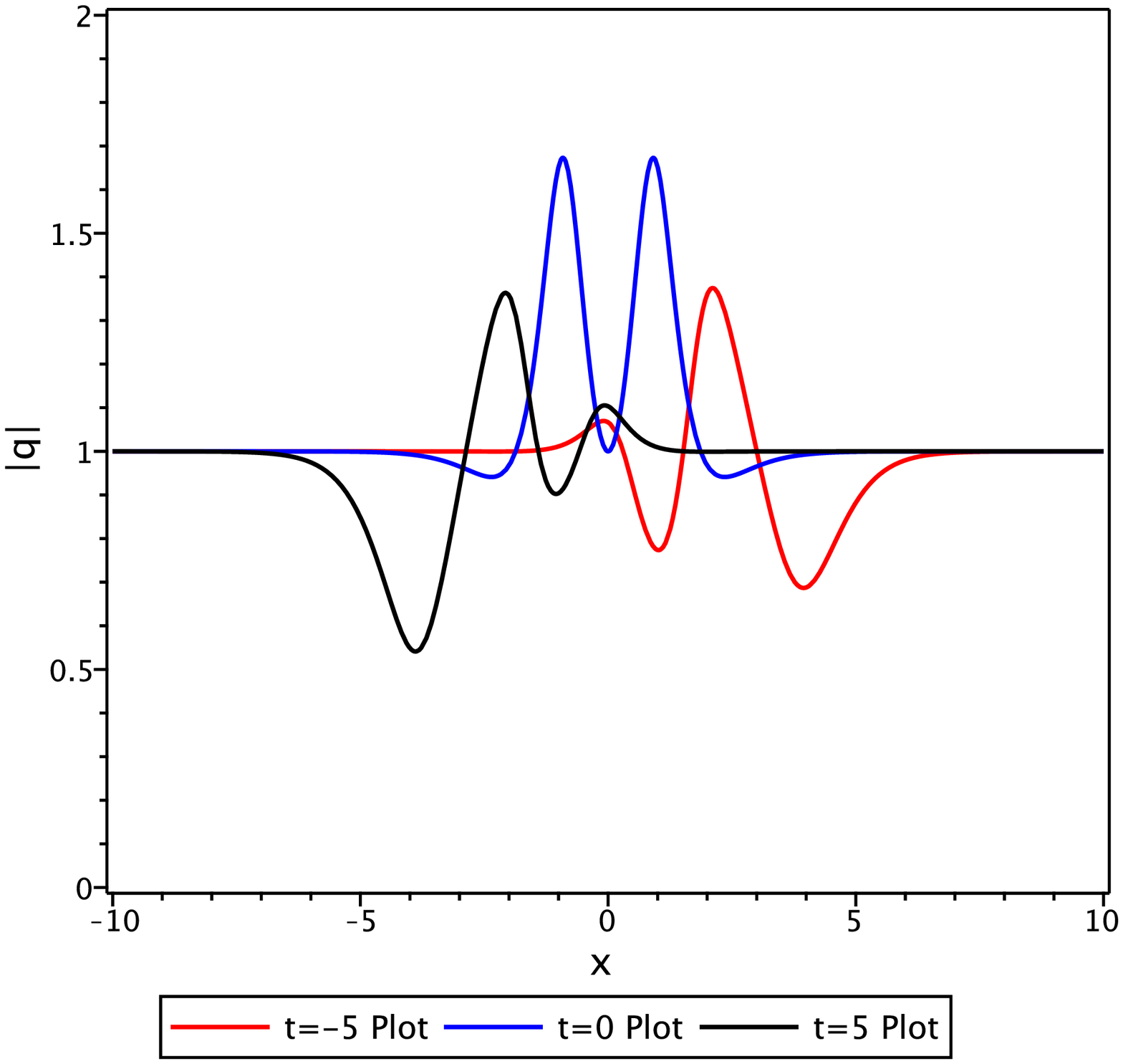}}}\\
$~~~~~~~~~~~~~~~(\textbf{a})~~~~~~~~~~~~~~~~
~~~~~~~~~~~~~~~~~~~~~~~~~(\textbf{b})~~~~~~~
~~~~~~~~~~~~~~~~~~~~~~~~~~~~(\textbf{c})$\\
\noindent { \small \textbf{Figure 6.} (Color online) The two-simple pole solution for Eq.\eqref{10} with the parameters: $b_{1}=-1, b_{2}=-1, q_{0}=1, \vartheta_{1}=\frac{\pi}{6}, \vartheta_{2}=\frac{\pi}{3}, \theta_{-}=0, \delta=\frac{1}{100}, \beta=\frac{1}{10}$.
$\textbf{(a)}$ Three dimensional plot;
$\textbf{(b)}$ The density plot;
$\textbf{(c)}$ The wave propagation along the $x$-axis at $t=-5$(red), $t=0$(blue), $t=5$(black).}\\
\\
{\rotatebox{0}{\includegraphics[width=4.6cm,height=4.0cm,angle=0]{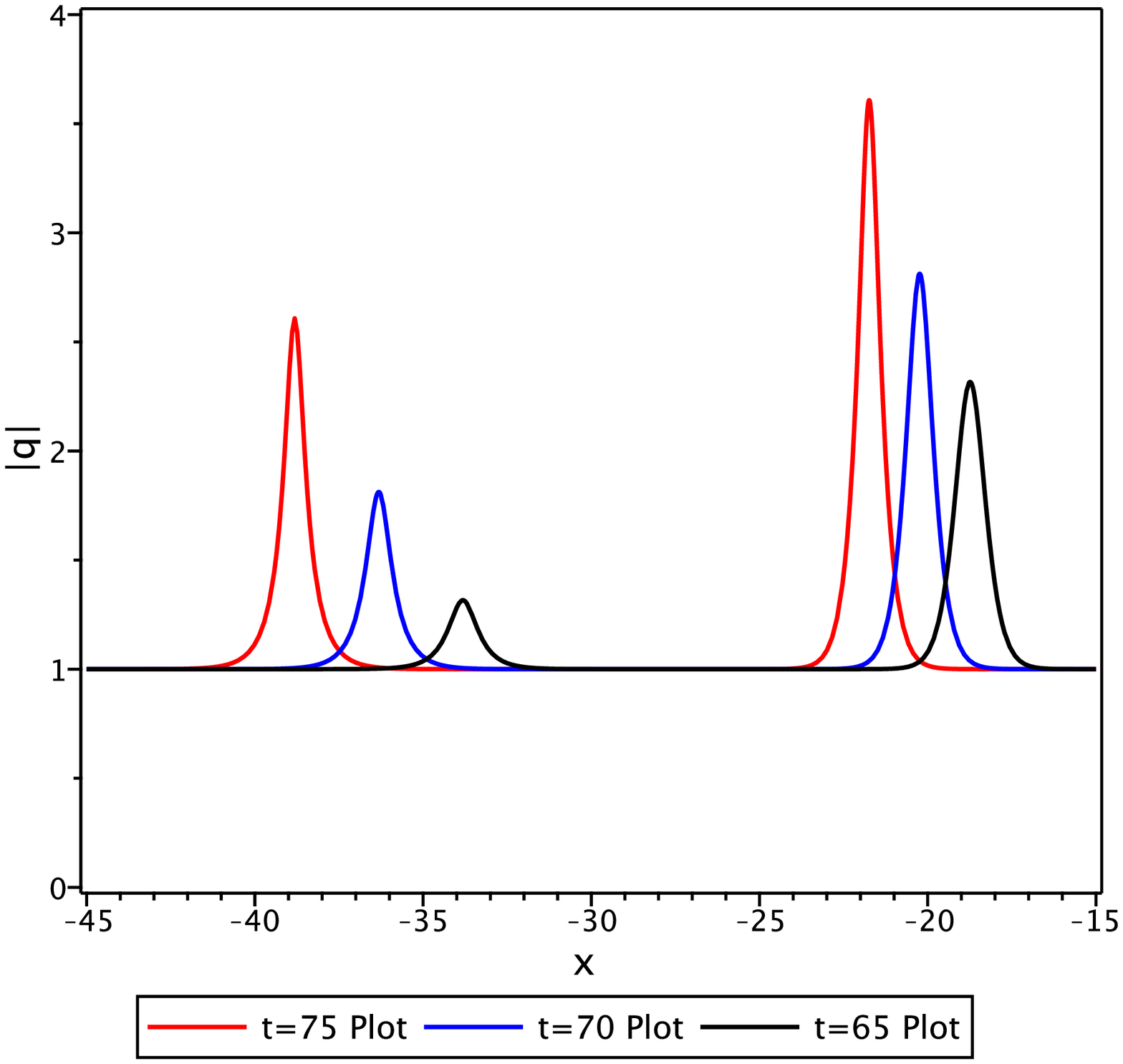}}}
~~~~
{\rotatebox{0}{\includegraphics[width=4.6cm,height=4.0cm,angle=0]{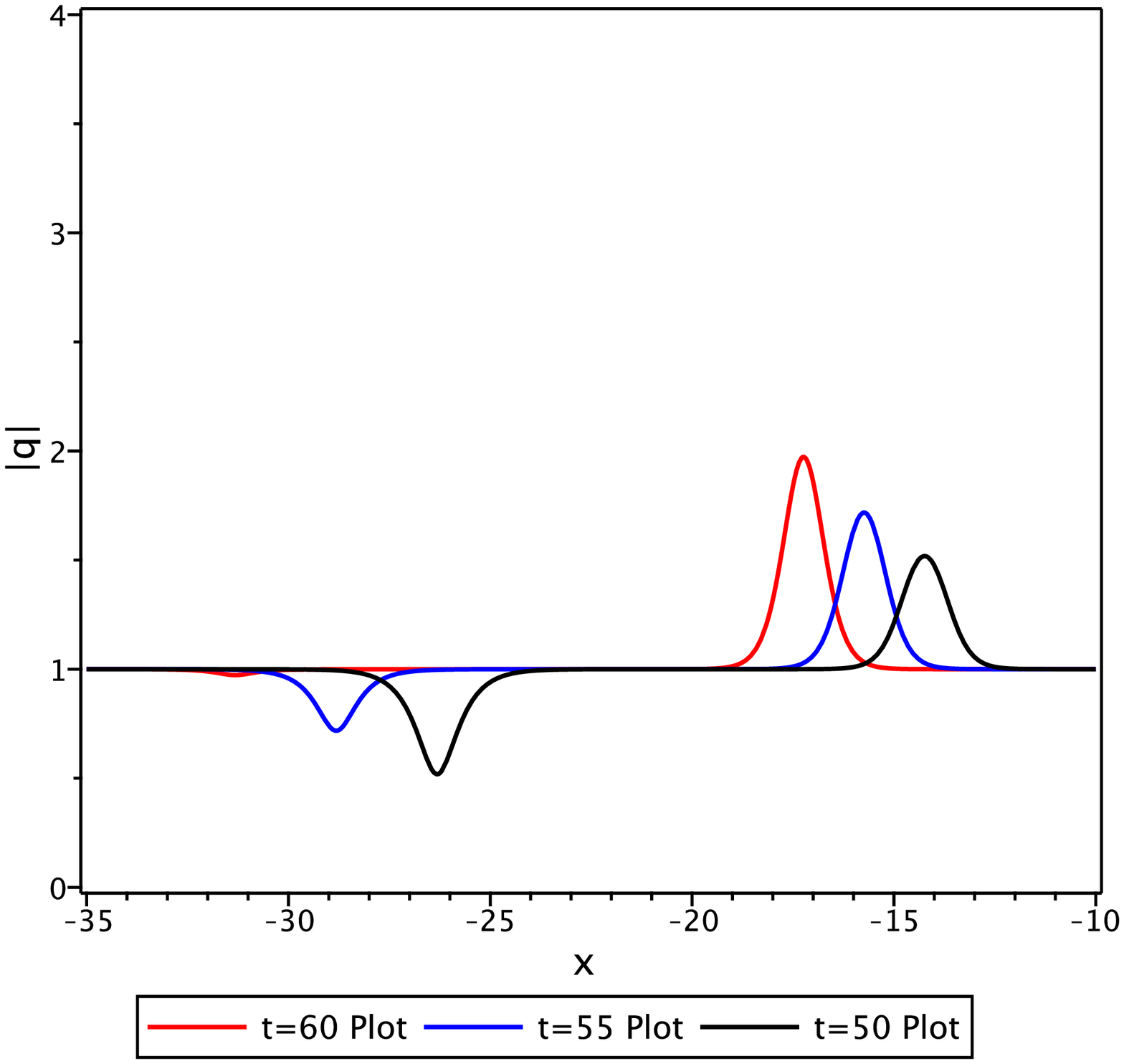}}}
~~~~
{\rotatebox{0}{\includegraphics[width=4.6cm,height=4.0cm,angle=0]{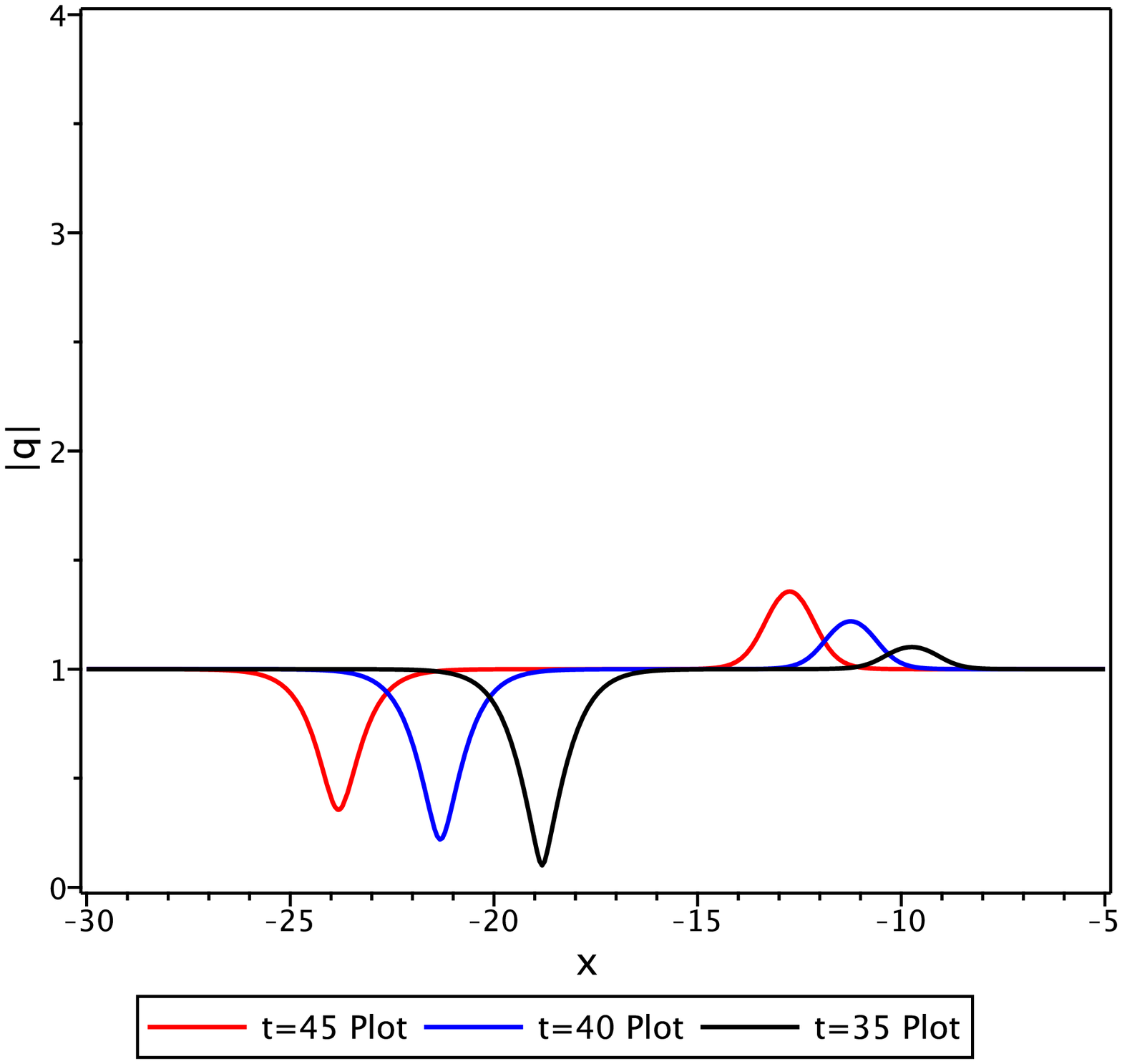}}}\\
$~~~~~~~~~~~~~~~(\textbf{a})~~~~~~~~~~~~~~~~
~~~~~~~~~~~~~~~~~~~~~~~~~(\textbf{b})~~~~~~~
~~~~~~~~~~~~~~~~~~~~~~~~~~~~(\textbf{c})$\\
{\rotatebox{0}{\includegraphics[width=4.6cm,height=4.0cm,angle=0]{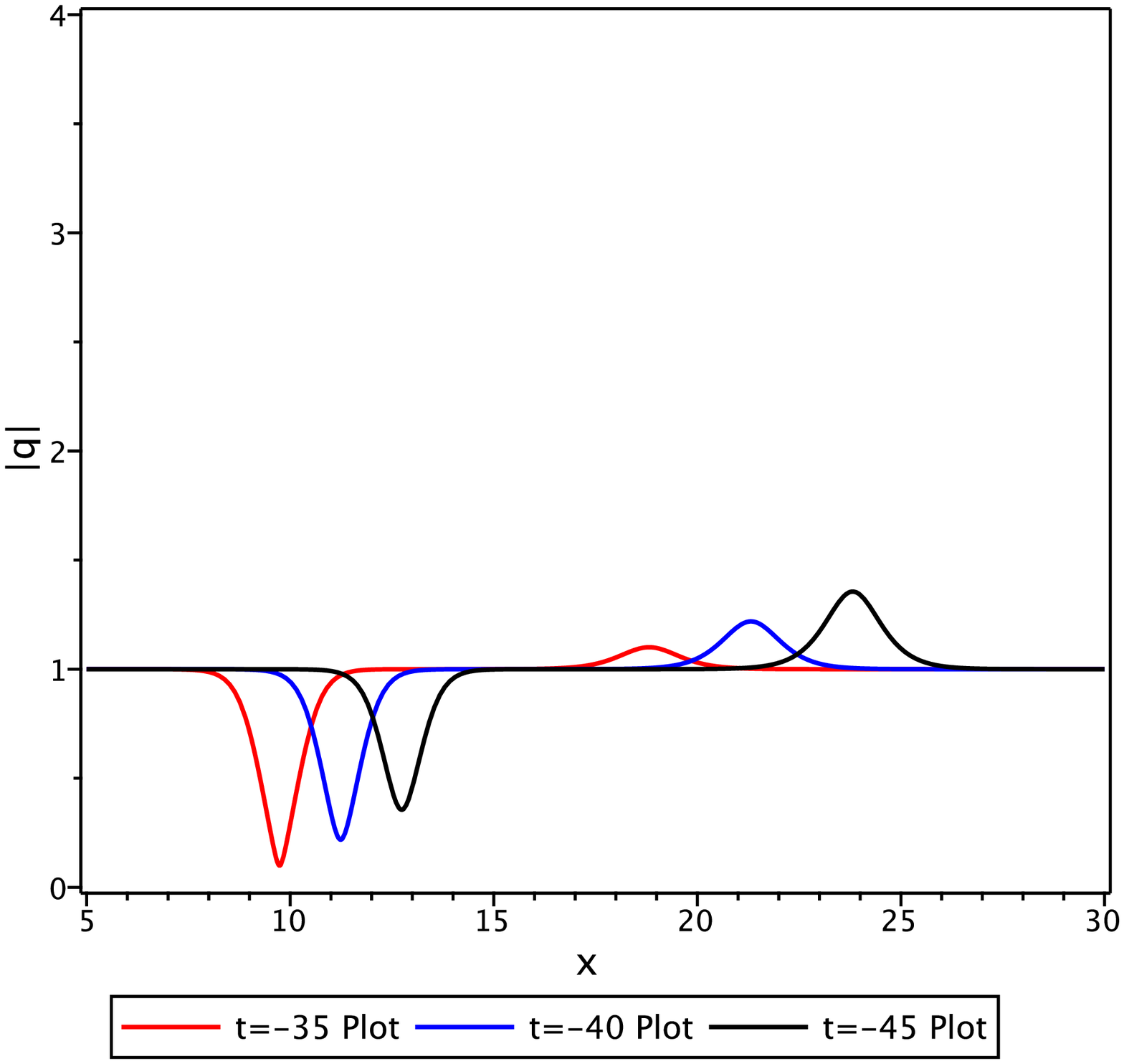}}}
~~~~
{\rotatebox{0}{\includegraphics[width=4.6cm,height=4.0cm,angle=0]{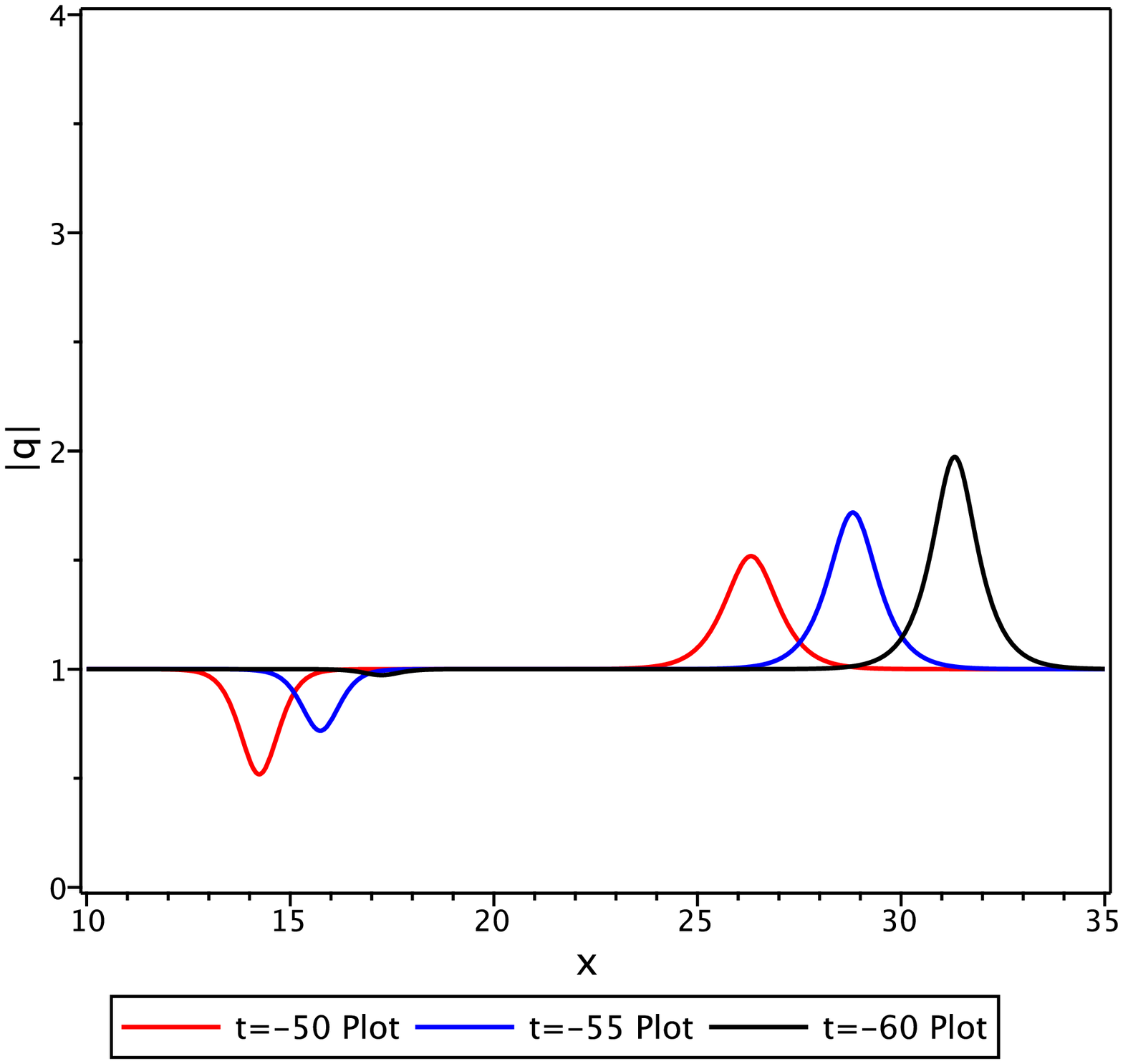}}}
~~~~
{\rotatebox{0}{\includegraphics[width=4.6cm,height=4.0cm,angle=0]{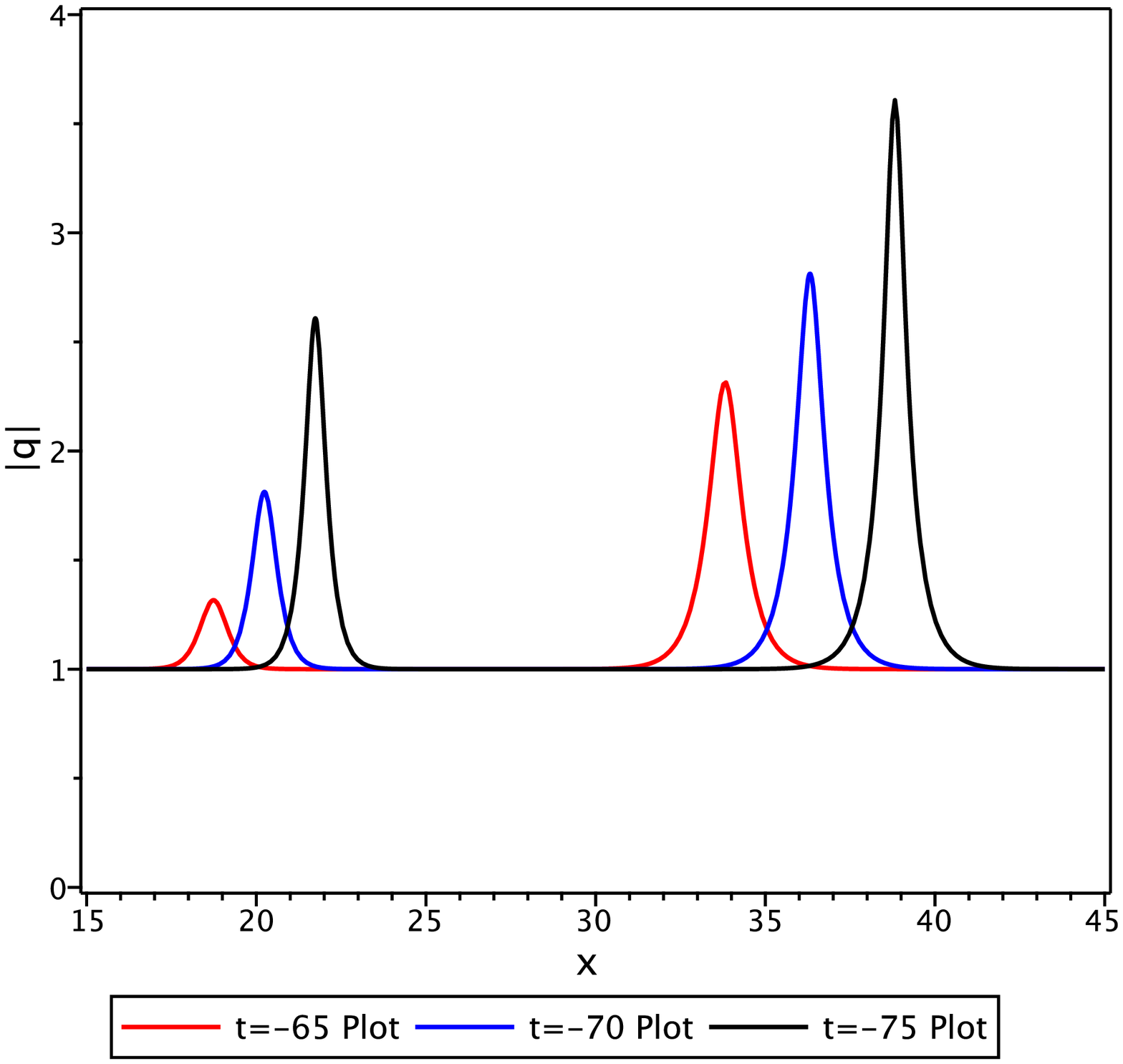}}}\\
$~~~~~~~~~~~~~~~(\textbf{d})~~~~~~~~~~~~~~~~
~~~~~~~~~~~~~~~~~~~~~~~~~(\textbf{e})~~~~~~~
~~~~~~~~~~~~~~~~~~~~~~~~~~~~(\textbf{f})$\\
\noindent { \small \textbf{Figure 7.} (Color online) The wave propagation of two-simple pole solution along the $x$-axis at different time. The parameters are $b_{1}=-1, b_{2}=-1, q_{0}=1, \vartheta_{1}=\frac{\pi}{6}, \vartheta_{2}=\frac{\pi}{3}, \theta_{-}=0, \delta=\frac{1}{100}, \beta=\frac{1}{10}$.}\\

\section{The nonlocal Hirota equation with NZBCs: double poles}

\subsection{Direct scattering problem}
In this part, we suppose that the $N$ discrete spectral points are double zeros of the scattering coefficients $s_{11}(z)$
and $s_{22}(z)$, that is, we have $s_{22}(z_{0})=s'_{22}(z_{0})=0$ and $s''_{22}(z_{0})\neq 0$ for $z_{0}\in D_{+}\cap\left\{z\in\mathbb{ C}: \mbox{Re} z>0\right\}$.

Define
\begin{align}\label{jia1}
b[z_{0}]=\frac{\Psi_{+2}(x, t, z_{0})}{\Psi_{-1}(x, t, z_{0})},\quad d[z_{0}]=\frac{\Psi'_{+2}(x, t; z_{0})-b[z_{0}]\Psi'_{-1}(x, t; z_{0})}{\Psi_{-1}(x, t; z_{0})}, \quad z_{0}\in \Upsilon\cap D_{+},\notag\\
b[z_{0}]=\frac{\Psi_{+1}(x, t, z_{0})}{\Psi_{-2}(x, t, z_{0})},\quad d[z_{0}]=\frac{\Psi'_{+1}(x, t; z_{0})-b[z_{0}]\Psi'_{-2}(x, t; z_{0})}{\Psi_{-2}(x, t; z_{0})}, \quad z_{0}\in \Upsilon\cap D_{-},
\end{align}
\begin{gather}
A[z_{0}]=\left\{
\begin{array}{lr}
\frac{2b[z_{0}]}{s''_{22}(z_{0})}, \quad z_{0}\in \Upsilon\cap D_{+},\\
\\
\frac{2b[z_{0}]}{s''_{11}(z_{0})}, \quad z_{0}\in \Upsilon\cap D_{-},
\end{array}
\right.\notag\\
B[z_{0}]=\left\{
\begin{array}{lr}
\frac{d[z_{0}]}{b[z_{0}]}-\frac{s'''_{22}(z_{0})}{3s''_{22}(z_{0})}, \quad z_{0}\in \Upsilon\cap D_{+},\\
\\
\frac{d[z_{0}]}{b[z_{0}]}-\frac{s'''_{11}(z_{0})}{3s''_{11}(z_{0})}, \quad z_{0}\in \Upsilon\cap D_{-}.
\end{array}
\right. \label{jia2}
\end{gather}
\begin{align}\label{jia3}
&\mathop{L_{-2}}_{z=z_{0}}\left[\frac{\Psi_{+2}(x, t; z)}{s_{22}(z)}\right]=A[z_{0}]\Psi_{-1}(x, t; z_{0}), \quad z_{0}\in \Upsilon\cap D_{+},\notag\\
&\mathop{L_{-2}}_{z=z_{0}}\left[\frac{\Psi_{+1}(x, t; z)}{s_{11}(z)}\right]=A[z_{0}]\Psi_{-2}(x, t; z_{0}), \quad z_{0}\in \Upsilon\cap D_{-},\notag\\
&\mathop{\mbox{Res}}_{z=z_{0}}\left[\frac{\Psi_{+2}(x, t; z)}{s_{22}(z)}\right]=A[z_{0}]\left[\Psi'_{-1}(x, t; z_{0})+B[z_{0}]\Psi_{-1}(x, t; z_{0})\right], \quad z_{0}\in \Upsilon\cap D_{+},\notag\\
&\mathop{\mbox{Res}}_{z=z_{0}}\left[\frac{\Psi_{+1}(x, t; z)}{s_{11}(z)}\right]=A[z_{0}]\left[\Psi'_{-2}(x, t; z_{0})+B[z_{0}]\Psi_{-2}(x, t; z_{0})\right], \quad z_{0}\in \Upsilon\cap D_{-},
\end{align}

\noindent \textbf{Proposition 4.1.} \emph{Two relations for $b[z_{0}]$, $d[z_{0}]$, $s_{22}^{(m)}(z_{0})$
 and $s_{11}^{(m)}(z_{0})$ are derived below:
}

\emph{$\bullet$ The first relation}
\begin{align}\label{jia4}
&b[z_{0}]=-\frac{1}{b[-z_{0}^{\ast}]^{\ast}}, \quad d[z_{0}]=-\frac{d[-z_{0}^{\ast}]^{\ast}}{b^{2}[-z_{0}^{\ast}]^{\ast}},  \notag\\ &s_{11}^{(m)}(z_{0})=(-1)^{m}s_{11}^{(m)}(-z_{0}^{\ast})^{\ast}, \quad s_{22}^{(m)}(z_{0})=(-1)^{m}s_{22}^{(m)}(-z_{0}^{\ast})^{\ast}, \ m\in \mathbb{N}.
\end{align}

\emph{$\bullet$ The second relation}
\begin{align}\label{jia5}
&b[z_{0}]=-\frac{q_{+}}{q_{-}^{\ast}}b[\frac{q_{0}^{2}}{z_{0}}], \quad d[z_{0}]=\frac{q_{+}q_{-}}{z_{0}^{2}}d[\frac{q_{0}^{2}}{z_{0}}], \quad s_{22}^{(m)}(z_{0})=(-\frac{q_{0}^{2}}{z_{0}^{2}})^{m}s_{11}^{(m)}(\frac{q_{0}^{2}}{z_{0}}),\quad z_{0}\in D_{+},\notag\\
&b[z_{0}]=-\frac{q_{+}^{\ast}}{q_{-}}b[\frac{q_{0}^{2}}{z_{0}}], \quad d[z_{0}]=\frac{q^{\ast}_{+}q^{\ast}_{-}}{z_{0}^{2}}d[\frac{q_{0}^{2}}{z_{0}}], \quad s_{11}^{(m)}(z_{0})=(-\frac{q_{0}^{2}}{z_{0}^{2}})^{m}s_{22}^{(m)}(\frac{q_{0}^{2}}{z_{0}}),\quad z_{0}\in D_{-}.
\end{align}
\subsection{Inverse problem with NZBCs and double poles}

By subtracting out the asymptotic values as $z\rightarrow\infty, z\rightarrow 0$, the residue, and the coefficient $L_{-2}$ from the original non-regular RHP, one can obtain the following regular RHP
\begin{gather}
M_{-}+\frac{i}{z}\sigma_{3}Q_{-}-I-\notag\\
\sum_{n=1}^{2N}\left[\frac{\mathop{L_{-2}}\limits_{z=\zeta_{n}}M_{+}}{(z-\zeta_{n})^{2}}
+\frac{\mathop{\mbox{Res}}\limits_{z=\zeta_{n}}M_{+}}{z-\zeta_{n}}+\frac{\mathop{L_{-2}}\limits_{z=\hat{\zeta}_{n}}M_{-}}
{(z-\hat{\zeta}_{n})^{2}}+\frac{\mathop{\mbox{Res}}\limits_{z=\hat{\zeta}_{n}}M_{-}}{z-\hat{\zeta}_{n}}\right]=\notag\\
M_{+}+\frac{i}{z}\sigma_{3}Q_{-}-I-\notag\\
\sum_{n=1}^{2N}\left[\frac{\mathop{L_{-2}}\limits_{z=\zeta_{n}}M_{+}}{(z-\zeta_{n})^{2}}
+\frac{\mathop{\mbox{Res}}\limits_{z=\zeta_{n}}M_{+}}{z-\zeta_{n}}+\frac{\mathop{L_{-2}}\limits_{z=\hat{\zeta}_{n}}M_{-}}
{(z-\hat{\zeta}_{n})^{2}}+\frac{\mathop{\mbox{Res}}\limits_{z=\hat{\zeta}_{n}}M_{-}}{z-\hat{\zeta}_{n}}\right]-M_{+}G.\label{jia7}
\end{gather}
Via the Plemelj's formulae, the above RHP can be solved as
\begin{gather}
M(x, t; z)=I-\frac{i}{z}\sigma_{3}Q_{-}+\frac{1}{2\pi i}\int_{\Sigma}\frac{M_{+}(x, t; \zeta)G(x, t; \zeta)}{\zeta-z}d\zeta\notag\\
+\sum_{n=1}^{2N}\left[\frac{\mathop{L_{-2}}\limits_{z=\zeta_{n}}M_{+}}{(z-\zeta_{n})^{2}}
+\frac{\mathop{\mbox{Res}}\limits_{z=\zeta_{n}}M_{+}}{z-\zeta_{n}}+\frac{\mathop{L_{-2}}\limits_{z=\hat{\zeta}_{n}}M_{-}}
{(z-\hat{\zeta}_{n})^{2}}+\frac{\mathop{\mbox{Res}}\limits_{z=\hat{\zeta}_{n}}M_{-}}{z-\hat{\zeta}_{n}}\right].\label{jia8}
\end{gather}
The residue and the coefficient $L_{-2}$ of $M(x, t; z)$ in Eq.\eqref{jia8}  can be written as
\begin{align}\label{jia6}
&\mathop{L_{-2}}_{z=\zeta_{n}}M_{+}=\left(0, A[\zeta_{n}]e^{-2i\theta(x, t; \zeta_{n})}\mu_{-1}(x, t; \zeta_{n})\right),\notag\\
&\mathop{L_{-2}}_{z=\hat{\zeta}_{n}}M_{-}=\left(A[\hat{\zeta}_{n}]e^{2i\theta(x, t; \hat{\zeta}_{n})}\mu_{-2}(x, t; \hat{\zeta}_{n}), 0\right),\notag\\
&\mathop{\mbox{Res}}_{z=\zeta_{n}}M_{+}=\left(0, A[\zeta_{n}]e^{-2i\theta(x, t; \zeta_{n})}\left[\mu'_{-1}(x, t; \zeta_{n})+\left[B[\zeta_{n}]-2i\theta'(x, t; \zeta_{n})\right]\mu_{-1}(x, t; \zeta_{n})\right]\right),\notag\\
&\mathop{\mbox{Res}}_{z=\hat{\zeta}_{n}}M_{-}=\left(A[\hat{\zeta}_{n}]e^{2i\theta(x, t; \hat{\zeta}_{n})}\left[\mu'_{-2}(x, t; \hat{\zeta}_{n})+\left[B[\hat{\zeta}_{n}]+2i\theta'(x, t; \hat{\zeta}_{n})\right]\mu_{-2}(x, t; \hat{\zeta}_{n})\right], 0\right).
\end{align}
Then, one has
\begin{gather}
\frac{\mathop{L_{-2}}\limits_{z=\zeta_{n}}M_{+}}{(z-\zeta_{n})^{2}}
+\frac{\mathop{\mbox{Res}}\limits_{z=\zeta_{n}}M_{+}}{z-\zeta_{n}}+\frac{\mathop{L_{-2}}\limits_{z=\hat{\zeta}_{n}}M_{-}}
{(z-\hat{\zeta}_{n})^{2}}+\frac{\mathop{\mbox{Res}}\limits_{z=\hat{\zeta}_{n}}M_{-}}{z-\hat{\zeta}_{n}}=\notag\\
\left(\hat{C}_{n}(z)\left[\mu'_{-2}(\hat{\zeta}_{n})+\left(\hat{D}_{n}+\frac{1}{z-\hat{\zeta}_{n}}\right)\mu_{-2}(\hat{\zeta}_{n})\right],
C_{n}(z)\left[\mu'_{-1}(\zeta_{n})+\left(D_{n}+\frac{1}{z-\zeta_{n}}\right)\mu_{-1}(\zeta_{n})\right]\right),\label{jia9}
\end{gather}
and
\begin{align}\label{jia10}
&C_{n}(z)=\frac{A[\zeta_{n}]e^{-2i\theta(\zeta_{n})}}{z-\zeta_{n}}, \quad D_{n}=B[\zeta_{n}]-2i\theta'(\zeta_{n}), \notag\\
&\hat{C}_{n}(z)=\frac{A[\hat{\zeta}_{n}]e^{2i\theta(\hat{\zeta}_{n})}}{z-\hat{\zeta}_{n}}, \quad \hat{D}_{n}=B[\hat{\zeta}_{n}]+2i\theta'(\hat{\zeta}_{n}).
\end{align}
Moreover, according to \eqref{26}, we obtain
\begin{gather}\label{jia16}
M^{(1)}(x, t; z)=-\frac{1}{2\pi i}\int_{\Sigma}M_{+}(x, t; \zeta)G(x, t; \zeta)d\zeta -i\sigma_{3}Q_{-} \notag\\
+\sum_{n=1}^{2N}\left(A[\hat{\zeta}_{n}]e^{2i\theta(\hat{\zeta}_{n})}\left(\mu'_{-2}(\hat{\zeta}_{n})+\hat{D}_{n}\mu_{-2}(\hat{\zeta}_{n})\right),
A[\zeta_{n}]e^{-2i\theta(\zeta_{n})}\left(\mu'_{-1}(\zeta_{n})+D_{n}\mu_{-1}(\zeta_{n})\right)\right).
\end{gather}
The potential $q(x, t)$ with double poles for the nonlocal Hirota equation with NZBCs is given by
\begin{gather}
q(x, t)=iM_{12}^{(1)}=q_{-}
-\frac{1}{2\pi}\int_{\Sigma}(M_{+}(x, t; \zeta)G(x, t; \zeta))_{12}d\zeta \notag\\
+i\sum_{n=1}^{2N}
A[\zeta_{n}]e^{-2i\theta(\zeta_{n})}\left(\mu'_{-11}(\zeta_{n})+D_{n}\mu_{-11}(\zeta_{n})\right).\label{jia17}
\end{gather}

\subsection{Trace formula and theta condition}
The scattering coefficients $s_{22}(z)$ and $s_{11}(z)$ respectively have simple zeros $\zeta_{n}$ and $\hat{\zeta}_{n}$, thus we can take
\begin{align}\label{jia18}
\beta^{+}(z)=s_{22}(z)\prod_{n=1}^{2N}(\frac{z-\hat{\zeta}_{n}}{z-\zeta_{n}})^{2},\ \beta^{-}(z)=s_{11}(z)\prod_{n=1}^{2N}(\frac{z-\zeta_{n}}{z-\hat{\zeta}_{n}})^{2},
\end{align}
that means $\beta^{+}(z)$ is analytic and has no zeros in $D_{+}$, and $\beta^{-}(z)$ is analytic and has no zeros in $D_{-}$.
Also, $\beta^{\pm}(z)\rightarrow o(1)$ as $z\rightarrow\infty$. According to the Plemelj's formulae, $\beta^{\pm}(z)$ can be written as
\begin{align}\label{jia19}
\log\beta^{\pm}(z)=\mp\frac{1}{2\pi i}\int_{\Sigma}\frac{\log(1-\rho\tilde{\rho})}{s-z}ds,\quad z\in D^{\pm}.
\end{align}
Using Eq.\eqref{jia18}, we derive the following trace formulae
\begin{align}\label{jia20}
&s_{22}(z)=\mbox{exp}\left[-\frac{1}{2\pi i}\int_{\Sigma}\frac{\log(1-\rho\tilde{\rho})}{s-z}ds\right] \prod_{n=1}^{2N}(\frac{z-\zeta_{n}}{z-\hat{\zeta}_{n}})^{2},\notag\\
&s_{11}(z)=\mbox{exp}\left[\frac{1}{2\pi i}\int_{\Sigma}\frac{\log(1-\rho\tilde{\rho})}{s-z}ds\right] \prod_{n=1}^{2N}(\frac{z-\hat{\zeta}_{n}}{z-\zeta_{n}})^{2}.
\end{align}
Let $z\rightarrow0$ in the first formula of Eq.\eqref{jia20}, one has
\begin{align}\label{jia21}
\mbox{exp}\left[\frac{i}{2\pi}\int_{\Sigma}\frac{\log(1-\rho\tilde{\rho})}{s}ds\right] \prod_{n=1}^{N}\frac{\mid z_{n}\mid^{8}}{q_{0}^{8}}=1.
\end{align}
In addition, taking the derivative of the Eq.\eqref{jia20} twice with respect to $z$, we obtain $s'_{22}(\zeta_{j})$ and $s'_{11}(\hat{\zeta}_{j})$, given by
\begin{align}\label{jia22}
&s''_{22}(\zeta_{j})=\mbox{exp}\left[-\frac{1}{2\pi i}\int_{\Sigma}\frac{\log(1-\rho\tilde{\rho})}{s-\zeta_{j}}ds\right] \frac{\prod_{m\neq j}2(\zeta_{j}-\zeta_{m})^{2}}{\prod_{m=1}^{2N}(\zeta_{j}-\hat{\zeta}_{m})^{2}},\notag\\
&s''_{11}(\hat{\zeta}_{j})=\mbox{exp}\left[\frac{1}{2\pi i}\int_{\Sigma}\frac{\log(1-\rho\tilde{\rho})}{s-\hat{\zeta}_{j}}ds\right] \frac{\prod_{m\neq j}2(\hat{\zeta}_{j}-\hat{\zeta}_{m})^{2}}{\prod_{m=1}^{2N}(\hat{\zeta}_{j}-\zeta_{m})^{2}}.
\end{align}
Notice that the general expressions of the $s'''_{22}(\zeta_{j}), s'''_{11}(\hat{\zeta}_{j})$ are very complicated, and we omit them here. However, with the aid of computer softwares such as Maple and Matlab, one can easily get the corresponding
exact expressions by taking the third derivative of the Eq.\eqref{jia20} with respect to $z$.
\subsection{Double poles soliton solutions of nonlocal Hirota equation under NZBCs}
In this subsection, for the case of double poles with reflectionless potential, we first need to evaluate $\mu'_{-1}(\zeta_{n}), \mu_{-1}(\zeta_{n}), \mu'_{-2}(\zeta_{n}), \mu_{-2}(\zeta_{n})$. When $\rho(z)=\tilde{\rho}(z)=0$, the second column of Eq.\eqref{jia8} meets
\begin{align}\label{jia11}
\mu_{-2}(z)=\left(\begin{array}{c}
    -\frac{i}{z}q_{-}\\
  1\\
\end{array}\right)+\sum_{n=1}^{2N}C_{n}(z)\left[\mu'_{-1}(\zeta_{n})+\left(D_{n}+\frac{1}{z-\zeta_{n}}\right)\mu_{-1}(\zeta_{n})\right],
\end{align}
\begin{align}\label{jia12}
\mu_{-2}'(z)=\left(\begin{array}{c}
    \frac{i}{z^{2}}q_{-}\\
  0\\
\end{array}\right)-\sum_{n=1}^{2N}\frac{C_{n}(z)}{z-\zeta_{n}}\left[\mu'_{-1}(\zeta_{n})+\left(D_{n}+\frac{2}{z-\zeta_{n}}\right)\mu_{-1}(\zeta_{n})\right].
\end{align}
We take the first-order derivative about $z$ in formula \eqref{37}, given by
\begin{align}\label{jia13}
\mu_{-2}'(z)=\frac{iq_{-}}{z^{2}}\mu_{-1}(\frac{q_{0}^{2}}{z})+\frac{iq_{0}^{2}q_{-}}{z^{3}}\mu'_{-1}(\frac{q_{0}^{2}}{z}).
\end{align}
Substituting Eqs.\eqref{jia13} and \eqref{37} into Eqs. \eqref{jia12} and \eqref{jia11},  and letting $z=\hat{\zeta}_{j}, j=1, 2,\cdots, 2N$, we obtain the following $4N$ linear system
\begin{align}\label{jia14}
\sum_{n=1}^{2N}\left\{C_{n}(\hat{\zeta}_{j})\mu'_{-1}(\zeta_{n})+\left[C_{n}(\hat{\zeta}_{j})\left(D_{n}+\frac{1}{\hat{\zeta}_{j}-\zeta_{n}}\right)+\frac{iq_{-}}{\hat{\zeta}_{j}}\delta_{j,n}\right]\mu_{-1}(\zeta_{n})
\right\}=\left(\begin{array}{c}
    \frac{i}{\hat{\zeta}_{j}}q_{-}\\
  -1\\
\end{array}\right),
\end{align}
\begin{gather}
\sum_{n=1}^{2N}\left\{\left(\frac{C_{n}(\hat{\zeta}_{j})}{\hat{\zeta}_{j}-\zeta_{n}}+\frac{iq_{0}^{2}q_{-}}{\hat{\zeta}_{j}^{3}}\delta_{j,n}\right)
\mu'_{-1}(\zeta_{n})\right.\notag\\
\left.+\left[\frac{C_{n}(\hat{\zeta}_{j})}{\hat{\zeta}_{j}-\zeta_{n}}\left(D_{n}+\frac{2}{\hat{\zeta}_{j}-\zeta_{n}}\right)
+\frac{iq_{-}}{\hat{\zeta}_{j}^{2}}\delta_{j,n}\right]\mu_{-1}(\zeta_{n})\right\}
=\left(\begin{array}{c}
    \frac{i}{\hat{\zeta}_{j}^{2}}q_{-}\\
  0\\
\end{array}\right).\label{jia15}
\end{gather}

\noindent \textbf{Theorem 4.1}  \emph{
The general formula of the double poles solution for the nonlocal Hirota \eqref{10} with NZBCs \eqref{11} is expressed as
\begin{align}\label{jia23}
q(x,t)=q_{-}-i\frac{\det \left(\begin{array}{cc}
    \mathcal{G}  &  \psi\\
  \omega^{T} &  0\\
\end{array}\right)}{\det (\mathcal{G})}.
\end{align}
}
\begin{proof}
According to Eqs.\eqref{jia14} and \eqref{jia15}, we easily obtain a $4N$ linear system with respect to $\mu_{-11}(\zeta_{n})$, $\mu'_{-11}(\zeta_{n})$
\begin{align}\label{jia24}
\sum_{n=1}^{2N}\left\{C_{n}(\hat{\zeta}_{j})\mu'_{-11}(\zeta_{n})+\left[C_{n}(\hat{\zeta}_{j})\left(D_{n}+\frac{1}{\hat{\zeta}_{j}-\zeta_{n}}\right)+\frac{iq_{-}}{\hat{\zeta}_{j}}\delta_{j,n}\right]\mu_{-11}(\zeta_{n})
\right\}=\frac{iq_{-}}{\hat{\zeta}_{j}}
\end{align}
\begin{gather}
\sum_{n=1}^{2N}\left\{\left(\frac{C_{n}(\hat{\zeta}_{j})}{\hat{\zeta}_{j}-\zeta_{n}}+\frac{iq_{0}^{2}q_{-}}{\hat{\zeta}_{j}^{3}}\delta_{j,n}\right)
\mu'_{-11}(\zeta_{n})\right.\notag\\
\left.+\left[\frac{C_{n}(\hat{\zeta}_{j})}{\hat{\zeta}_{j}-\zeta_{n}}\left(D_{n}+\frac{2}{\hat{\zeta}_{j}-\zeta_{n}}\right)
+\frac{iq_{-}}{\hat{\zeta}_{j}^{2}}\delta_{j,n}\right]\mu_{-11}(\zeta_{n})\right\}
=\frac{iq_{-}}{\hat{\zeta}_{j}^{2}},\label{jia25}
\end{gather}
the above linear system can be denoted in the matrix form:
\begin{align}\label{jia26}
\mathcal{G}\eta=\psi,
\end{align}
where
\begin{align}\label{jia27}
&\psi=\left(\begin{array}{c}
     \psi^{(1)}\\
  \psi^{(2)}\\
\end{array}\right),\ \psi^{(1)}=(\frac{iq_{-}}{\hat{\zeta}_{1}}, \frac{iq_{-}}{\hat{\zeta}_{2}},\cdots, \frac{iq_{-}}{\hat{\zeta}_{2N}})^{T}, \psi^{(2)}=(\frac{iq_{-}}{\hat{\zeta}_{1}^{2}}, \frac{iq_{-}}{\hat{\zeta}_{2}^{2}},\cdots, \frac{iq_{-}}{\hat{\zeta}_{2N}^{2}})^{T}, \notag\\
&\eta=\left(\begin{array}{c}
     \eta^{(1)}\\
  \eta^{(2)}\\
\end{array}\right),\ \eta^{(1)}=(\mu_{-11}(\zeta_{1}), \mu_{-11}(\zeta_{2}),\cdots, \mu_{-11}(\zeta_{2N}))^{T}, \notag\\
&\eta^{(2)}=(\mu'_{-11}(\zeta_{1}), \mu'_{-11}(\zeta_{2}),\cdots, \mu'_{-11}(\zeta_{2N}))^{T}, \mathcal{G}=\left(\begin{array}{cc}
    \mathcal{G}^{(11)} &  \mathcal{G}^{(12)}\\
  \mathcal{G}^{(21)} &  \mathcal{G}^{(22)}\\
\end{array}\right),
\end{align}
with $\mathcal{G}^{(im)}=\left(\mathcal{G}^{(im)}_{jn}\right)_{2N\times 2N}(i, m=
1, 2)$ given by
\begin{align}\label{jia28}
&\mathcal{G}^{(11)}_{jn}=C_{n}(\hat{\zeta}_{j})\left(D_{n}+\frac{1}{\hat{\zeta}_{j}-\zeta_{n}}\right)+\frac{iq_{-}}{\hat{\zeta}_{j}}\delta_{j,n}, \quad
\mathcal{G}^{(12)}_{jn}=C_{n}(\hat{\zeta}_{j}), \notag\\
&\mathcal{G}^{(21)}_{jn}=\frac{C_{n}(\hat{\zeta}_{j})}{\hat{\zeta}_{j}-\zeta_{n}}\left(D_{n}+\frac{2}{\hat{\zeta}_{j}-\zeta_{n}}\right)
+\frac{iq_{-}}{\hat{\zeta}_{j}^{2}}\delta_{j,n}, \quad
\mathcal{G}^{(22)}_{jn}=\frac{C_{n}(\hat{\zeta}_{j})}{\hat{\zeta}_{j}-\zeta_{n}}+\frac{iq_{0}^{2}q_{-}}{\hat{\zeta}_{j}^{3}}\delta_{j,n}.
\end{align}
At the case of reflectionless potential, Eq.\eqref{jia17} can be defined as
\begin{align}\label{jia29}
q=q_{-}+i\omega^{T} \eta,
\end{align}
where
\begin{align}\label{jia30}
&\omega=\left(\begin{array}{c}
     \omega^{(1)}\\
  \omega^{(2)}\\
\end{array}\right),
\omega^{(2)}=(A[\zeta_{1}]e^{-2i\theta(\zeta_{1})}, A[\zeta_{2}]e^{-2i\theta(\zeta_{2})},\cdots, A[\zeta_{2N}]e^{-2i\theta(\zeta_{2N})})^{T},\notag\\
&\omega^{(1)}=(A[\zeta_{1}]e^{-2i\theta(\zeta_{1})}D_{1}, A[\zeta_{2}]e^{-2i\theta(\zeta_{2})}D_{2},\cdots, A[\zeta_{2N}]e^{-2i\theta(\zeta_{2N})}D_{2N})^{T}.
\end{align}
From Eqs. \eqref{jia26}, we get the expression of the double poles soliton solution.
\end{proof}

For example, we have the one-double poles soliton solution of the nonlocal Hirota equation with NZBCs when $N=1$. Let $z_{1}=q_{0}e^{i\vartheta_{1}}, \vartheta_{1}\in(0, \frac{\pi}{2}).$ Then, from \eqref{28}, we have $\zeta_{1}=q_{0}e^{i\vartheta_{1}}, \zeta_{2}=-q_{0}e^{-i\vartheta_{1}}, \hat{\zeta}_{1}=q_{0}e^{-i\vartheta_{1}}, \hat{\zeta}_{2}=-q_{0}e^{i\vartheta_{1}}.$  Let $q_{-}=q_{0}e^{i\theta_{-}}, \theta_{-}\in \{0, \pi \}$, $b[\zeta_{1}]=b_{1}, d[\zeta_{1}]=d_{1}$, where $b_{1}, d_{1}$ are all arbitrary parameters. From Proposition 4.1, we have $b[\zeta_{2}]=-\frac{1}{b_{1}^{\ast}}, d[\zeta_{2}]=-(\frac{d_{1}}{b_{1}^{2}})^{\ast}$. From Eq.\eqref{jia22}, one has
\begin{align}\label{jia31}
&s_{22}''(\zeta_{1})=\frac{2(\zeta_{1}-\zeta_{2})^{2}}{(\zeta_{1}-\hat{\zeta}_{1})^{2}(\zeta_{1}-\hat{\zeta}_{2})^{2}}
=\frac{(\cos(\vartheta_{1}))^{2}}{2(q_{0}ie^{i\vartheta_{1}}\sin(\vartheta_{1}))^{2}},\notag\\ &s_{22}''(\zeta_{2})=\frac{2(\zeta_{2}-\zeta_{1})^{2}}{(\zeta_{2}-\hat{\zeta}_{1})^{2}(\zeta_{2}-\hat{\zeta}_{2})^{2}}
=\frac{(\cos(\vartheta_{1}))^{2}}{2(q_{0}ie^{-i\vartheta_{1}}\sin(\vartheta_{1}))^{2}},
\end{align}
and through Maple computation, we get
\begin{align}\label{jia32}
&s_{22}'''(\zeta_{1})=\frac{12(\zeta_{1}-\zeta_{2})(2\zeta_{1}\zeta_{2}-\zeta_{1}^{2}-(\hat{\zeta}_{1}+\hat{\zeta}_{2})\zeta_{2}+\hat{\zeta}_{1}\hat{\zeta}_{2})}
{(\zeta_{1}-\hat{\zeta}_{1})^{3}(\zeta_{1}-\hat{\zeta}_{2})^{3}},\notag\\ &s_{22}'''(\zeta_{2})=\frac{12(\zeta_{2}-\zeta_{1})(2\zeta_{1}\zeta_{2}-\zeta_{2}^{2}-(\hat{\zeta}_{1}+\hat{\zeta}_{2})\zeta_{1}+\hat{\zeta}_{1}\hat{\zeta}_{2})}
{(\zeta_{2}-\hat{\zeta}_{1})^{3}(\zeta_{2}-\hat{\zeta}_{2})^{3}},
\end{align}
then using Eq.\eqref{jia2}, we can derive the $A[\zeta_{1}], A[\zeta_{2}], B[\zeta_{1}], B[\zeta_{2}]$.
Thus the one-double poles solution of the nonlocal Hirota equation \eqref{10} is deduced as
\begin{align}\label{jia33}
q(x,t)=q_{-}-i\frac{\det \left(\begin{array}{ccccc}
    \mathcal{G}^{(11)}_{11}  &  \mathcal{G}^{(11)}_{12}&  \mathcal{G}^{(12)}_{11}&  \mathcal{G}^{(12)}_{12}&  \psi_{1}^{(1)}\\
     \mathcal{G}^{(11)}_{21}  &  \mathcal{G}^{(11)}_{22}&  \mathcal{G}^{(12)}_{21}&  \mathcal{G}^{(12)}_{22}&  \psi_{2}^{(1)}\\
      \mathcal{G}^{(21)}_{11}  &  \mathcal{G}^{(21)}_{12}&  \mathcal{G}^{(22)}_{11}&  \mathcal{G}^{(22)}_{12}&  \psi_{1}^{(2)}\\
       \mathcal{G}^{(21)}_{21}  &  \mathcal{G}^{(21)}_{22}&  \mathcal{G}^{(22)}_{21}&  \mathcal{G}^{(22)}_{22}&  \psi_{2}^{(2)}\\
        \omega^{(1)}_{1}  &  \omega^{(1)}_{2}&  \omega^{(2)}_{1}&  \omega^{(2)}_{2}&  0\\
\end{array}\right)}{\det \left(\begin{array}{cccc}
 \mathcal{G}^{(11)}_{11}  &  \mathcal{G}^{(11)}_{12}&  \mathcal{G}^{(12)}_{11}&  \mathcal{G}^{(12)}_{12}\\
     \mathcal{G}^{(11)}_{21}  &  \mathcal{G}^{(11)}_{22}&  \mathcal{G}^{(12)}_{21}&  \mathcal{G}^{(12)}_{22}\\
      \mathcal{G}^{(21)}_{11}  &  \mathcal{G}^{(21)}_{12}&  \mathcal{G}^{(22)}_{11}&  \mathcal{G}^{(22)}_{12}\\
       \mathcal{G}^{(21)}_{21}  &  \mathcal{G}^{(21)}_{22}&  \mathcal{G}^{(22)}_{21}&  \mathcal{G}^{(22)}_{22}\\
\end{array}\right)},
\end{align}
where
\begin{align}\label{jia34}
&\mathcal{G}^{(11)}_{jn}=C_{n}(\hat{\zeta}_{j})\left(D_{n}+\frac{1}{\hat{\zeta}_{j}-\zeta_{n}}\right)+\frac{iq_{-}}{\hat{\zeta}_{j}}\delta_{j,n}, \quad
\mathcal{G}^{(12)}_{jn}=C_{n}(\hat{\zeta}_{j}), \notag\\
&\mathcal{G}^{(21)}_{jn}=\frac{C_{n}(\hat{\zeta}_{j})}{\hat{\zeta}_{j}-\zeta_{n}}\left(D_{n}+\frac{2}{\hat{\zeta}_{j}-\zeta_{n}}\right)
+\frac{iq_{-}}{\hat{\zeta}_{j}^{2}}\delta_{j,n}, \quad
\mathcal{G}^{(22)}_{jn}=\frac{C_{n}(\hat{\zeta}_{j})}{\hat{\zeta}_{j}-\zeta_{n}}+\frac{iq_{0}^{2}q_{-}}{\hat{\zeta}_{j}^{3}}\delta_{j,n}.\notag\\
&C_{n}(\hat{\zeta}_{j})=\frac{A[\zeta_{n}]e^{-2i\theta(\zeta_{n})}}{\hat{\zeta}_{j}-\zeta_{n}}, \quad D_{n}=B[\zeta_{n}]-2i\theta'(\zeta_{n}), \notag\\
&\omega^{(1)}_{j}=A[\zeta_{j}]e^{-2i\theta(\zeta_{j})}D_{j}, \ \omega^{(2)}_{j}=A[\zeta_{j}]e^{-2i\theta(\zeta_{j})}, \
\psi_{j}^{(1)}=\frac{iq_{-}}{\hat{\zeta}_{j}},\  \psi_{j}^{(2)}=\frac{iq_{-}}{\hat{\zeta}_{j}^{2}},\notag\\
&\theta(\zeta_{j})=\frac{(\zeta_{j}^{2}-q_{0}^{2})\left[\beta t(q_{0}^{4}+4q_{0}^{2}\zeta_{j}^{2}+\zeta_{j}^{4})+i\delta \zeta_{j}t(q_{0}^{2}+\zeta_{j}^{2})+x\zeta_{j}^{2}\right]}{2\zeta_{j}^{3}},\ j, n=1,  2.\notag\\
&\theta'(\zeta_{j})=\frac{3\beta t(q_{0}^{6}+q_{0}^{4}\zeta_{j}^{2}+q_{0}^{2}\zeta_{j}^{4}+\zeta_{j}^{6})+2i\delta \zeta_{j}t(q_{0}^{4}+\zeta_{j}^{4})+\zeta_{j}^{2}x(q_{0}^{2}+\zeta_{j}^{2})}{2\zeta_{j}^{4}}.
\end{align}

As a matter of convenience, we take $b_{1}=1, d_{1}=1, q_{0}=1, \vartheta_{1}=\frac{\pi}{4}, \theta_{-}=0, \delta=\frac{1}{100}, \beta=1$ as a example to illustrate the correlative dynamic behavior for the one-double poles solution for nonlocal Hirota equation
with NZBCs  via image simulation. As displayed in Fig. 8, before the interaction $t=-5$, the wave profile contains two
dark solitons, then they happen the strong collision at $t=0$.
After that, the wave profile becomes a two dark soliton again. Moreover, in Fig. 9, when $|t|$ increases, the two dark soliton degrades into the two anti-dark soliton.
\\
{\rotatebox{0}{\includegraphics[width=4.6cm,height=4.0cm,angle=0]{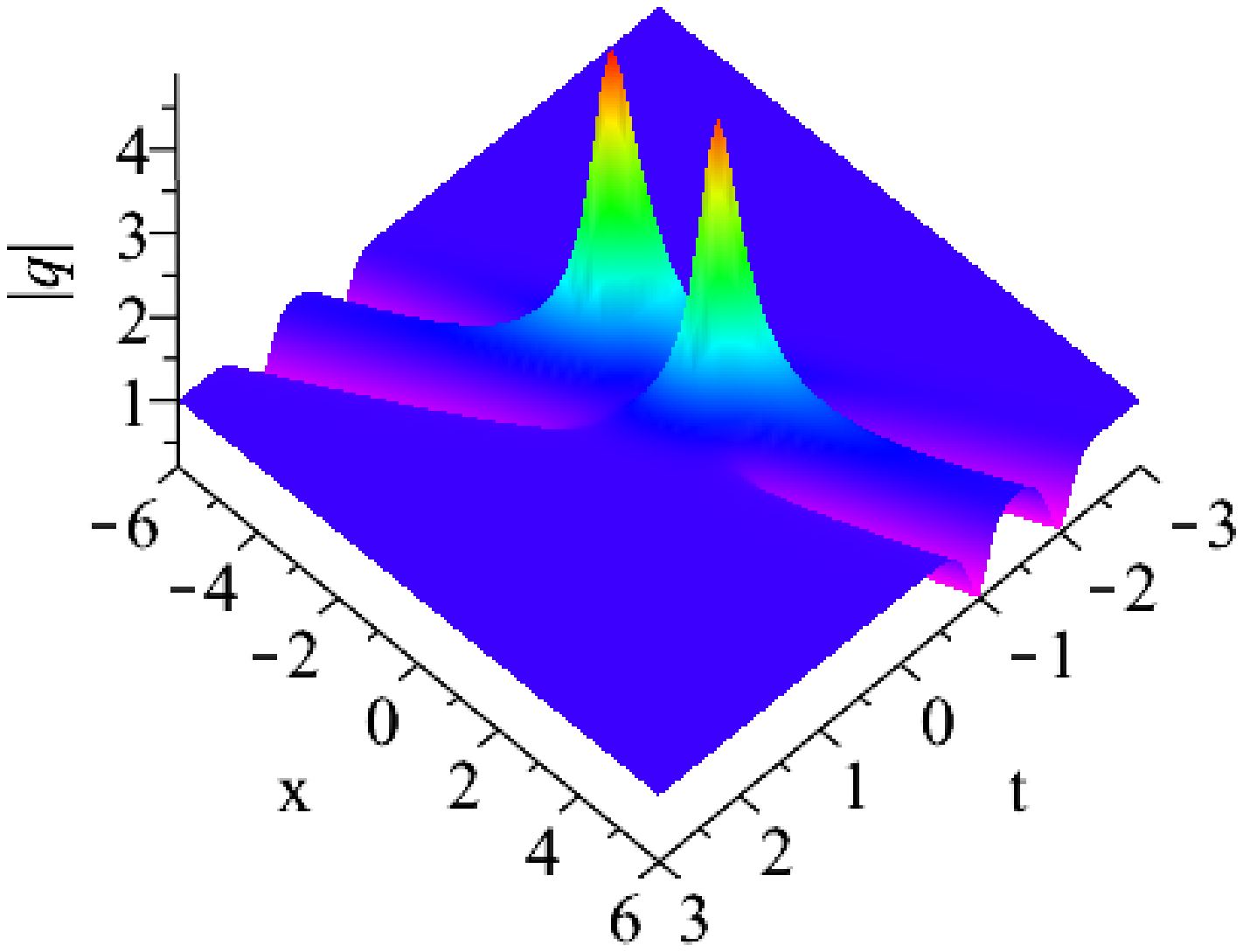}}}
~~~~
{\rotatebox{0}{\includegraphics[width=4.6cm,height=4.0cm,angle=0]{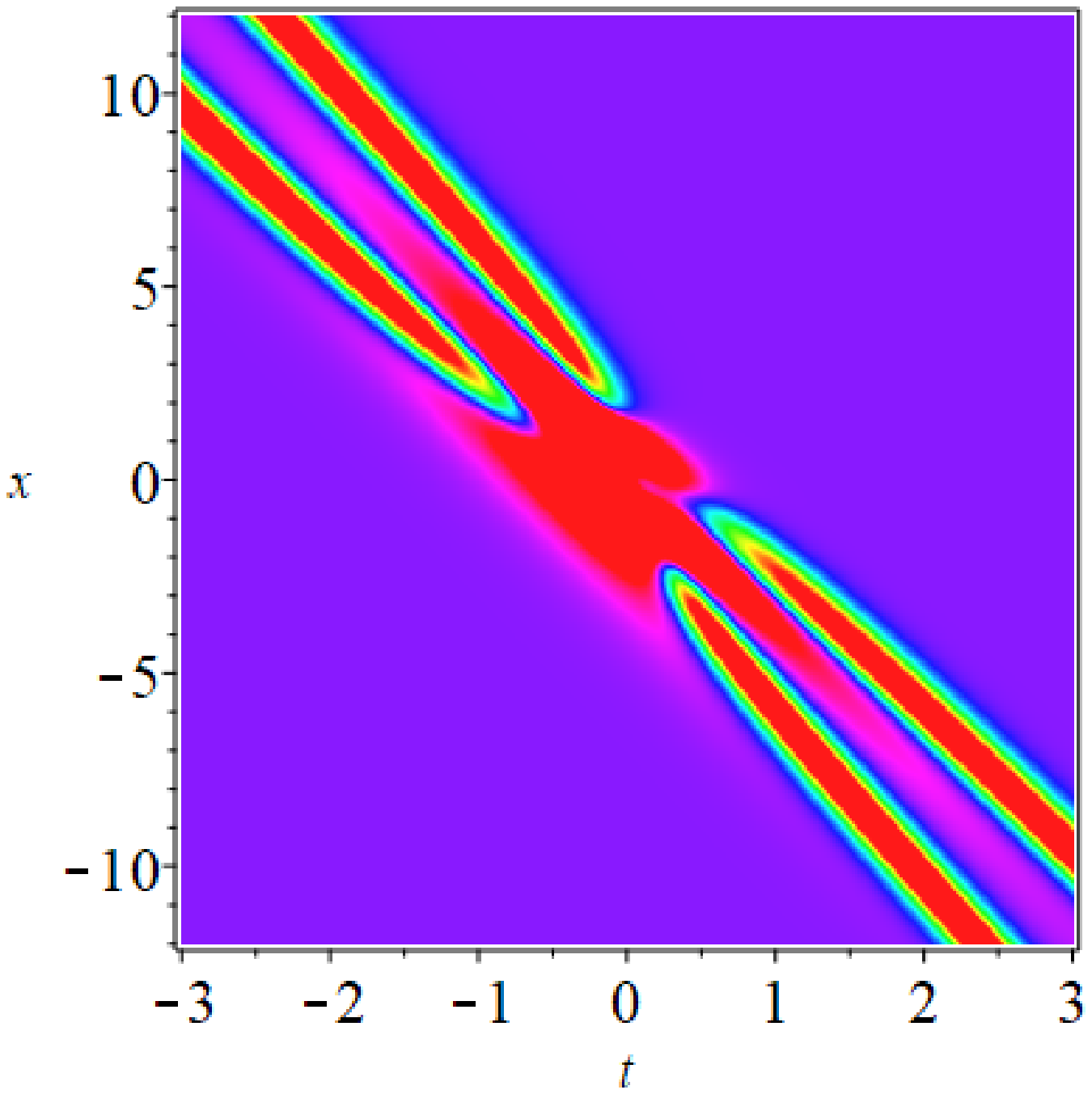}}}
~~~~
{\rotatebox{0}{\includegraphics[width=4.6cm,height=4.0cm,angle=0]{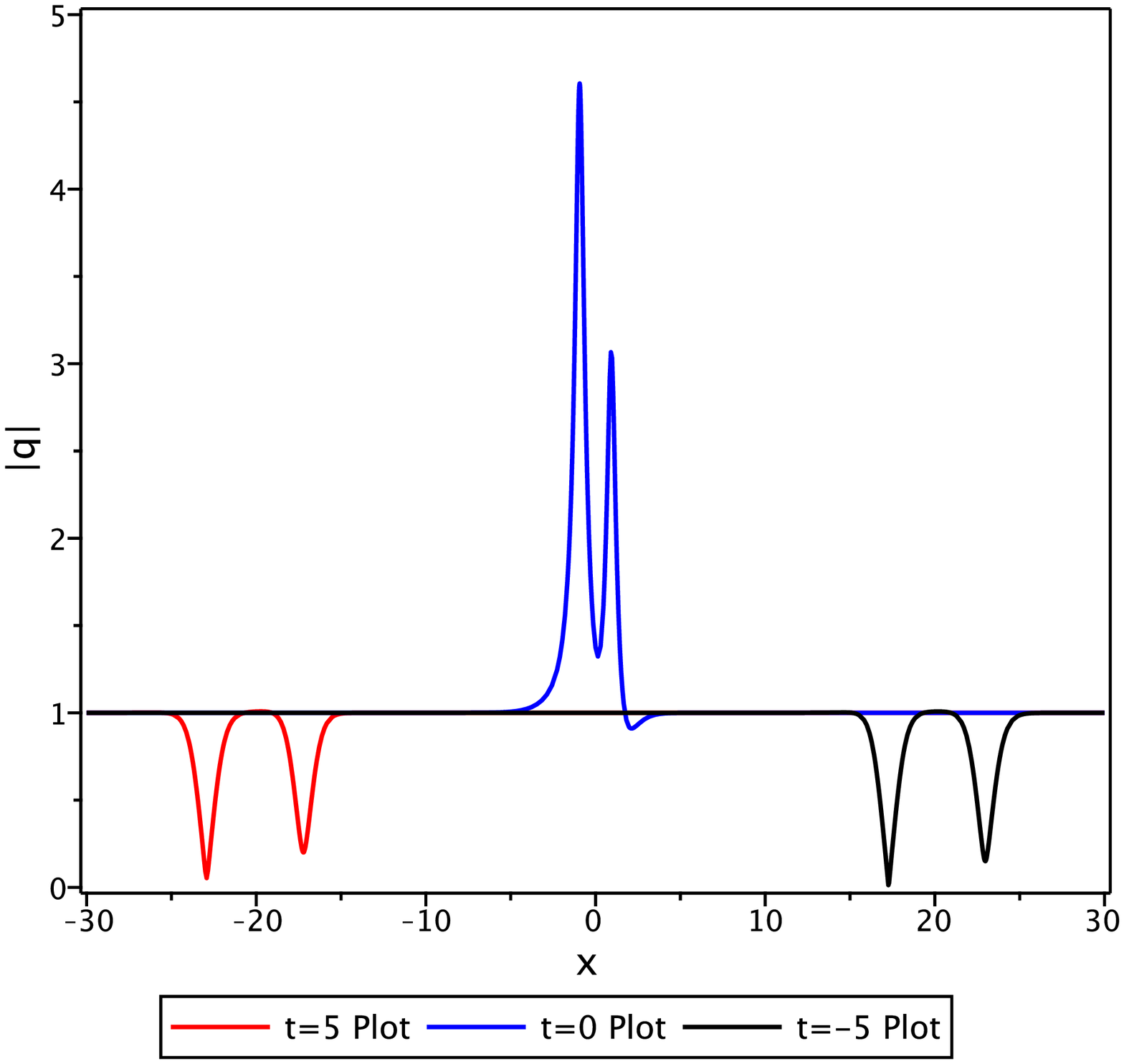}}}\\
$~~~~~~~~~~~~~~~(\textbf{a})~~~~~~~~~~~~~~~~
~~~~~~~~~~~~~~~~~~~~~~~~~(\textbf{b})~~~~~~~
~~~~~~~~~~~~~~~~~~~~~~~~~~~~(\textbf{c})$\\
\noindent { \small \textbf{Figure 8.} (Color online) The one-double poles solution for Eq.\eqref{10} with the parameters: $b_{1}=1, d_{1}=1, q_{0}=1, \vartheta_{1}=\frac{\pi}{4}, \theta_{-}=0, \delta=\frac{1}{100}, \beta=1$.
$\textbf{(a)}$ Three dimensional plot;
$\textbf{(b)}$ The density plot;
$\textbf{(c)}$ The wave propagation along the $x$-axis at $t=-5$(black), $t=0$(blue), $t=5$(red).}\\
\\
{\rotatebox{0}{\includegraphics[width=4.6cm,height=4.0cm,angle=0]{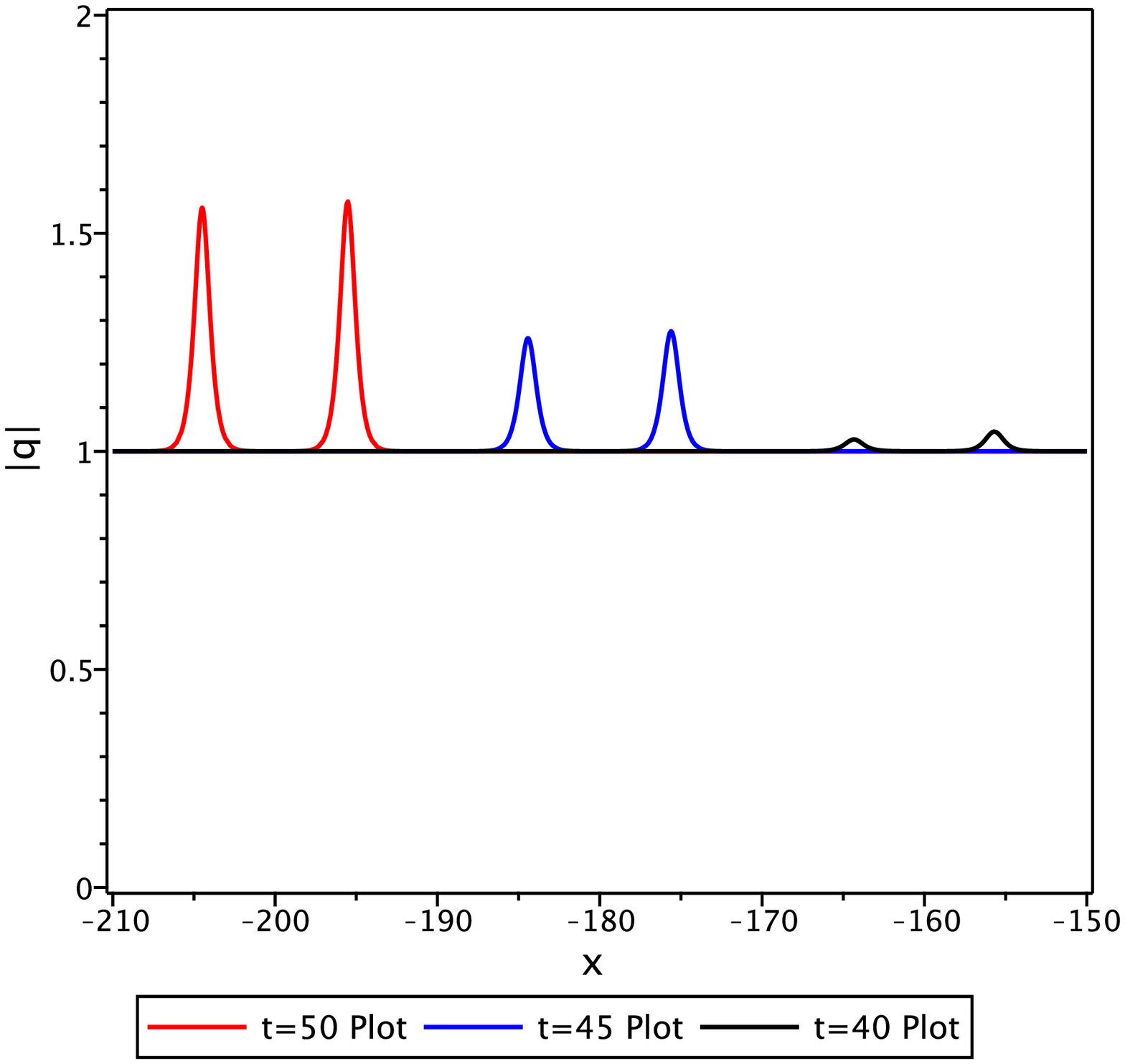}}}
~~~~
{\rotatebox{0}{\includegraphics[width=4.6cm,height=4.0cm,angle=0]{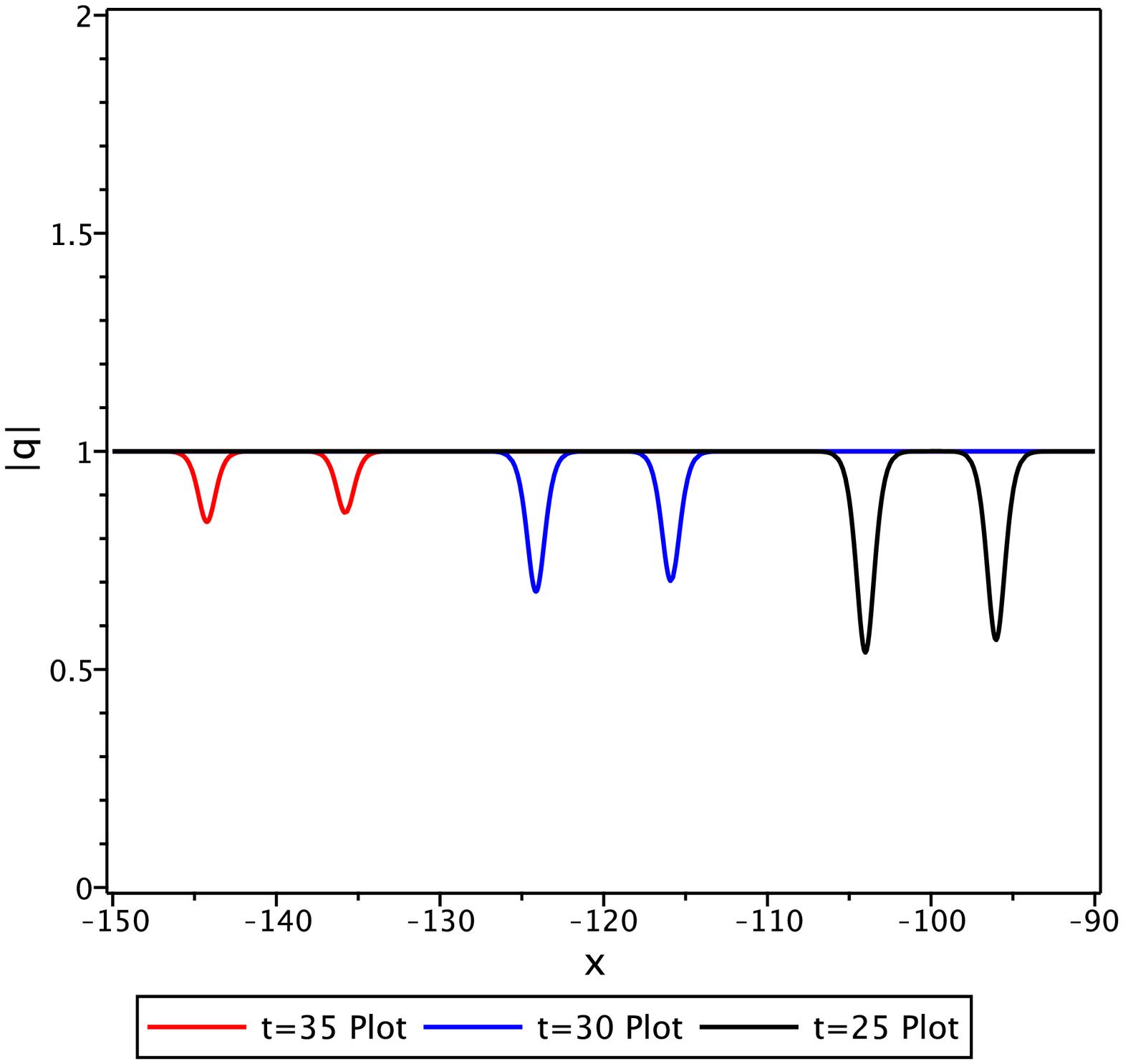}}}
~~~~
{\rotatebox{0}{\includegraphics[width=4.6cm,height=4.0cm,angle=0]{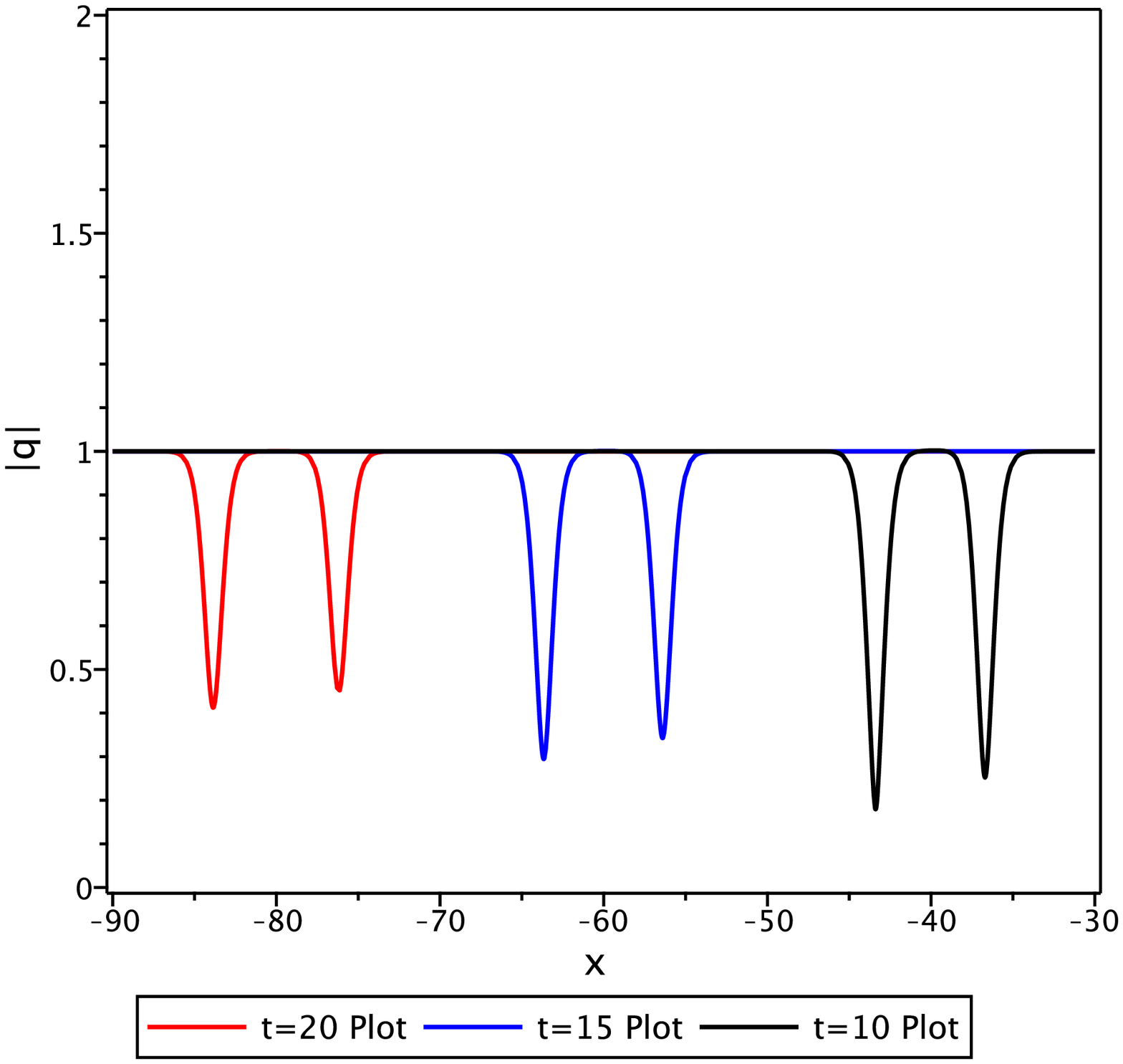}}}\\
$~~~~~~~~~~~~~~~(\textbf{a})~~~~~~~~~~~~~~~~
~~~~~~~~~~~~~~~~~~~~~~~~~(\textbf{b})~~~~~~~
~~~~~~~~~~~~~~~~~~~~~~~~~~~~(\textbf{c})$\\
{\rotatebox{0}{\includegraphics[width=4.6cm,height=4.0cm,angle=0]{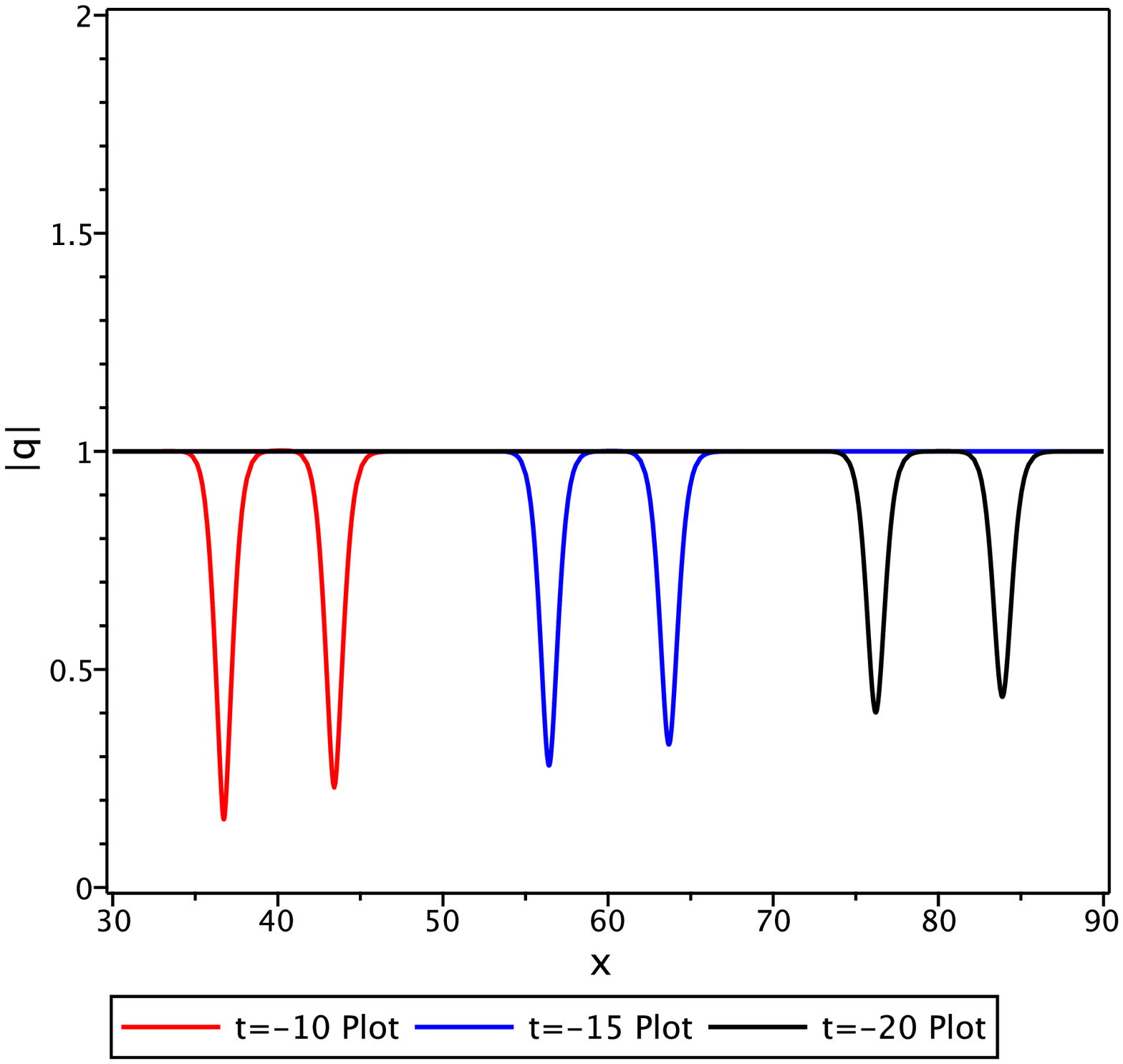}}}
~~~~
{\rotatebox{0}{\includegraphics[width=4.6cm,height=4.0cm,angle=0]{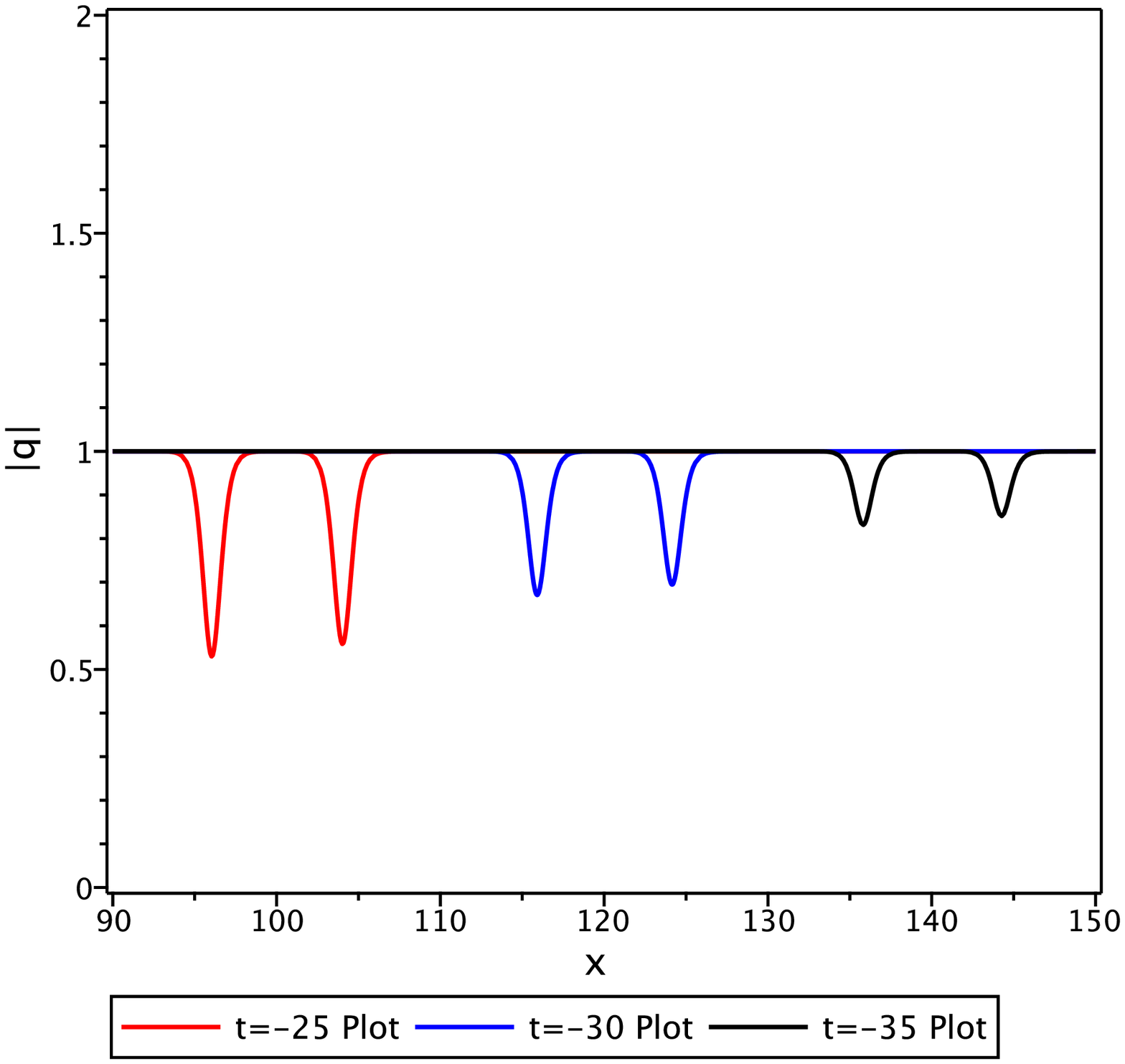}}}
~~~~
{\rotatebox{0}{\includegraphics[width=4.6cm,height=4.0cm,angle=0]{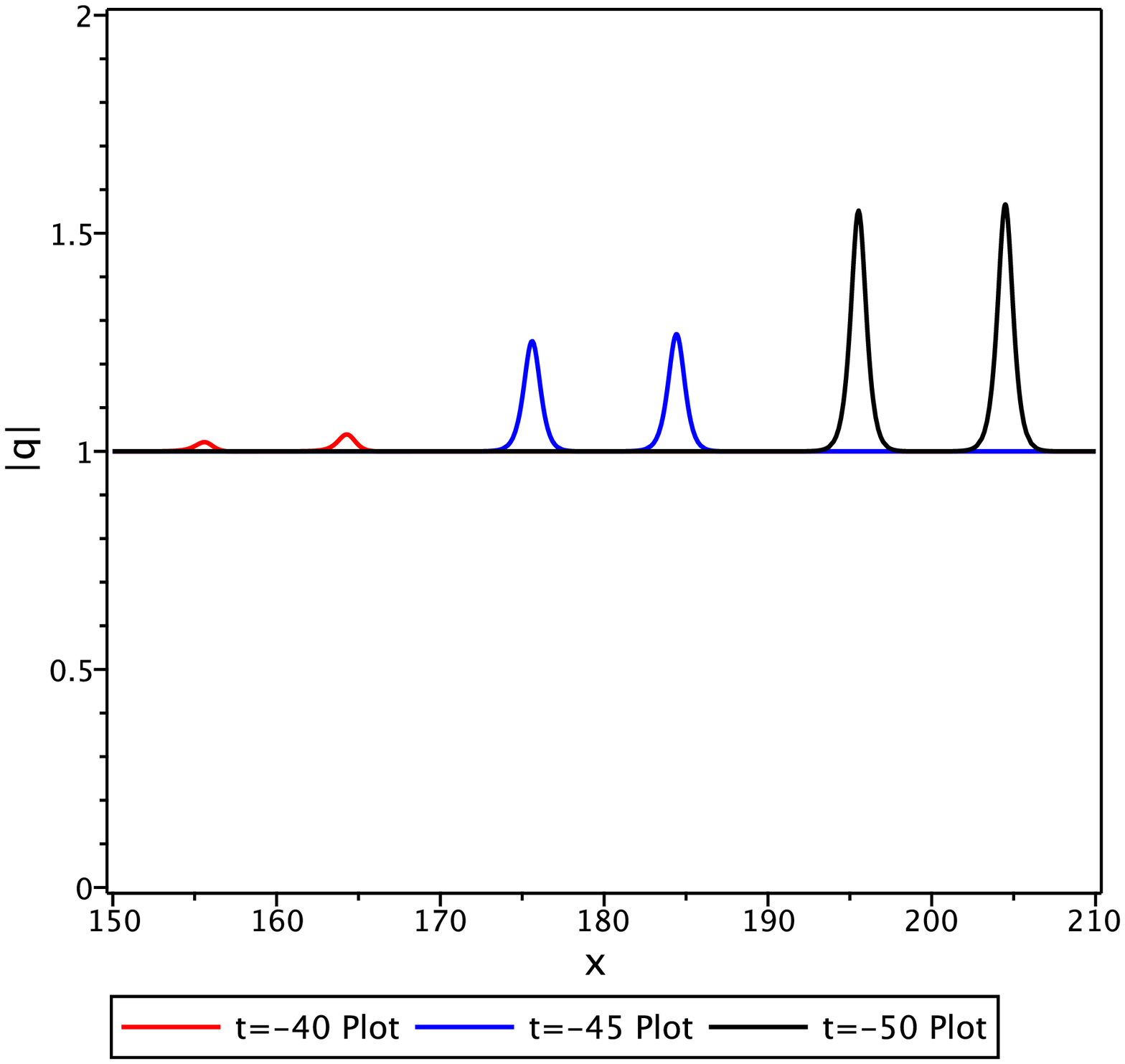}}}\\
$~~~~~~~~~~~~~~~(\textbf{d})~~~~~~~~~~~~~~~~
~~~~~~~~~~~~~~~~~~~~~~~~~(\textbf{e})~~~~~~~
~~~~~~~~~~~~~~~~~~~~~~~~~~~~(\textbf{f})$\\
\noindent { \small \textbf{Figure 9.} (Color online) The wave propagation of one-double poles solution along the $x$-axis at different time. The parameters are $b_{1}=1, d_{1}=1, q_{0}=1, \vartheta_{1}=\frac{\pi}{4}, \theta_{-}=0, \delta=\frac{1}{100}, \beta=1$.}\\

Furthermore, we analyze the asymptotic states of the one-double poles solution as $t\rightarrow \pm \infty$ under $b_{1}=1, d_{1}=1, q_{0}=1, \vartheta_{1}=\frac{\pi}{4}, \theta_{-}=0, \delta=0, \beta=1$. Through analysing the expression of the solution, we know the two characteristic curves are $x=-4t+\frac{\sqrt{2}}{2}\log(t)$ and
$x=-4t-\frac{\sqrt{2}}{2}\log(t)$, respectively. Using maple symbol calculations, we can derive the long-time asymptotic state of the one-double poles solution as moving along these two characteristic curves, given by
\begin{align}
&\mid q(x, t)\mid^{2}\rightarrow
\frac{(128-e^{2\sqrt{2}(x+4t-\frac{\sqrt{2}}{2}\log(t))})^{2}}
{(128+e^{2\sqrt{2}(x+4t-\frac{\sqrt{2}}{2}\log(t))})^{2}+512e^{2\sqrt{2}(x+4t-\frac{\sqrt{2}}{2}\log(t))}}+\notag\\
&\frac{(128-e^{-2\sqrt{2}(x+4t+\frac{\sqrt{2}}{2}\log(t))})^{2}}
{(128+e^{-2\sqrt{2}(x+4t+\frac{\sqrt{2}}{2}\log(t))})^{2}+512e^{-2\sqrt{2}(x+4t+\frac{\sqrt{2}}{2}\log(t))}},\ \mbox{as} \ t\rightarrow \infty.
\end{align}
From the above expression, It is not hard to see that the one-double poles solution reduces into the
two dark one-soliton solution as $t\rightarrow\infty$, and when $t\rightarrow\infty$, the position shift of two dark one-soliton
solution is $\sqrt{2}\log(t)$, which depends on $t$. We also select three different time $t=20, t=40, t=60$ to verify the above asymptotic expressions by numerical plotting in Fig 10. Numerical results show that the exact solution and the asymptotic solution are almost identical. 
\\
{\rotatebox{0}{\includegraphics[width=4.6cm,height=4.0cm,angle=0]{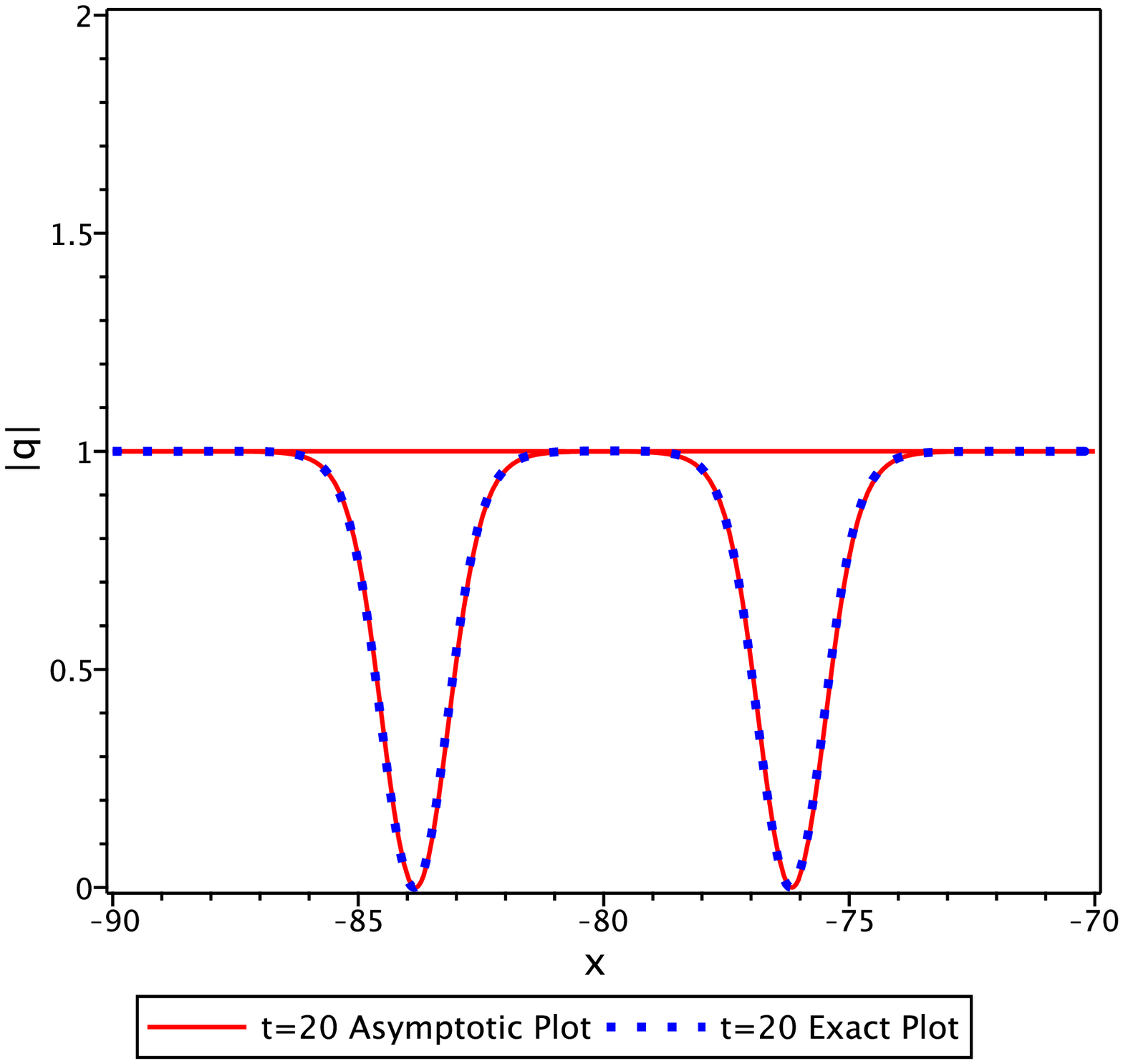}}}
~~~~
{\rotatebox{0}{\includegraphics[width=4.6cm,height=4.0cm,angle=0]{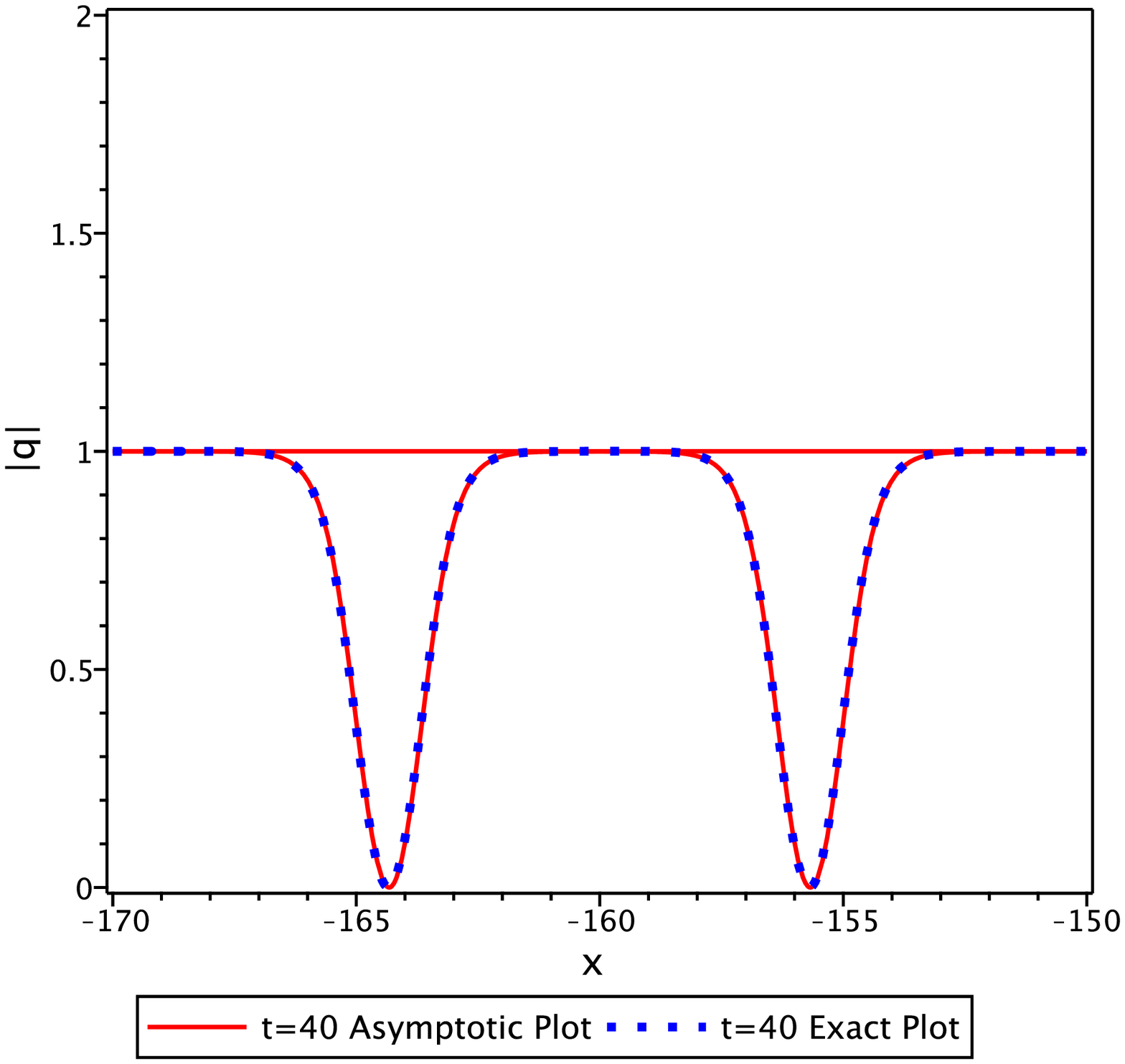}}}
~~~~
{\rotatebox{0}{\includegraphics[width=4.6cm,height=4.0cm,angle=0]{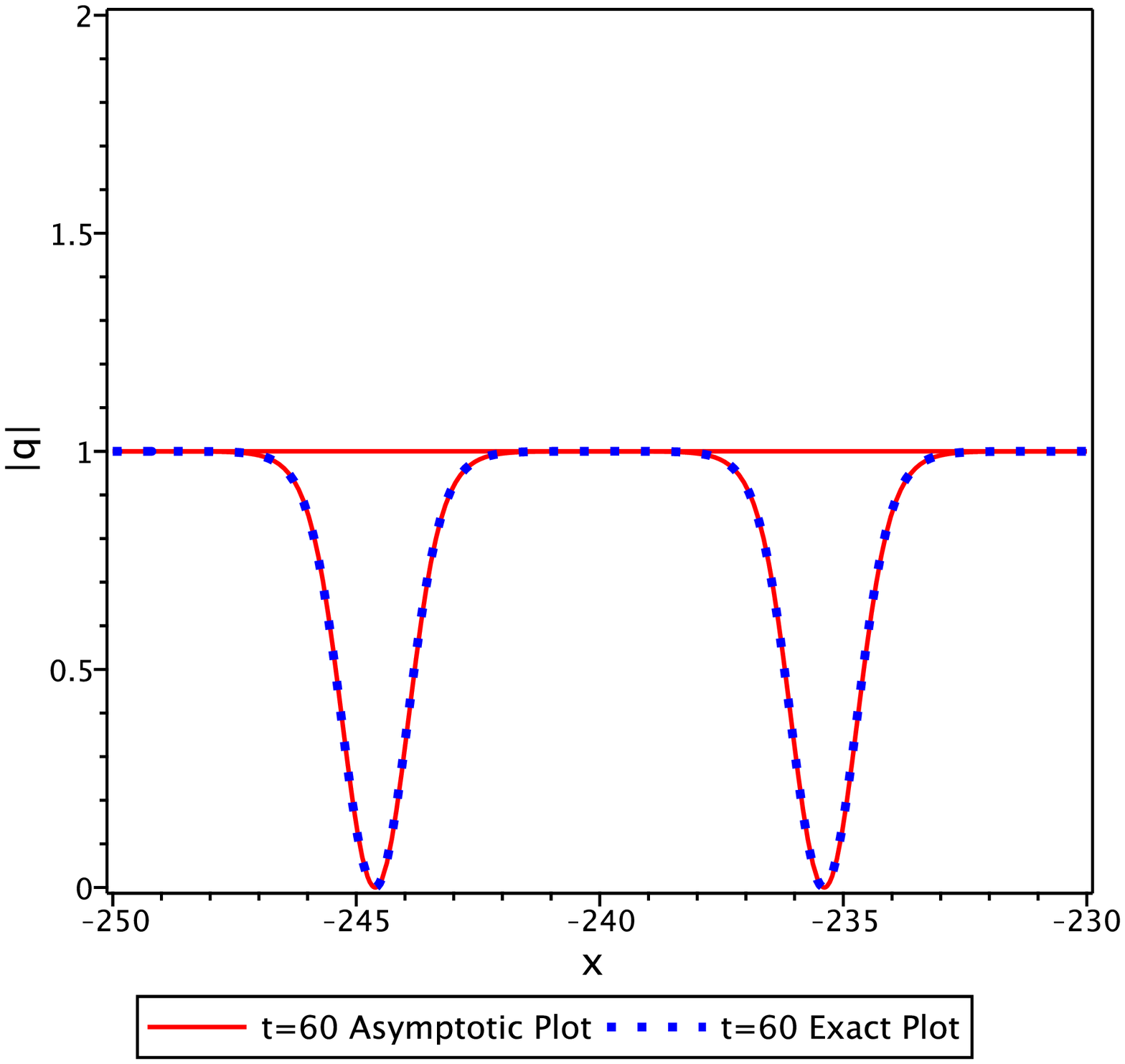}}}\\
$~~~~~~~~~~~~~~~(\textbf{a})~~~~~~~~~~~~~~~~
~~~~~~~~~~~~~~~~~~~~~~~~~(\textbf{b})~~~~~~~
~~~~~~~~~~~~~~~~~~~~~~~~~~~~(\textbf{c})$\\
\noindent { \small \textbf{Figure 10.} (Color online) The comparison between the exact solution and the
asymptotic solution at time $t=20, t=40, t=60$.}\\

\section{The data-driven soliton solutions for the nonlocal Hirota equation via PINN algorithm}
In this section, the PINN algorithm is used to learn the soliton  solutions for the nonlocal Hirota equation \eqref{10} with Dirichlet boundary conditions, given by
\begin{align}\label{47}
\left\{
\begin{array}{lr}
q_{t}+\delta\left[q_{xx}-2q^{2}q^{\ast}(-x, -t)+2q_{0}^{2}q\right]+\beta\left[q_{xxx}-6qq^{\ast}(-x,-t)q_{x}\right]=0, \\
x\in [x_{0}, x_{1}], \quad t\in [t_{0}, t_{1}], \quad q(x, t_{0})=q_{0}(x),\\
q(x_{0}, t)=q_{lb}(t),\ q(x_{1},t)=q_{ub}(t),
\end{array}
\right.
\end{align}
where $x_{0}, x_{1}$ denote the corresponding boundaries of $x, t_{0}, t_{1}$ are initial and final times of $t$. The $q_{0}(x)$ defines the initial
condition. We would apply PINN scheme
to investigate the data-driven soliton solutions of Eq.\eqref{47} with $\delta=0.01, \beta=1$.
The soliton  solutions include the dark soliton solution  and bright-dark soliton wave. These solutions have be derived in section 3,  and their corresponding dynamic behavior also have been discussed above. As presented in \eqref{46}, the precise form of these solutions can be written as follows:

Using the same parameters as Fig. 2,  the exact dark soliton solution admits
\begin{align}\label{48}
q(x,t)=\frac{-2\sqrt{2}ie^{\sqrt{2}(4t+x)}\sin(\frac{t}{50})-e^{2\sqrt{2}(4t+x)}+1}{
2\sqrt{2}ie^{\sqrt{2}(4t+x)}\cos(\frac{t}{50})-e^{2\sqrt{2}(4t+x)}+1}.
\end{align}

Using the same parameters as Fig. 4, the exact bright-dark soliton solution admits
\begin{align}\label{49}
q(x,t)=\frac{\sqrt{2}e^{4\sqrt{2}t+\sqrt{2}x-\frac{it}{50}}-
100\sqrt{2}e^{4\sqrt{2}t+\sqrt{2}x+\frac{it}{50}}-10e^{2\sqrt{2}(4t+x)}+10}{\sqrt{2}ie^{4\sqrt{2}t+\sqrt{2}x-\frac{it}{50}}+
100\sqrt{2}ie^{4\sqrt{2}t+\sqrt{2}x+\frac{it}{50}}-10e^{2\sqrt{2}(4t+x)}+10
}.
\end{align}

\subsection{The PINN  algorithm}
In this subsection, we commit  to introduce the PINN  algorithm \cite{Peng-4} for the data-driven
solutions in detail. The main idea of the PINN algorithm is to use a deep neural network to find the solutions
of Eq.\eqref{47}. Let $q(x, t)=u(x, t)+iv(x, t), q^{\ast}(-x, -t)=u(-x, -t)-iv(-x, -t)$ being its real and imaginary parts, respectively, and
then substituting them into Eq.\eqref{47}, we have
\begin{align}\label{50}
\left\{
\begin{array}{lr}
u_{t}+\beta u_{xxx}+\delta u_{xx}-6\beta[uu(-x, -t)+vv(-x,-t)]u_{x}-2\delta u^{2}u(-x, -t)\\
-6\beta[uv(-x, -t)-vu(-x,-t)]v_{x}
+2\delta[1-2vv(-x,-t)]u+2\delta v^{2}u(-x,-t)=0,\\
v_{t}+\beta v_{xxx}+\delta v_{xx}+6\beta[uv(-x, -t)-vu(-x,-t)]u_{x}-2\delta v^{2}v(-x, -t)\\
-6\beta[uu(-x, -t)+vv(-x,-t)]v_{x}
+2\delta[1-2uu(-x,-t)]v+2\delta u^{2}v(-x,-t)=0.
\end{array}
\right.
\end{align}
Then the physics-informed neural networks $f_{u}(x, t), f_{v}(x, t)$ can be defined as
\begin{align}\label{51}
\left\{
\begin{array}{lr}
f_{u}:=u_{t}+\beta u_{xxx}+\delta u_{xx}-6\beta[uu(-x, -t)+vv(-x,-t)]u_{x}-2\delta u^{2}u(-x, -t)\\
-6\beta[uv(-x, -t)-vu(-x,-t)]v_{x}
+2\delta[1-2vv(-x,-t)]u+2\delta v^{2}u(-x,-t),\\
f_{v}:=v_{t}+\beta v_{xxx}+\delta v_{xx}+6\beta[uv(-x, -t)-vu(-x,-t)]u_{x}-2\delta v^{2}v(-x, -t)\\
-6\beta[uu(-x, -t)+vv(-x,-t)]v_{x}
+2\delta[1-2uu(-x,-t)]v+2\delta u^{2}v(-x,-t),
\end{array}
\right.
\end{align}
of which $u(x, t; w, b), v(x, t; w, b)$ represent the output of the neural network, which is an approximation of the solution $q(x, t)$. Applying automatic differentiation mechanism in $u(x, t; w, b), v(x, t; w, b)$, the residual PINN $f_{u}(x, t), f_{v}(x, t)$ are given\cite{PuChen420}.
Then, the multi-hidden-layer deep neural network is used to train  the network parameters $w, b$. To updates training parameters, we construct the following Loss functions which can be minimized via  using L-BFGS optimization method \cite{wangya44}
\begin{align}\label{52}
Loss_{\Theta}=Loss_{u}+Loss_{v}+Loss_{f_{u}}+Loss_{f_{v}},
\end{align}
with
\begin{align}\label{53}
\left\{
\begin{array}{lr}
Loss_{u}=\frac{1}{N_{q}}\sum_{i=1}^{N_{q}}|\hat{u}(x_{q}^{i},t_{q}^{i})-u^{i}|^{2},\\ Loss_{v}=\frac{1}{N_{q}}\sum_{i=1}^{N_{q}}|\hat{v}(x_{q}^{i},t_{q}^{i})-v^{i}|^{2},\\
%Loss_{u_{2}}=\frac{1}{N_{q}}\sum_{i=1}^{N_{q}}|\hat{u}_{2}(x_{q}^{i},t_{q}^{i})-u_{2}^{i}|^{2},\\ %Loss_{v_{2}}=\frac{1}{N_{q}}\sum_{i=1}^{N_{q}}|\hat{v}_{2}(x_{q}^{i},t_{q}^{i})-v_{2}^{i}|^{2},\\
Loss_{f_{u}}=\frac{1}{N_{f}}\sum_{l=1}^{N_{f}}|f_{u}(x_{f}^{l},t_{f}^{l})|^{2},\\
Loss_{f_{v}}=\frac{1}{N_{f}}\sum_{l=1}^{N_{f}}|f_{v}(x_{f}^{l},t_{f}^{l})|^{2},\\
%Loss_{f_{u_{2}}}=\frac{1}{N_{f}}\sum_{l=1}^{N_{f}}|f_{u_{2}}(x_{f}^{l},t_{f}^{l})|^{2},\\
%Loss_{f_{v_{2}}}=\frac{1}{N_{f}}\sum_{l=1}^{N_{f}}|f_{v_{2}}(x_{f}^{l},t_{f}^{l})|^{2},
\end{array}
\right.
\end{align}
where $\{x_{q}^{i}, t_{q}^{i}, u^{i}\}_{i=1}^{N_{q}}$ and $\{x_{q}^{i}, t_{q}^{i}, v^{i}\}_{i=1}^{N_{q}}$  denote the sampling initial and boundary value training data, respectively. $\{x_{f}^{l}, t_{f}^{l}\}_{l=1}^{N_{f}}$  denote the sampling collocation points for $f_{u}$ and $f_{v}$. On the one hand, the loss function \eqref{52} makes the learning solution approximate the exact one, on the other hand, it makes the hidden $\hat{u}, \hat{v}$ satisfy the target nonlinear partial differential equation \eqref{47}. To understand PINN algorithm more intuitively, the flow chart of PINN algorithm for nonlocal Hirota equation is shown in Fig. 11, in which neural network and physical information part can be seen. The aim is to optimize the loss function using the neural network part as well as the physics information part.\\
{\centerline{\includegraphics[width=14.0cm,height=6.0cm,angle=0]{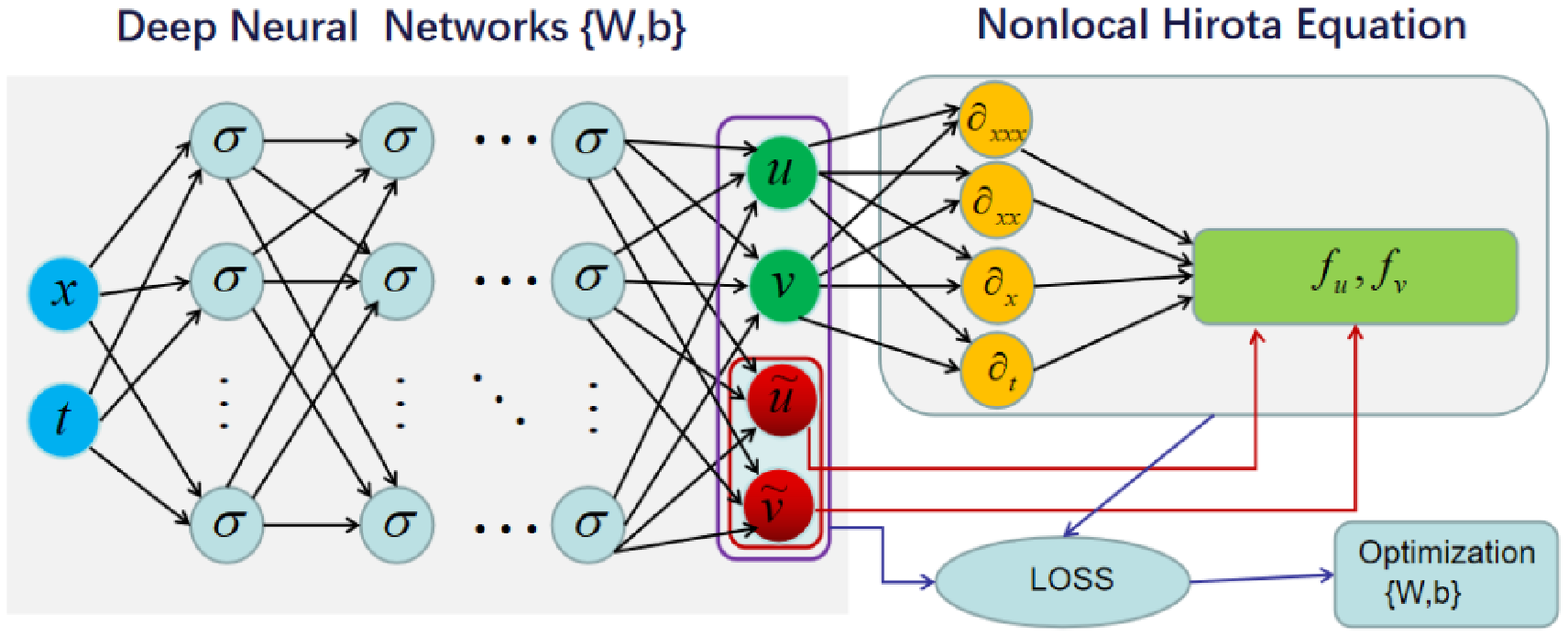}}}\\
\noindent { \small \textbf{Figure 11.} (Color online) The PINN scheme solving the nonlocal Hirota equation, $\tilde{u}=u(-x, -t), \tilde{v}=v(-x, -t)$.}\\

To obtain the data-driven soliton solution for the nonlocal Hirota equation \eqref{47},  we choose the PINN which contains 9-hidden-layer  neural network with each layer having 40 neurons.  Their activation functions  both are the hyperbolic tangent (tanh). The all codes are written by Python 3.7 and run on Tensorflow 1.15, and the corresponding hardware is a HP Laptop 14s-dr2xxx with 2.40 GHz 4-core 11th Gen Intel Core i5-1135G7 and 16 GB memory.

\subsection{The data-driven dark soliton solution}
First, we take $[x_{0}, x_{1}]=[-3, 3]$ and $[t_{0}, t_{1}]=[-3,3]$ in Eq.\eqref{47} as the boundary conditions, and select the following initial condition arising from the dark soliton solution \eqref{48}
\begin{align}\label{54}
q_{0}(x)=q(x, -3),
\end{align}
and the Dirichlet boundary conditions for Eq.\eqref{47}
\begin{align}\label{55}
q_{lb}(t)=q(-3,t), \ q_{ub}(t)=q(3,t), \ t\in \left[-3, 3\right].
\end{align}

In terms of MATLAB software, the traditional finite difference method can be used to capture the original training data by dispersing Eq.\eqref{48} with dividing spatial region $[-3, 3]$ into 1500 points and time region $[-3, 3]$ into 1000 points.  The original training data contains the initial boundary data and the inner points. Here, we choose $N_{q}=1500$  as the random sample points from initial boundary data and $N_{f}=30000$ as random collocation points from the inner points based on the Latin hypercube sampling (LHS) method \cite{PuChen470}. Processing these obtained training data in the PINN scheme,  the  data-driven dark soliton solution $q(x, t)$  can be eventually learned, which has a 2.695814e-04 $\mathbb{L}_{2}$-norm error compared with the exact one. The total learning process executes 927 times and takes about 1262.4172 seconds. The corresponding dynamic behaviors are displayed in following Figs. 12 and 13.\\
{\centerline{\includegraphics[width=14.0cm,height=7.0cm,angle=0]{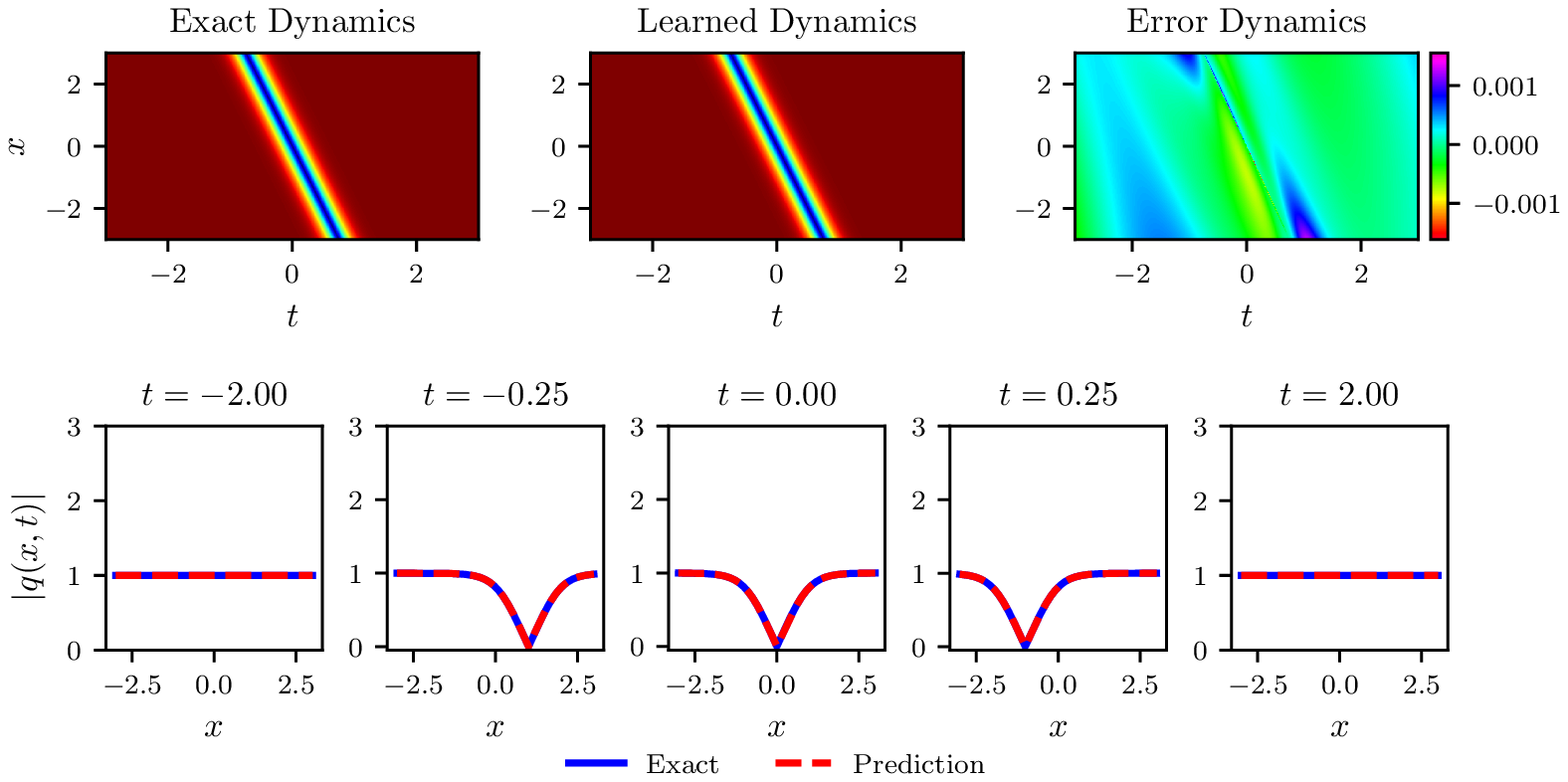}}}\\
\noindent { \small \textbf{Figure 12.} (Color online) The data-driven dark soliton solution $q(x, t)$ for nonlocal Hirota equation \eqref{47}:
The exact, learned and error dynamics density plots, and the sectional drawings which contain the learned and explicit dark soliton solution $q(x, t)$ at the  five distinct times $t=-2, t=-0.25, t=0, t=0.25, t=2$.}\\

{\rotatebox{0}{\includegraphics[width=5.6cm,height=5.2cm,angle=0]{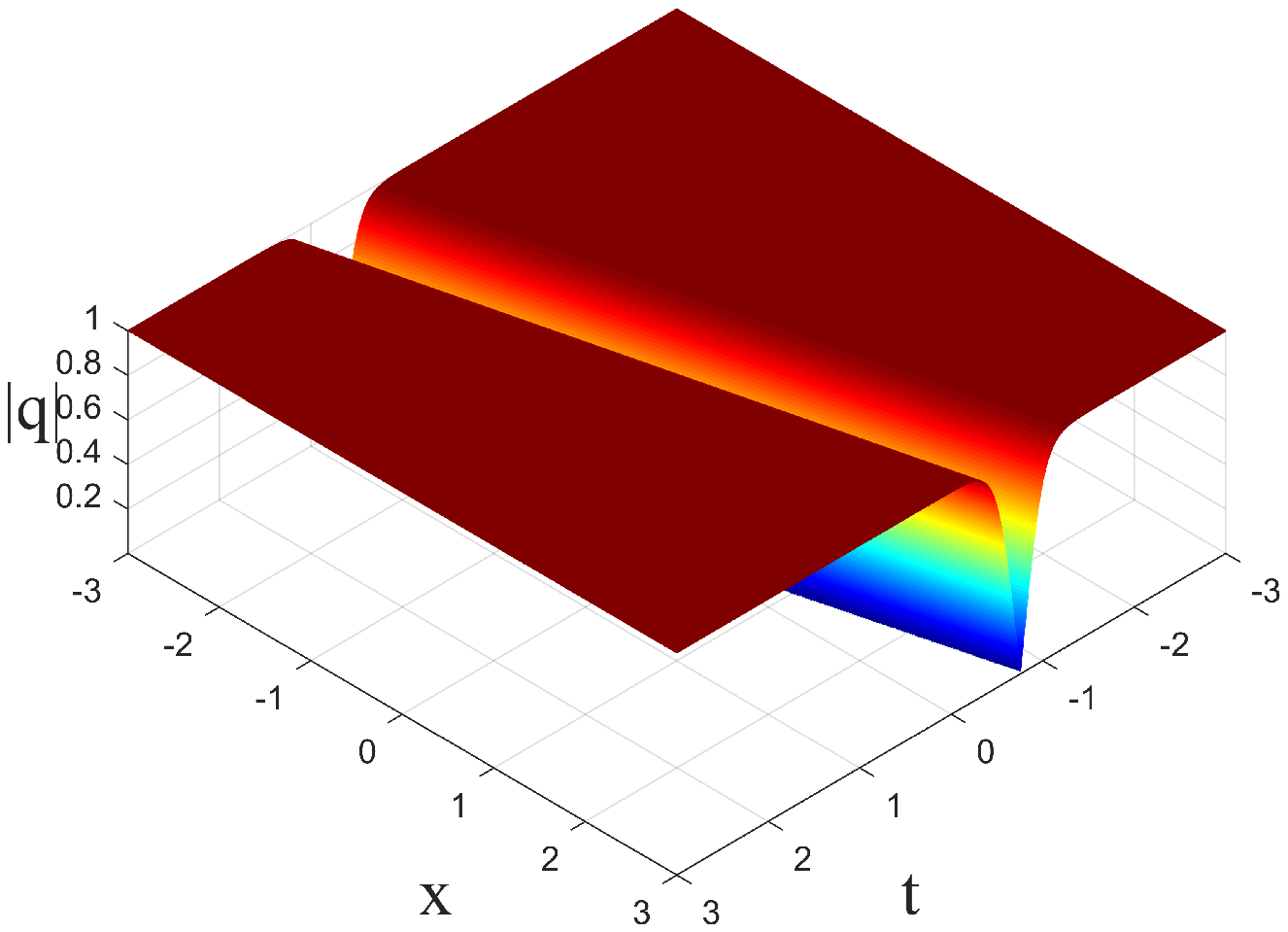}}}
~~~~~~~~~~~~~~~~
{\rotatebox{0}{\includegraphics[width=5.0cm,height=4.4cm,angle=0]{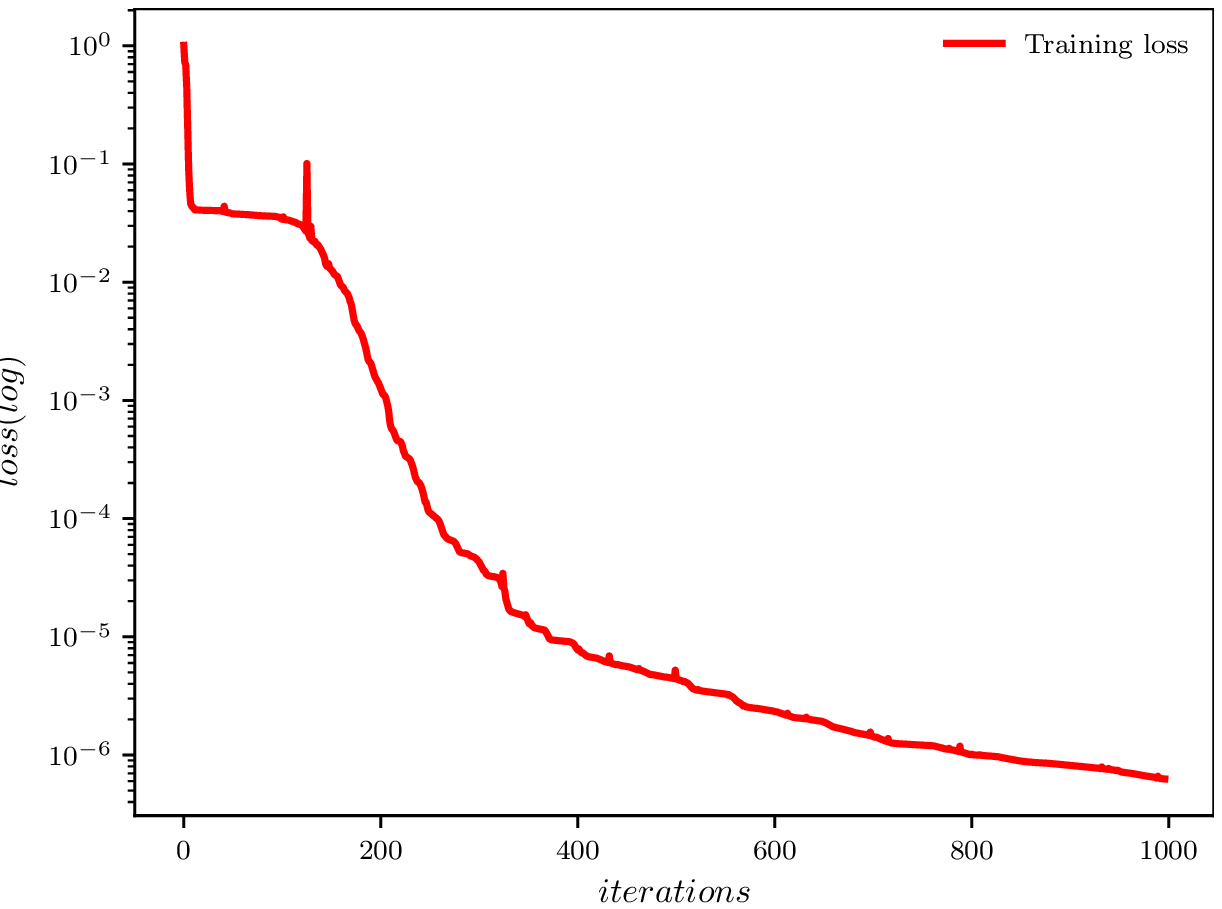}}}\\
$~~~~~~~~~~~~~~~~~~~~~~~~~~~(\textbf{a})~~~~~~~~~~~~~~~~~~~~~~~~~~~~~
~~~~~~~~~~~~~~~~~~~~~~~~~~~~(\textbf{b})$\\
\noindent { \small \textbf{Figure 13.} (Color online) The data-driven dark soliton solution $q(x, t)$ for nonlocal Hirota equation \eqref{47}:
$\textbf{(a)}$ The three-dimensional plot;
$\textbf{(b)}$ The loss curve figure.}\\

In Fig. 12, the wave propagation pattern along the $x$-axis and the density plots for the data-driven dark soliton solution are
shown respectively. From the Fig.12, it is easy to find that the
error range is about -0.001 to 0.001 between  the learned dynamics
and error dynamics. This fact can verify the simulation is pretty good. The Fig. 13 is the three-dimensional plot and loss curve figure of  data-driven dark soliton solution.

\subsection{The data-driven bright-dark soliton solution}
For data-driven periodic wave solution, we let $[x_{0}, x_{1}]=[-5, 5]$ and $[t_{0}, t_{1}]=[-5,5]$ in Eq.\eqref{47} as the boundary conditions, and select the following initial condition arising from the bright-dark wave solution\eqref{49}
\begin{align}\label{56}
q_{0}(x)=q(x, -5),
\end{align}
and the Dirichlet boundary conditions
\begin{align}\label{57}
q_{lb}(t)=q(-5,t), \ q_{ub}(t)=q(5,t), \ t\in \left[-5, 5\right].
\end{align}

Carrying out the same process as subsection 4.2,  the data-driven bright-dark soliton solution is generated successfully. The results of the experiment show that the $\mathbb{L}_{2}$-norm error between learning solution $q(x, t)$ and exact one is 8.243617e-04, and the whole learning process iterates 1807 times with costing 2737.0690 seconds. The main dynamic behaviors for the data-driven bright-dark soliton solution are plotted in Figs.14 and 15. From these plots, we also find the learning effect is pretty good with a very small error and a rapidly decaying Loss curve.
\\
{\centerline{\includegraphics[width=14.0cm,height=7.0cm,angle=0]{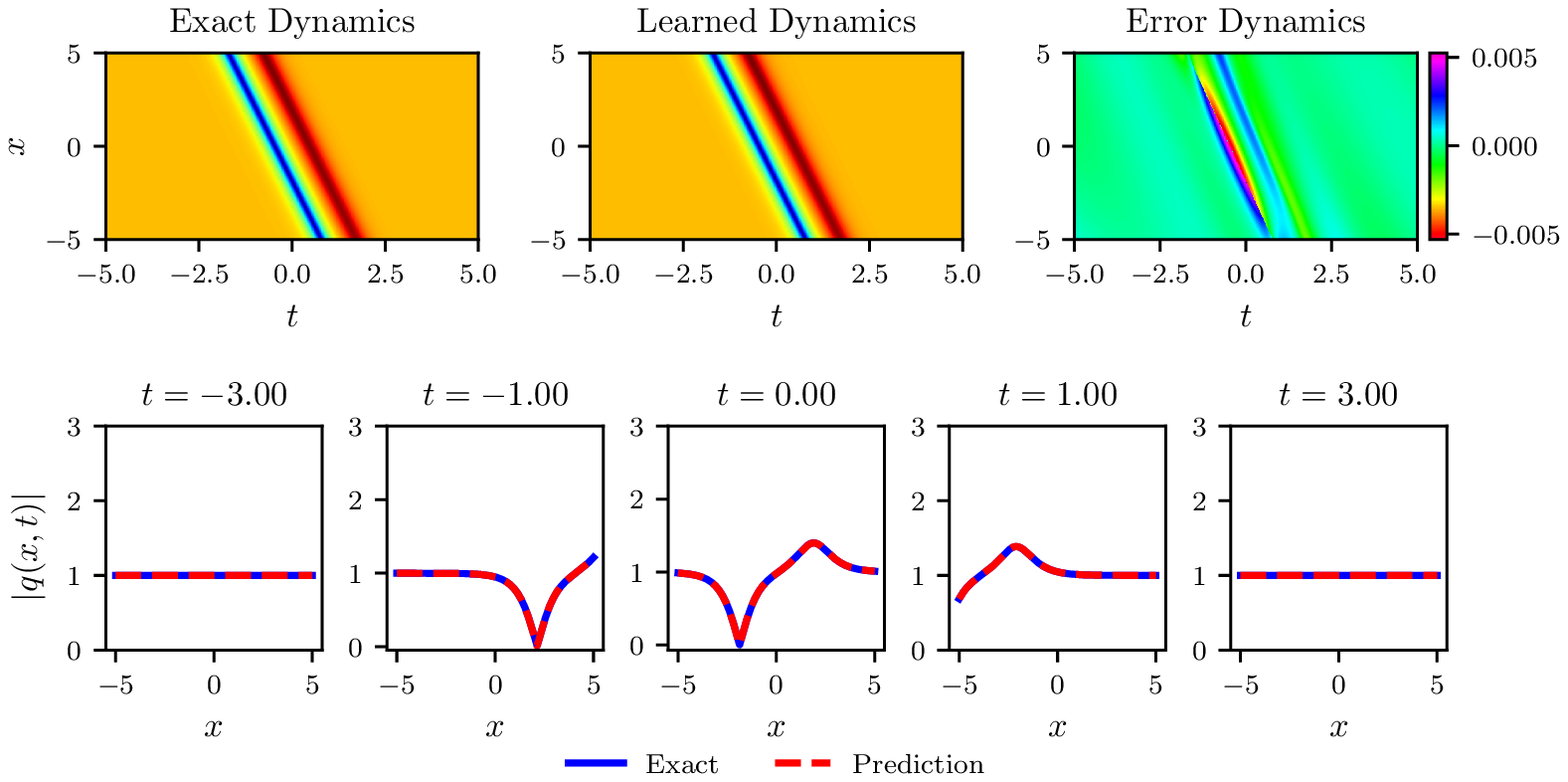}}}\\
\noindent { \small \textbf{Figure 14.} (Color online) The data-driven bright-dark soliton solution $q(x, t)$ for nonlocal Hirota equation \eqref{47}:
The exact, learned and error dynamics density plots, and the sectional drawings which contain the learned and explicit bright-dark soliton solution $q(x, t)$ at the  five distinct times $t=-3, t=-1, t=0, t=1, t=3$.}\\

{\rotatebox{0}{\includegraphics[width=5.6cm,height=5.2cm,angle=0]{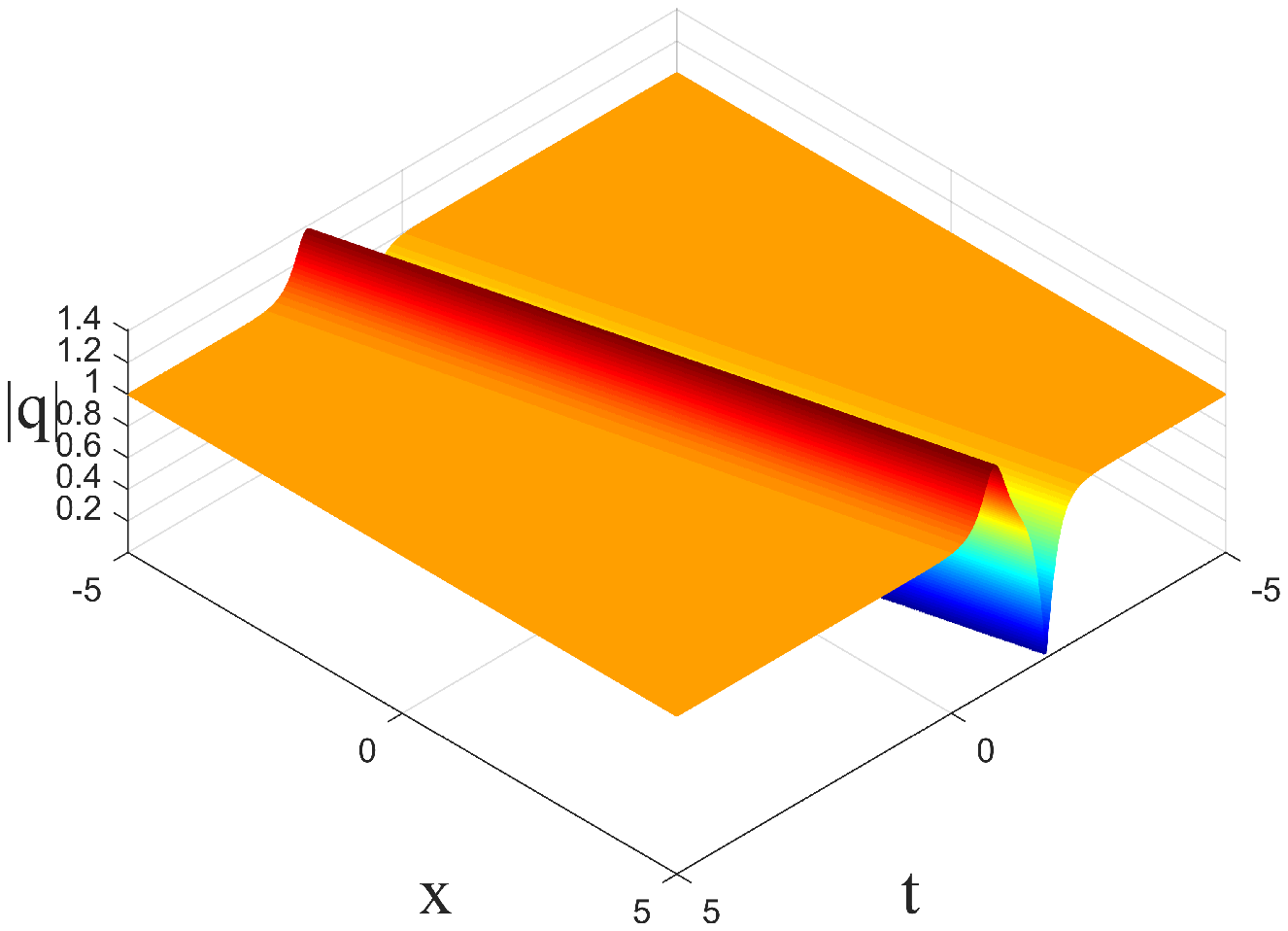}}}
~~~~~~~~~~~~~~~~
{\rotatebox{0}{\includegraphics[width=5.0cm,height=4.4cm,angle=0]{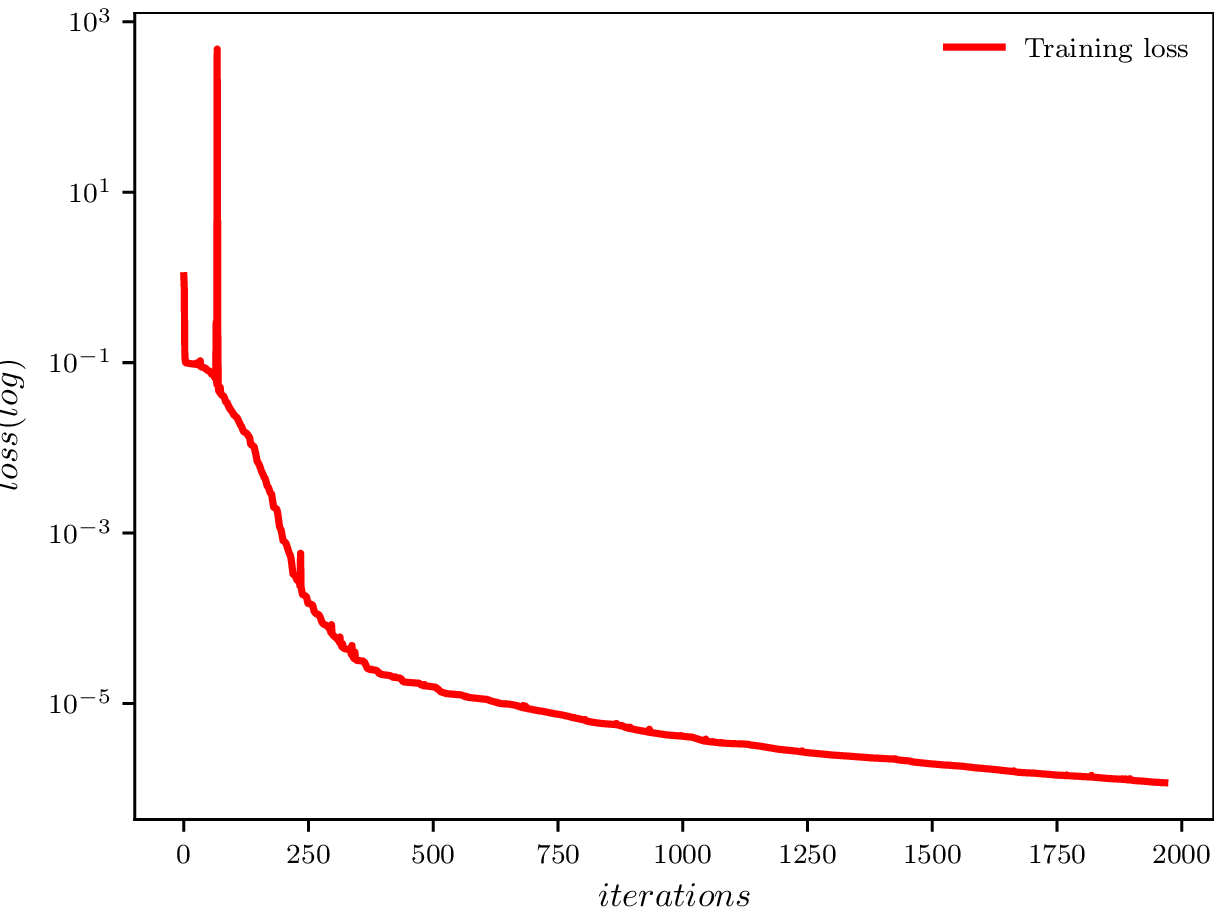}}}\\
$~~~~~~~~~~~~~~~~~~~~~~~~~~~(\textbf{a})~~~~~~~~~~~~~~~~~~~~~~~~~~~~~
~~~~~~~~~~~~~~~~~~~~~~~~~~~~(\textbf{b})$\\
\noindent { \small \textbf{Figure 15.} (Color online) The data-driven bright-dark soliton solution $q(x, t)$ for nonlocal Hirota equation \eqref{47}:
$\textbf{(a)}$ The three-dimensional plot;
$\textbf{(b)}$ The loss curve figure.}\\

\subsection{The PINN algorithm for the data-driven parameter discovery}
In this section,  we put our attention to the problem of data-driven discovery of nonlocal Hirota equation \eqref{47} by using PINN  algorithm. Our goal is to identify the parameters $\delta, \beta$  in terms of the dark soliton solution \eqref{48}. In the same way, the physics-informed neural networks $f_{u}(x, t), f_{v}(x, t)$ for the equation \eqref{47} are given in \eqref{51}.

Using the Latin Hypercube Sampling,  a training data set can be generated through  selecting randomly $N_{q}=1500$  as the initial boundary data and $N_{f}=30000$  as the collocation points with the aid of the exact soliton solution \eqref{48} with $\delta=0.01, \beta=1$ and $(x, t) \in [-3, 3]\times [-3, 3]$. In terms of the obtained training data set, using a 9-hidden-layer deep PINN with 40 neurons
per layer, the data-driven parameters $\delta, \beta$ can be discovered. The corresponding  results are summarized in Table 1. We observe that the PINN is able to correctly identify the unknown parameters with very high accuracy when the training data was
corrupted without noise. Specifically, as we can see in Table 1, at the case of 0.005 noise and 0.01 noise, the error of parameters $\delta$ and $\beta$ is still receivable, which illustrates that the predictions remain robust. Of course, we can also find that noise has a bad effect on the error value of parameters.

\begin{table}[htbp]
  \caption{Data-driven parameter discovery of $\delta, \beta$ in the sense of dark soliton}
  \label{Tab:bookRWCal}
  \centering
  \begin{tabular}{l|cccc}
  \toprule
  \diagbox{\textbf{Noise}}{\textbf{Parameter}} & $\delta$ & error of $\delta$ & $\beta$ & error of $\beta$\\
  \hline
  Correct parameter   & 0.01& 0& 1& 0\\
  Without noise   &0.00976523 & 2.34770$\times 10^{-2}$ & 0.9998441 &1.559$\times 10^{-4}$\\
  With a 0.005  noise   &0.01055261&5.52606$\times 10^{-2}$ & 0.9998072&1.928$\times 10^{-4}$\\
  With a 0.01  noise   &0.00934824&6.51762$\times 10^{-2}$ & 0.9998599&1.401$\times 10^{-4}$\\
  \bottomrule
  \end{tabular}
\end{table}

The variation of unknown parameters and loss functions with iteration is analyzed when different noises are used in inverse problems. Figs. 16(a) and (b) show the changes of unknown parameters with iteration under different noises.
We find that the unknown parameters fluctuate less in the absence of noise, but the parameters fluctuate more as the noise increases.
Fig. 16(c) depicts the changes of loss functions for different noises as the number of iterations increases. It indicates
that the convergence effect becomes worse and worse with the increase of noise. Therefore, we can conclude that when
discovering the physical parameters of the model, the less noise the better the training effect.
\\
{\rotatebox{0}{\includegraphics[width=4.6cm,height=4.0cm,angle=0]{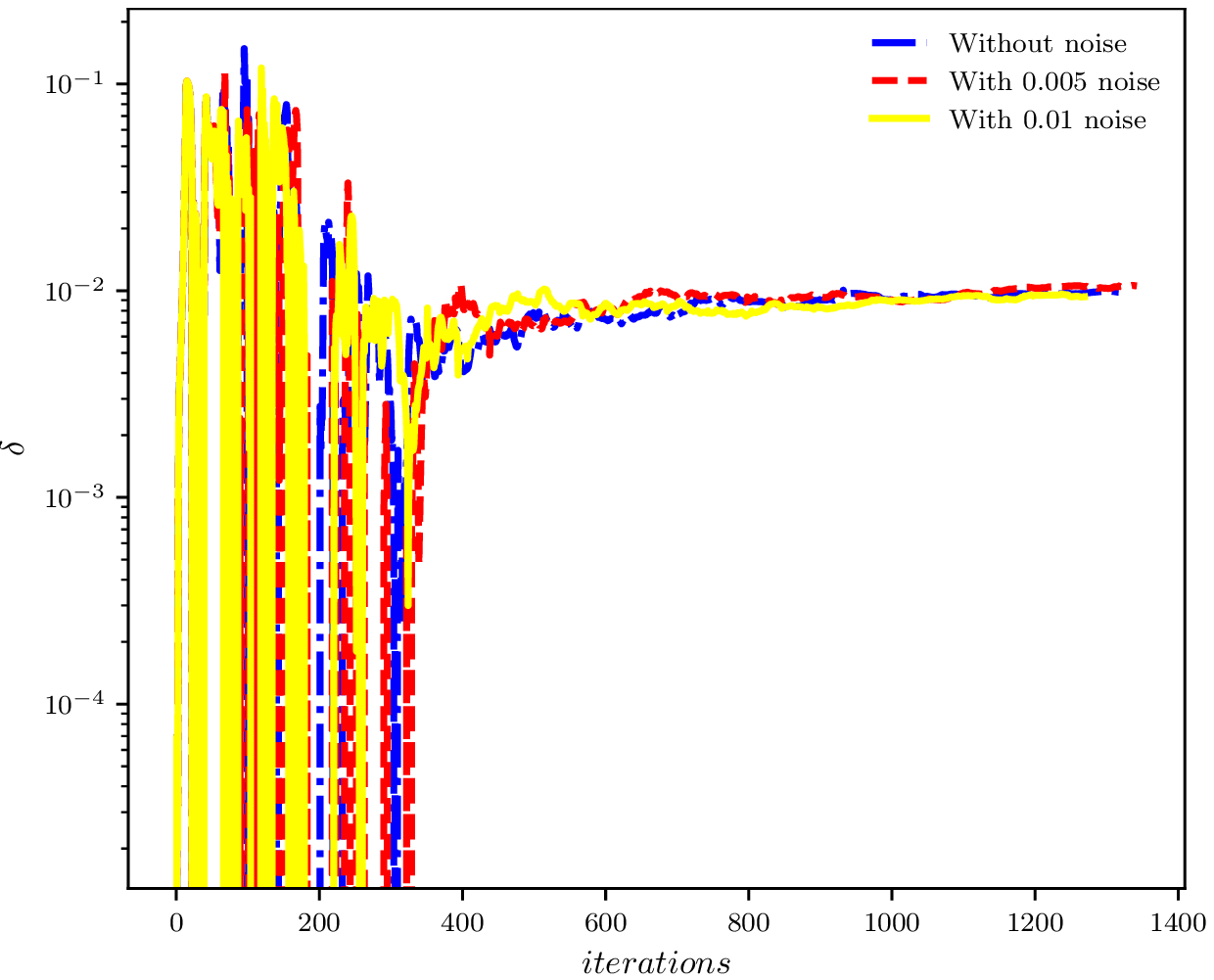}}}
~~~~
{\rotatebox{0}{\includegraphics[width=4.6cm,height=4.0cm,angle=0]{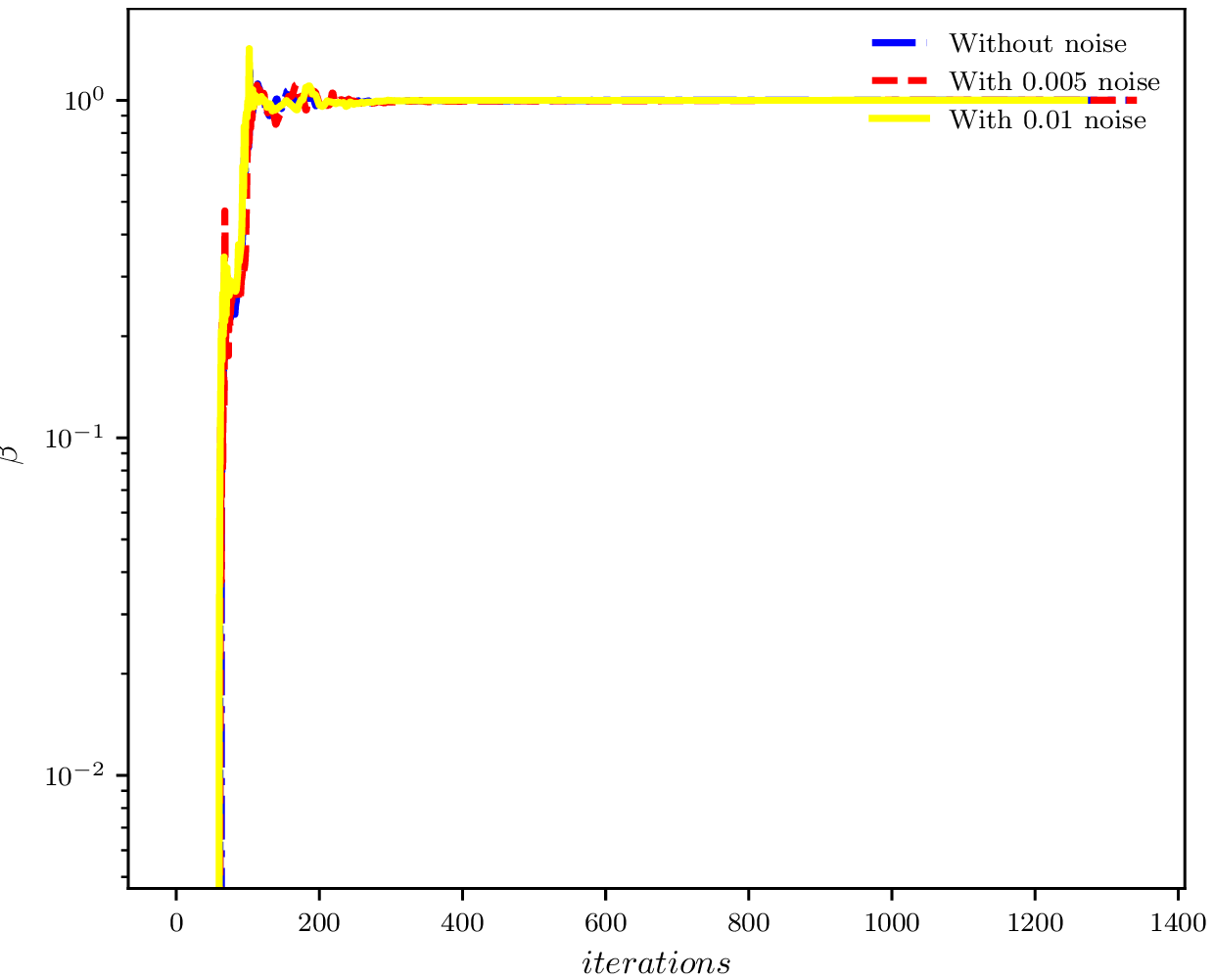}}}
~~~~
{\rotatebox{0}{\includegraphics[width=4.6cm,height=4.0cm,angle=0]{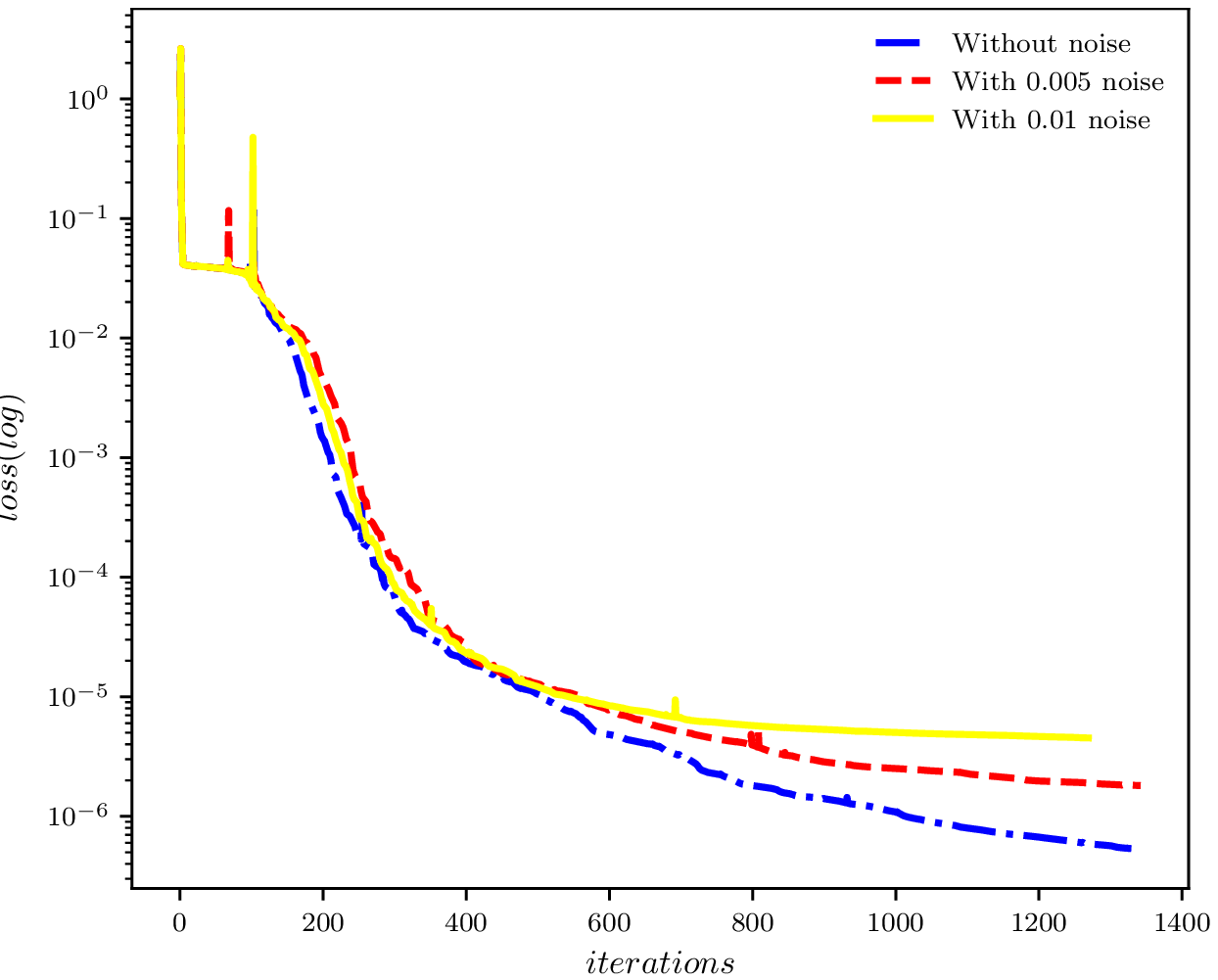}}}\\
$~~~~~~~~~~~~~~~(\textbf{a})~~~~~~~~~~~~~~~~
~~~~~~~~~~~~~~~~~~~~~~~~~(\textbf{b})~~~~~~~
~~~~~~~~~~~~~~~~~~~~~~~~~~~~(\textbf{c})$\\
\noindent { \small \textbf{Figure 16.} (Color online)$\textbf{(a, b)}$ the variation of unknown parameters $\delta, \beta$ and $\textbf{(c)}$  the variation of loss function with the different noise.}\\

\section{Conclusion}
In this paper, we have applied the RH method to discuss the nonlocal Hirota equation with NZBCs. Through solving the RH problem at the case of simplify poles, we have given out the general $N$-soliton solutions for the nonlocal Hirota equation under NZBCs.  The critical technique shown in this work is to eliminate the properties of singularities via subtracting the residue from the original non-regular RH problem when reflection coefficients have simplify poles. For the case of double poles, we also need to subtract the extra coefficient $L_{2}$. Additionally, the asymptotic value of jump matrix is subtracted from the original non-regular RH problem. Then the regular RH problem can be displayed, which can be solved by Plemelj formula. Finally, the $N$-simplify poles and  $N$-double poles solutions can be derived by using the solution of RH problem to reconstruct potential function. Compared with the local Hirota equation, the Symmetry reductions of Just solutions and scattering matrix is different, which results in a disparate discrete spectra distribution. The dynamical patterns of one-simplify pole solution with different parameters and one-double pole solution have been discussed in detail. Especially, some novel dynamic behaviors have been found for the nonlocal Hirota equation and the asymptotic state of one-double poles solution was discussed.  In addition, we will study the long time asymptotic behaviors for the nonlocal Hirota equation with NZBCs via the Deift-Zhou method in another paper.

Additionally, we have also studied the data-driven soliton solutions and parameters discovery
for the nonlocal Hirota equation with Dirichlet boundary conditions via the PINN method. Remarkably, due to the nonlocal Hirota equation has
$\mathcal{PT}$ symmetry term, it is quite different with the local Hirota equation. Through adding the nonlocal term into the NN, we can successfully handle the $\mathcal{PT}$ symmetry term and give out the data-driven soliton solutions and the parameter prediction
for the nonlocal Hirota equation.
Our results indicate that the deep learning can be applied to solve nonlocal  integrable systems.

\section*{Acknowledgements}
\hspace{0.3cm}
This work was supported by the project is supported by National Natural Science Foundation of China(No.12175069)
and Science and Technology Commission of Shanghai Municipality (No.21JC1402500 and
No.18dz2271000).

\end{document}